\documentclass[aps,pre,twocolumn,reprint,amsmath,amssymb,showpacs,nofootinbib,
floatfix,eqsecnum]{revtex4-1}
\pdfoutput=1

\usepackage{mathbbol}
\usepackage[pdftex]{graphicx}
\usepackage{hyperref}

\newlength{\onepicwidth} 
\newlength{\threepicwidth} 

\DeclareMathOperator\diag{diag}
\DeclareMathOperator\arccot{arccot}
\DeclareMathOperator\arccsch{arccsch}
\DeclareMathOperator\erf{erf}
\DeclareMathOperator\erfc{erfc}
\DeclareMathOperator\erfi{erfi}
\DeclareMathOperator\Ei{Ei}
\DeclareMathOperator\Si{Si}

\DeclareMathOperator\EVD{EVD}

\newcommand\1{\mathbb 1}
\newcommand\0{\mathbb 0}
\newcommand\mI{\mathcal I}
\newcommand\ev[1]{\left\langle#1\right\rangle}
\newcommand{\two}{$2 \times 2$}
\newcommand{\four}{$4 \times 4$}
\newcommand{\eq}{Eq.~}
\newcommand{\eqs}{Eqs.~}
\newcommand{\fig}{Fig.~}
\newcommand{\sect}{Sec.~}
\newcommand{\sects}{Secs.~}
\newcommand{\app}{App.~}
\newcommand{\selfdual}{self-dual}
\newcommand\s{s}
\newcommand{\f}{F}
\newcommand{\g}{d}
\newcommand{\A}{\hat{f}}
\newcommand{\B}{g}
\newcommand{\re}{^\text{re}}
\newcommand{\im}{^\text{im}}

\begin{document}

\title{Wigner surmise for mixed symmetry classes in random
  matrix theory}
\author{Sebastian Schierenberg}
\author{Falk Bruckmann}
\author{Tilo Wettig}
\affiliation{Institute for Theoretical Physics, University of
  Regensburg, 93040 Regensburg, Germany}
\date{\today}

\begin{abstract}
  We consider the nearest-neighbor spacing distributions of mixed
  random matrix ensembles interpolating between different symmetry
  classes, or between integrable and non-integrable systems.  We
  derive analytical formulas for the spacing distributions of \two\ or
  \four\ matrices and show numerically that they provide very good
  approximations for those of random matrices with large dimension.
  This generalizes the Wigner surmise, which is valid for pure
  ensembles that are recovered as limits of the mixed ensembles. We
  show how the coupling parameters of small and large matrices must be
  matched depending on the local eigenvalue density.
\end{abstract}

\maketitle

\allowdisplaybreaks[4]

\section{Introduction}

Random matrix theory (RMT) is a powerful mathematical tool which can
be used to describe the statistical behavior of quantities arising in
a wide variety of complex systems.  It has been applied to many
mathematical and physical problems with great success, see
\cite{Guhr:1997ve,Verbaarschot:2000dy,rmthandbook} for reviews.  This
wide range of applications is based on the fact that RMT describes
universal quantities that do not depend on the detailed dynamical
properties of a given system but rather are determined by global
symmetries that are shared by all systems in a given symmetry class.

In RMT the operator governing the behavior of the system, such as the
Hamilton or Dirac operator, is replaced by a random matrix with
suitable symmetries.  One then studies statistical properties of the
eigenvalue spectrum of such random matrices, typically in the limit of
large matrix dimension.  To compare different systems in the same
symmetry class with RMT, the eigenvalues of the physical system as
well as those of the random matrices need to be ``unfolded''
\cite{Brody:1981cx}.  The purpose of such an unfolding procedure is to
separate the average behavior of the spectral density (which is not
universal) from the spectral fluctuations (which are universal).
Unfolding is essentially a local rescaling of the eigenvalues,
resulting in an unfolded spectrum with mean level spacing equal to
unity.  How the rescaling is to be done is not unique and may depend
on the system under study.

In this paper we focus on the so-called nearest-neighbor spacing
distribution $P(s)$, i.e., the probability density to find two
adjacent (unfolded) eigenvalues at a distance $s$.  This quantity
probes the strength of the eigenvalue repulsion due to interactions
and can be computed analytically for the classical RMT ensembles,
resulting in rather complicated expressions given in terms of prolate
spheroidal functions \cite{Mehta:2004}.  However, it was realized
early on that the level spacing distribution of large random matrices
is very well approximated by that of $2\times2$ matrices in the same
symmetry class.\footnote{This does not work for non-Hermitian complex
  matrices \cite{Grobe:1988, Akemann:2009my}.}  For most practical
purposes it is sufficient to use this so-called Wigner surmise
\cite{Wigner:1957} instead of the exact analytical result.  It
is given by
\begin{equation}
  \label{eq:wigner}
  P_\beta(\s)=a_\beta\s^\beta e^{-b_\beta\s^2}
\end{equation}
with $\beta=1,2,4$ corresponding to the Gaussian orthogonal (GOE),
unitary (GUE), and symplectic (GSE) ensemble of RMT, respectively.
The quantities $a_\beta$ and $b_\beta$ are chosen such that
\begin{align}
\label{eq:norm}
  \int_0^\infty d\s\,P_\beta(\s)=1\;\text{ and }\;
  \ev{\s}=\int_0^\infty d\s\,P_\beta(\s)\,\s=1
\end{align}
in all three cases.  Explicit formulas will be given in
\sect\ref{derspacsec}.

RMT describes quantum systems whose classical counterparts are chaotic
\cite{Bohigas:1983er} and correctly predicts the strong short-range
correlations of the eigenvalues due to interactions.  In contrast, the
level spacing distribution of a quantum system whose classical
counterpart is integrable is given by that of a Poisson process,
\begin{equation}
  \label{eq:poisson}
  P_0(\s)=e^{-\s}\,,
\end{equation}
corresponding to uncorrelated eigenvalues.  We assign the Dyson index
$\beta = 0$ to ensembles of this kind, which is a consistent extension
of the generalized Gaussian ensembles with arbitrary real $\beta>0$
introduced in \cite{Edelman:2002}.

Often physical systems consist of parts with different symmetries, or
of a classically integrable and a chaotic part.  Changing a parameter
of the system may then result in transitions between different
symmetry classes.  Now, the question is whether a symmetry transition
in a given physical system can be described by a transition between
RMT ensembles (or Poisson).  It has been shown in numerous studies
that this is indeed the case.  For example, billiards are showcases
for the interplay of chaos and integrability, and certain billiards
exhibit Poisson-GOE transitions \cite{Cheon:1991, Shigehara:1993,
  Csordas:1994, Abul-Magd:2008}.  A transition between GOE and GUE
behavior takes place in the spectrum of a kicked top \cite{Lenz:1991}
or kicked rotor \cite{Shukla:1997} when time-reversal symmetry is
gradually broken.  Furthermore, a transition from Poisson to GOE
statistics was found for random points on fractals as the dimension is
changed \cite{Sakhr:2005}.  In the spectrum of the hydrogen atom in a
magnetic field, transitions were observed from Poisson to GOE
\cite{Wintgen:1987} as well as from GOE to GUE \cite{Goldberg:1991}.
Transitions from Poisson to GOE or GUE statistics also occur in condensed
matter physics, e.g., in the metal-insulator (Anderson) transition
\cite{Shklovskii:1993, Shukla:2005} whose properties are similar to those of the
Brownian motion model introduced in Ref.~\cite{Dyson:1962}.  In
relativistic particle physics the Dirac operator shows transitions
between different chiral symmetry classes \cite{Follana:2006} or an
Anderson-type transition \cite{Garcia-Garcia:2007, Kovacs:2010,
  Bruckmann:2011}.  In the spectra of nuclei a transition between GOE
and Poisson spectral statistics takes place when levels sequences with
different exact quantum numbers are mixed \cite{Brody:1981cx}.  We
thus conclude that RMT is broadly applicable not only to pure systems
but also to mixed systems.

In this paper, we assume the Hamiltonian describing the mixed system
to be of the form\footnote{Other possibilities have also been
  investigated, see, e.g., \cite{Berry:1984,Hussein:1993,Mehta:2004},
  but will not be considered in this paper.}
\begin{equation}
  \label{eq:trans}
  H = H_\beta + \lambda H_{\beta'}\,,
\end{equation}
where $H_\beta$ represents the original system whose
symmetry/integrability is broken by the perturbation $H_{\beta'}$ for
small coupling parameter $\lambda$, and vice versa for large
$\lambda$. For the quantities we analyze the absolute scale of $H$ is
irrelevant, only the relative scale between the different parts
matters.

From the level statistics point of view, $H_\beta$ and $H_{\beta'}$
correspond either to a Poisson process or to one of the three RMT
ensembles.  Hence, there are $\binom42=6$ possibilities for a
transition between two of these four cases in \eq\eqref{eq:trans},
i.e., Poisson-GOE, Poisson-GUE, Poisson-GSE, GOE-GUE, GOE-GSE, and
GUE-GSE.  If a GSE matrix is involved in the transition, there are two
possibilities for the other matrix: \selfdual\ or not.\footnote{An
  even-dimensional matrix $A$ is called \selfdual\ if $JA^TJ^T=A$ with
  $J$ given in \eq\eqref{eq:J}.} This leads to an even larger variety
of mixed ensembles.  Many transitions of this kind have been studied
in earlier works, usually for large matrix dimension.  Transitions
between Gaussian ensembles are considered in \cite{Mehta:2004}, but
closed forms for the spacing distribution could not be obtained, and
\selfdual\ symmetry was not conserved in the transitions involving the
GSE.  Mixtures of Gaussian ensembles with conserved \selfdual\
symmetry and small matrix size are considered in \cite{Nieminen:2009},
but only numerical results are given for the spacing distributions.
Other examples include the heuristic Brody distribution
\cite{Brody:1973} interpolating between Poisson and the GOE, the
spacing distribution of a generalized Gaussian ensemble of \two\ real
random matrices \cite{Berry:2009}, and a complete study of the
transition between Poisson and the GUE \cite{Guhr:1996wn}.  The
two-point correlation function of the latter case is also studied in
\cite{Kunz:1998}.

Note that an exact analytical calculation of $P(s)$ for systems
described by an Ansatz of the form \eqref{eq:trans} is much harder
than, e.g., the analytical calculation of low-order spectral
correlation functions, which are already difficult to obtain.  Here,
we do not attempt an analytical calculation of $P(s)$ for large matrix
dimension.  Rather, motivated by the reliability of the Wigner
surmise, we study the possible transitions in \eq\eqref{eq:trans} for
$2\times2$ matrices (or, in the symplectic case, $4\times4$ matrices,
because the smallest non-trivial \selfdual\ matrix has this size) and
compare the resulting level spacing distributions with that of large
random matrices, the latter obtained numerically.  The cases of
Poisson-GOE and GOE-GUE were worked out earlier by Lenz and Haake
\cite{Lenz:1991}, and the spacing distribution of a \two\ matrix
interpolating between Poisson and GUE is given in \cite{kota:1999}.
These cases will briefly be reviewed below, and the remaining ones are
the main subject of this work.

This paper is organized as follows. In \sect\ref{derspacsec} we derive
analytical results for $P(s)$ for small matrix sizes.  If $H_{\beta'}$
is from the GSE (i.e., $H_{\beta'}$ is \selfdual) we construct in
\sects\ref{poissympsec}, \ref{ortsympsec}, and \ref{unisympsec}
\selfdual\ matrices $H_\beta$ to maintain the Kramers degeneracy.  In
\sect\ref{sympgaussec} we consider the case where a \four\ GSE matrix
is perturbed by a non-\selfdual\ GUE matrix.
Section~\ref{largespectraapp} provides strong numerical evidence that
the results obtained in \sect\ref{derspacsec} approximate the spacing
distributions of large random matrices very well. We give a
perturbative argument for the matching of the couplings used for the
Wigner surmise and for large matrices, respectively, and derive an
approximate result that involves the eigenvalue density.  This result
describes the numerical data rather well.  We also show that the
transitions from the GSE to either a non-\selfdual\ Poissonian ensemble
or the GOE proceed via an intermediate transition to the GUE and can
also be described by the surmises calculated in \sect\ref{derspacsec}.
We summarize our findings and conclude in \sect\ref{summarysec}.
Technical details are worked out in several appendices.

\section{Spacing distributions for small matrices}
\label{derspacsec}

\subsection{Preliminaries}

\begin{figure*}[t]
  \includegraphics[width=\threepicwidth,clip=true]
  {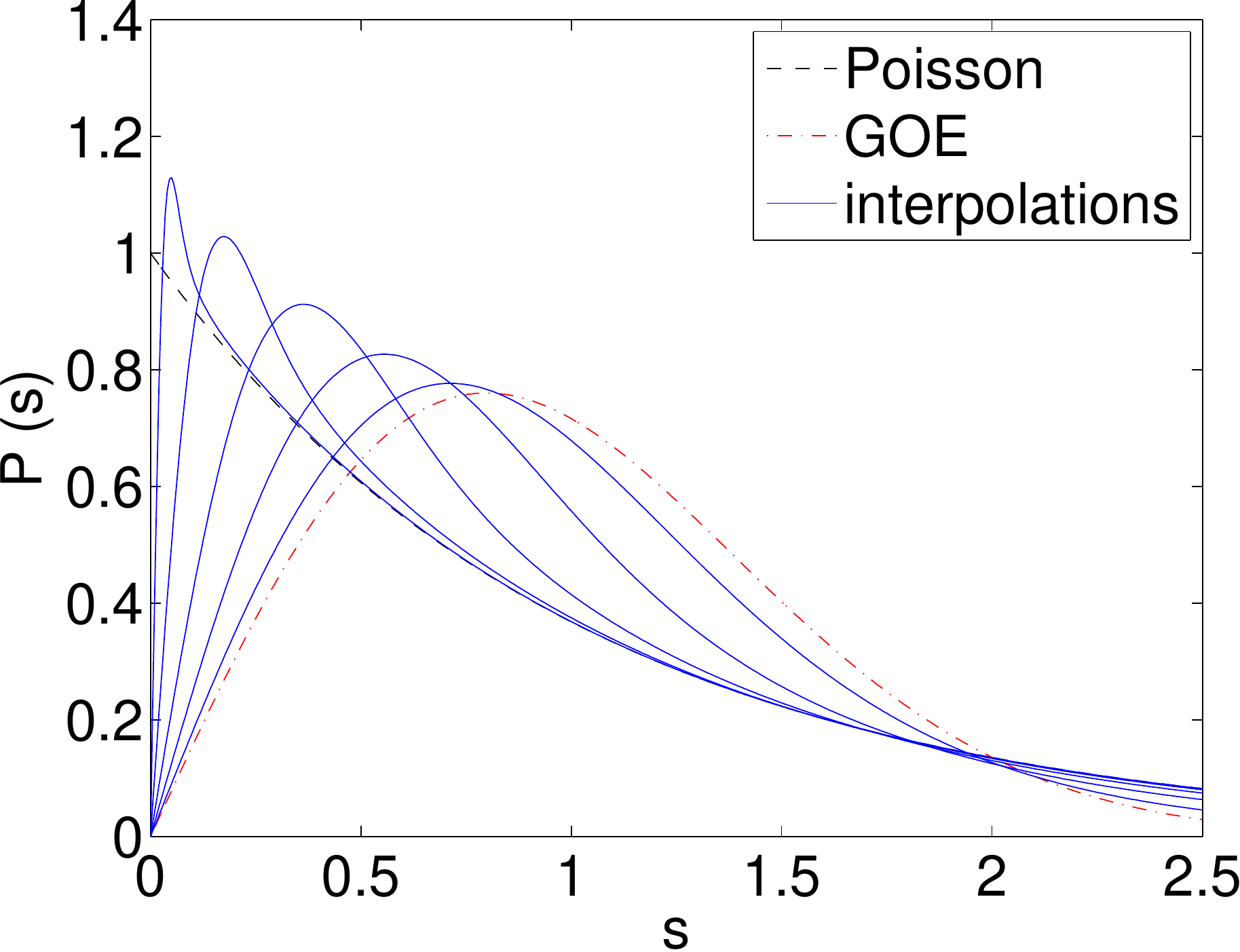}\hfill
  \includegraphics[width=\threepicwidth,clip=true]
  {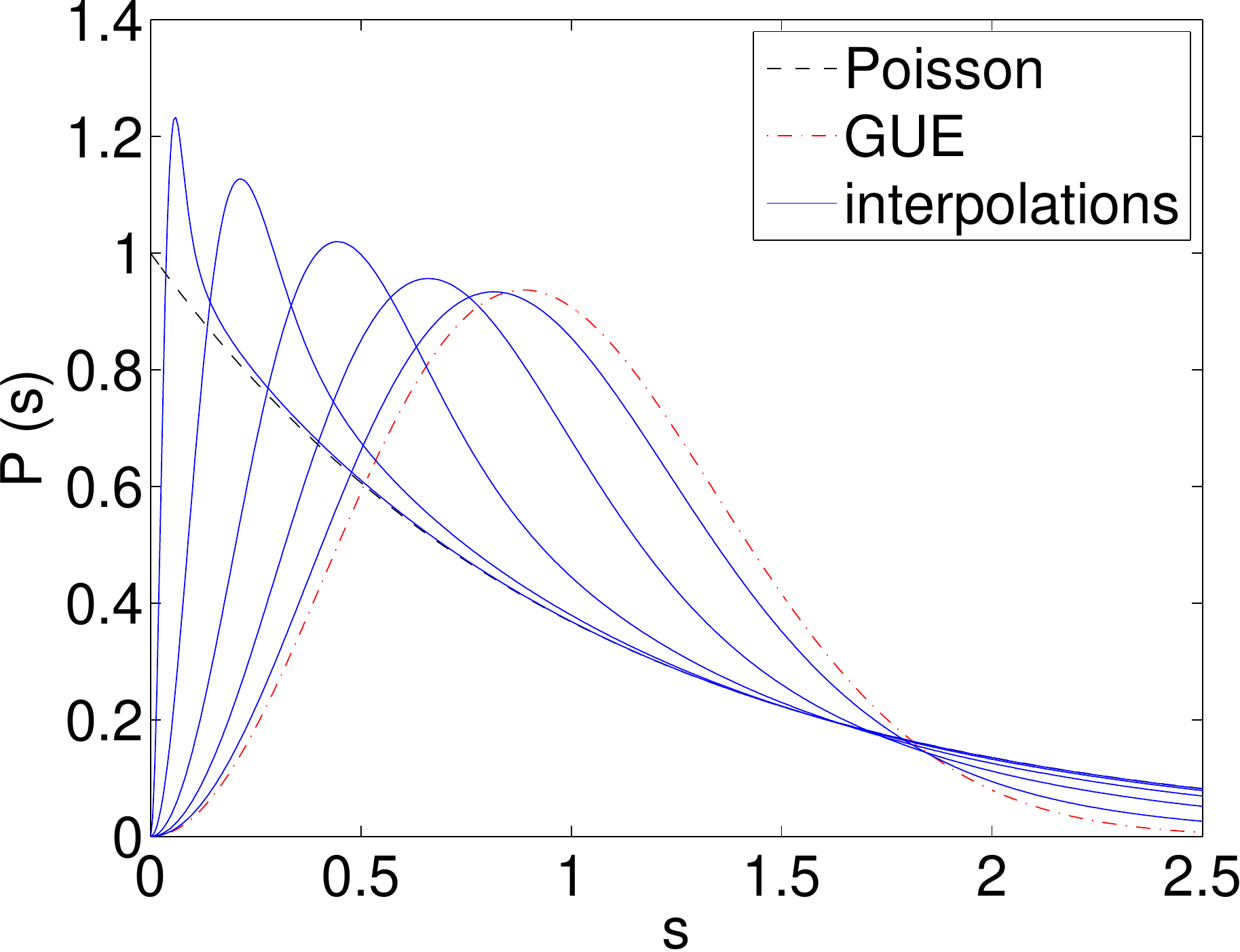}\hfill
  \includegraphics[width=\threepicwidth,clip=true]
  {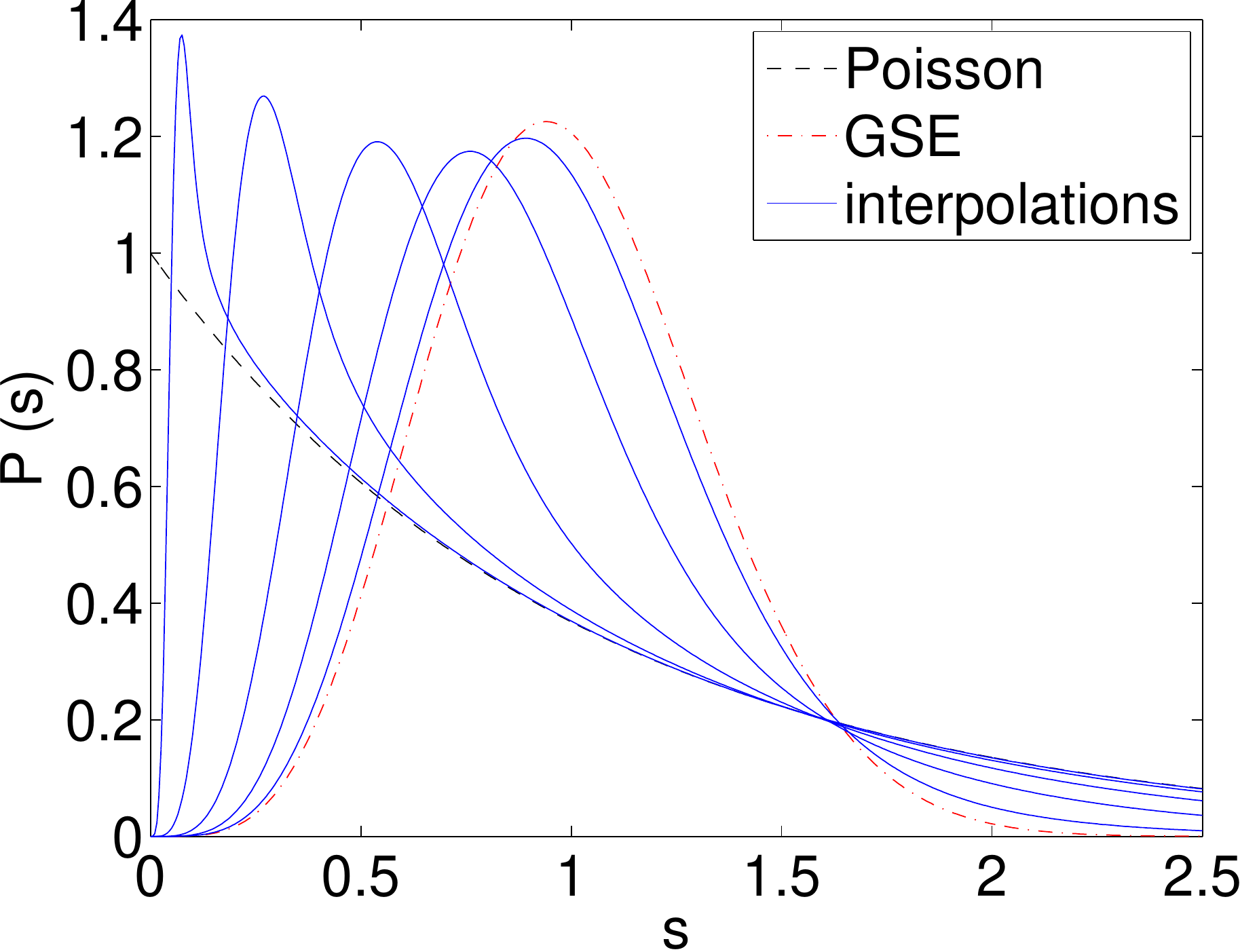}
  \caption{(Color online) Spacing distributions $P_{0\to\beta'}(\s; \lambda)$ of the
    transitions Poisson $\to$ GOE (left,
    \eq\eqref{poisortresultform}), GUE (middle,
    \eq\eqref{poisuniresultform}) and GSE (right,
    \eq\eqref{poissympresultform}) for \two\ or \four\ matrices with
    coupling parameters $\lambda = 0.02$, $0.08$, $0.2$, $0.4$, and
    $0.8$ (maxima moving from left to right as $\lambda$ increases).
    In the GSE case the matrix representing the Poisson process was
    made \selfdual.  All formulas were verified by comparison to
    numerically obtained spacing distributions of \two\ or \four\
    random matrices.}
  \label{poisgaussimplefigure}
\end{figure*}

In the spirit of the Wigner surmise, we now calculate the
distributions $P(\s)$ of eigenvalue spacings $\s$ of mixed ensembles
for the smallest nontrivial (i.e., \two\ or \four) matrices, with
$P(\s)$ normalized as in \eq\eqref{eq:norm}.  Unfolding is not needed
for these matrices since they have only two independent eigenvalues
(except for \sect\ref{sympgaussec}).  We first study the transitions
from the integrable to the chaotic case for the three Gaussian
ensembles and then proceed to the transitions between different
symmetry classes.

We define the \two\ Poisson process by a matrix
\begin{equation}\label{poisensform}
  H_0 = \begin{pmatrix} 0 & 0 \\ 0 & p\end{pmatrix},
\end{equation}
where $p$ is a Poisson distributed non-negative random number with
unit mean value, i.e., its probability density is $P_0(p)=e^{-p}$.
The eigenvalue spacing of this matrix is obviously Poissonian, as the
spacing is just $p$, and therefore we obtain \eq\eqref{eq:poisson}.

The choice of $H_0$ may look like a special case, but it suffices for
our purposes. The most general Hermitian \two\ matrix with spacing $p$
can be obtained from \eq\eqref{poisensform} by a common shift of the
eigenvalues (which does not influence the spacing) and a basis
transformation. This transformation can be absorbed in the perturbing
matrix since it does not change the probability distribution of the
latter.  To see this suppose we had started with a general nondiagonal
$H_0$, also with eigenvalues $0$ and $p$, instead of
\eq\eqref{poisensform}.  When added to a random matrix $H_{\beta'}$
with $\beta' = 1, 2, 4$, we choose it to be real symmetric, Hermitian,
or \selfdual, respectively, in order to preserve the symmetry
properties of $H_{\beta'}$.  Then $H_0$ is diagonalized by a suitable
matrix $\Omega$, i.e., $\diag(0, p) = \Omega^{-1} H_0 \Omega$, where
$\Omega$ is orthogonal ($\beta=1$), unitary ($\beta=2$), or symplectic
($\beta=4$).  In the total matrix $H$ this is equivalent to
$\Omega^{-1} H_{\beta'} \Omega$ perturbing $\diag(0, p)$, but the
probability distribution of the perturbation is invariant under the
transformation $\Omega$.

For matrices $H_{1,2,4}$ from the GOE, GUE, and GSE, respectively, we
choose the mean values of the matrix elements to be $0$ and the
normalization
\begin{equation}\label{standardnormform}
  \Big\langle\big[\big(H_{1,2,4}\big)_{ii}\big]^2\Big\rangle
  = 2\,\Big\langle\big[\big(H_{1,2,4}^{(\nu)}\big)_{i\neq j}\big]^2
  \Big\rangle = 1 \,.
\end{equation}
The index $\nu=0,\ldots,\beta-1$ distinguishes the components of the
complex/quaternion GUE/GSE matrix elements, while the GOE matrix
elements possess only a real part.

All results we derive from \eq\eqref{eq:trans} will be symmetric in
$\lambda$ since the distribution of the elements of $H_{\beta'}$ is
symmetric about zero (the perturbation will be taken from one of the
Gaussian ensembles in each case).  This means that our results should
be expressed in terms of $|\lambda|$.  To avoid such cumbersome
notation we restrict ourselves to non-negative $\lambda$.

\subsection{Poisson to GOE}
\label{poisortsec}

We first consider the case that corresponds to a classically
integrable system perturbed by a chaotic part with anti-unitary
symmetry squaring to $\1$. The integrable part is represented by a
Poisson process, and the chaotic one by the GOE. The spacing
distribution for this case has been derived in \cite{Lenz:1991}, and
we state it here for the sake of completeness.

The \two\ random matrix 
\begin{equation}\label{poisortensform}
  H = H_0 + \lambda H_1 = \begin{pmatrix}0&0\\0&p\end{pmatrix} +
\lambda\begin{pmatrix}a&c\\c&b\end{pmatrix}
\end{equation}
consists of $H_0$ from \eqref{poisensform} and $H_1$ from the GOE,
i.e., a real symmetric matrix with normalization given in
\eq\eqref{standardnormform}.  The calculations are very similar to the
ones for the transition from Poisson to the GSE, which are presented
in \sect\ref{poissympsec} (see also \app\ref{app:large_s}).  The
resulting spacing distribution of $H$ reads
\begin{equation}\label{poisortresultform}
  P_{0\to 1}(s;\lambda)
  = Cs\,e^{-D^2s^2} \int_{0}^{\infty}dx\,
  e^{-\frac{x^2}{4\lambda^2}-x}\, I_0\left(\frac{xDs}\lambda\right)
\end{equation}
with
\begin{align}\label{eq:poisgoeD}
  D(\lambda) &= \frac{\sqrt{\pi}}{2\lambda}\,
  U\left(-\tfrac12,0,\lambda^2\right), \\
  C(\lambda) &= 2 D(\lambda)^2\,,
\end{align}
where $U$ is the Tricomi confluent hypergeometric function (or Kummer
function) \cite[\eq(13.1.3)]{Abram:1964} and $I_0$ is a modified
Bessel function \cite[\eq(9.6.3)]{Abram:1964}.  $P_{0\to
  1}(s;\lambda)$ is plotted in \fig\ref{poisgaussimplefigure} (left)
for various values of $\lambda$.  The formula is equivalent to the one
given in \cite{Lenz:1991}, but our integration variable $x$ is scaled
differently.

In the limiting cases of $\lambda\to0$ and $\lambda\to\infty$ we
have
\begin{align}
  D(\lambda) \sim 
  \begin{cases}
    1/(2\lambda) & \text{for }\lambda\to0\,,\\
    \sqrt\pi/2 & \text{for }\lambda\to\infty\,.
  \end{cases}
\end{align}
Using the asymptotic expansion of the Bessel function, it is
straightforward to show that for $\lambda\to0$ we obtain the Poisson
result $e^{-s}$.  It is even simpler to show that the Wigner surmise
$(\pi s/2)\,e^{-\pi s^2/4}$ for the GOE is obtained for
$\lambda\to\infty$.

The small-$s$ behavior of $P_{0\to 1}(\s;\lambda)$ shows interesting
features.  To investigate this behavior, we consider separately the
cases $\lambda=0$ and $\lambda>0$.  For $\lambda=0$ we have by
construction
\begin{equation}
  P_{0\to 1}(\s;0) = e^{-\s}=1-\s+\mathcal O(\s^2)\,.
\end{equation}
For $\lambda>0$ we obtain from \eq\eqref{poisortresultform}
\begin{equation}\label{eq:poisgoelambdatozero}
  P_{0\to 1}(\s;\lambda ) = c(\lambda)\s + \mathcal O(\s^3)
\end{equation}
with
\begin{equation}\label{eq:poisgoelambdatozero2}
  c(\lambda)\sim\frac{\sqrt\pi}{2\lambda} \quad\text{for }\lambda\to0\,.
\end{equation}
which means that we recover the linear level repulsion of the GOE for
arbitrarily small $\lambda$, i.e., for arbitrarily small admixture of
the chaotic part as also observed in \cite{Robnik:1987, Hasegawa:1988,
  Caurier:1990}.  This implies that for $\lambda\to0$ the
distribution, viewed as a function of $\lambda$, develops a
discontinuity at $\s=0$, since $P_{0\to 1}(\s=0;\lambda=0)=1$ while
$P_{0\to1}(\s=0;\lambda>0)=0$.  This effect is clearly seen in
\fig\ref{poisgaussimplefigure} (left).

For small values of $\lambda$ and $\s$, we observe something
reminiscent of the Gibbs phenomenon, i.e., the interpolation
overshoots the Poisson curve considerably. In the limit of
$\lambda\to0$, one can show (see \app\ref{gibbsapprmt}) that the
maximum of $P_{0\to 1}$ is at $\s_\text{max} = 2.51393\,\lambda$ with
a finite value of $P_{0\to 1}(\s_\text{max};\lambda\to0) = 1.17516$.
This implies an overshoot of $17.5\%$ compared to the Poisson curve.
Such an effect also occurs in the transitions from Poisson to GUE and
GSE that are treated in \sects\ref{poisunisec} and \ref{poissympsec}
below, with a quadratic/quartic level repulsion in the small-$\s$
regime.

The large-$s$ behavior of $P_{0\to 1}(\s;\lambda)$ is analyzed in
\app\ref{app:large_s}, and we obtain Poisson-like behavior for any
finite $\lambda$, see \eq\eqref{eq:large_s}.  This is in contrast to
the small-$s$ behavior, which is GOE-like for any nonzero $\lambda$.

\subsection{Poisson to GUE}
\label{poisunisec}

We now consider the transition from Poisson to the GUE.  This
corresponds to a classically integrable system with a chaotic
perturbation without anti-unitary symmetry. The \two\ random matrix
\begin{equation}\label{poisuniensform}
  H = H_0 + \lambda H_2 = \begin{pmatrix}0&0\\0&p\end{pmatrix} +
  \lambda\begin{pmatrix}a&c_0+ic_1\\c_0-ic_1&b\end{pmatrix} 
\end{equation}
contains $H_2$ from the GUE, i.e., a complex Hermitian matrix with
normalization \eqref{standardnormform}.  The spacing distribution of
an equivalent setup with different normalizations of the random matrix
elements was already considered in \cite{kota:1999}, so we just state
the result,
\begin{equation}\label{poisuniresultform}
  P_{0\to 2}(\s;\lambda) = C\s^2\,
  e^{-D^2s^2}\int_0^\infty dx\,e^{-\frac{x^2}{4\lambda^2}-x}\,
  \frac{\sinh z}{z}
\end{equation}
with $z=xDs/\lambda$ and
\begin{align}\label{eq:poisgueD}
  D(\lambda) &= \frac1{\sqrt{\pi}} 
  + \frac1{2\lambda}e^{\lambda^2}\erfc(\lambda)
  -\frac\lambda2\Ei\left(\lambda^2\right)\notag\\  
  &\quad + \frac{2\lambda^2}{\sqrt{\pi}}\, {}_2F_2\left(\tfrac12 ,1;
    \tfrac32 ,\tfrac32 ;\lambda^2\right),\\ 
  C(\lambda) &= \frac{4D(\lambda)^3}{\sqrt{\pi}}\,.
\end{align}
Here, $\erfc$ is the complementary error function
\cite[Eq.~(7.1.2)]{Abram:1964}, $\Ei$ is the exponential integral
\cite[Eq.~(5.1.2)]{Abram:1964}, and ${}_2F_2$ is a generalized
hypergeometric function \cite[Eq.~(9.14.1)]{Grad:1994}.  We could also
have written the result in the form of \eqs\eqref{eq:intp} and
\eqref{eq:Xbeta} since $\sinh z=\sqrt{\pi z/2}\,I_{1/2}(z)$.

To check the validity of \eq\eqref{poisuniresultform} and to see the
emergence of the limiting spacing distributions, we now consider the
limits $\lambda\to 0$ and $\lambda\to\infty$.  First note that for
$\lambda\to0$ we have
\begin{align}\label{eq:cdlambdatozero}
  D\sim\frac1{2\lambda}\quad\text{and}\quad
  C\sim\frac1{2\lambda^3\sqrt\pi}
\end{align}
so that \eq\eqref{poisuniresultform} becomes for $s>0$
\begin{align}
  &P_{0\to 2}(\s;0) = \lim_{\lambda\to0}\frac{\s^2}{2\lambda^3\sqrt\pi}
  \int_{0}^{\infty}dx\,e^{-\frac1{4\lambda^2}(\s^2+x^2)-x}\,
  \frac{\sinh z}{z}\notag\\
  &=\frac \s{2\sqrt\pi}\int_{0}^{\infty}dx\,\frac{e^{-x}}x
  \underset{=2\sqrt\pi[\delta(\s-x)-\delta(\s+x)]}
  {\underbrace{\lim_{\lambda\to0}\frac1\lambda
      \left(e^{-\frac{(\s-x)^2}{4\lambda^2}}
        -e^{-\frac{(\s+x)^2}{4\lambda^2}}\right)}}\notag\\
  &=e^{-\s}\,, \label{eq:poissonlimit}
\end{align}
which is the Poisson distribution as required.  For $\lambda\to
\infty$ we have
\begin{align}
  D\sim\frac2{\sqrt\pi}\quad\text{and}\quad
  C\sim\frac{32}{\pi^2}
\end{align}
so that \eq\eqref{poisuniresultform} becomes
\begin{align}
  P_{0\to 2}(\s;\infty) &= 
  \frac{32\s^2}{\pi^2} e^{-\frac{4s^2}{\pi}}
  \lim_{\stackrel{\lambda\to\infty}{z\to0}}
  \int_{0}^{\infty}dx\,e^{-\frac{x^2}{4\lambda^2}-x}\,
  \frac{\sinh z}{z}\notag\\
  &=\frac{32\s^2}{\pi^2}e^{-\frac{4s^2}{\pi}}\,,
\end{align}
which is the Wigner surmise for the GUE.

The integral in \eq\eqref{poisuniresultform} can be computed
numerically without difficulties as the integrand decays like a
Gaussian for large $x$ and becomes constant for small
$x$.\footnote{Note that the integral can be expressed in terms of
  imaginary error functions, but for increasing $\s$ delicate
  cancellations occur that make it impractical to use this form for
  numerical evaluation.  This is why we present
  \eq\eqref{poisuniresultform} as the final formula, which is well
  suited for numerical integration.}  The resulting distribution
$P_{0\to 2}(\s;\lambda)$ is plotted in \fig\ref{poisgaussimplefigure}
(middle).

As in \sect\ref{poisortsec}, a discontinuity is found at $s=0$ towards
the Poisson result.  For $\lambda>0$ we obtain from
\eq\eqref{poisuniresultform}
\begin{equation}\label{eq:poisguelambdatozero}
  P_{0\to 2}(\s;\lambda ) = c(\lambda)\s^2 + \mathcal O(\s^4)
\end{equation}
with
\begin{equation}\label{eq:poisguelambdatozero2}
  c(\lambda)\sim\frac1{2\lambda^2} \quad\text{for }\lambda\to0\,.
\end{equation}
Hence we obtain the quadratic level repulsion of the GUE for
arbitrarily small coupling parameter. For $\lambda\to0$, the maximum
of the function is at $\s_\text{max} = 3.00395\,\lambda$, with a value
of $P_{0\to 2}(\s_\text{max}; \lambda\to0) = 1.28475$ (see
\app\ref{gibbsapprmt}).

The large-$s$ behavior of $P_{0\to 2}(\s;\lambda)$ is given by
\eq\eqref{eq:large_s}, i.e., it is Poisson-like.

\subsection{Poisson to GSE}
\label{poissympsec}

In this case, a classically integrable system is perturbed by a
chaotic part with anti-unitary symmetry squaring to $-\1$ and hence
represented by the \selfdual\ matrices of the GSE. One has to consider
\four\ matrices here, because a \selfdual\ \two\ matrix is
proportional to $\1_2$ and has only one non-degenerate eigenvalue. As
mentioned in the introduction, there are now two possibilities: The
Poisson process could be represented by a \selfdual\ or a
non-\selfdual\ matrix.  Here we only consider the former possibility,
while the latter will be discussed in \sect\ref{sec:gsetoother}.
A \selfdual\ Poisson matrix is obtained by taking a tensor product of
\eq\eqref{poisensform} with $\1_2$.  Thus the transition matrix is
\begin{align}
&H = H_0 \otimes \1_2+ \lambda H_4 = 
   \begin{pmatrix}
    0&0&0&0\\
    0&0&0&0\\
    0&0&p&0\\
    0&0&0&p
  \end{pmatrix}\notag\\\label{poissympensform}
  &+\lambda
  \begin{pmatrix}
    a&0&c_0+ic_3&c_1+ic_2\\
    0&a&-c_1+ic_2&c_0-ic_3\\
    c_0-ic_3&-c_1-ic_2&b&0\\
    c_1-ic_2&c_0+ic_3&0&b
  \end{pmatrix},
\end{align}
where the GSE matrix $H_4$ is Hermitian and \selfdual, and can be
represented by a \two\ matrix whose elements are real quaternions, see
\cite{Mehta:2004} for details.

We now explain the calculation of the spacing distribution for this
transition. The computation of the previous cases, Poisson to GOE and
Poisson to GUE, can be done in a similar fashion.

Due to the \selfdual\ structure of $H$, the spacing $S$ between its
non-degenerate eigenvalues spacing can be computed analytically and
reads
\begin{equation}\label{spaceform}
  S = \lambda\sqrt{(a-b-p/\lambda)^2 + 4c_\mu c_\mu}\,,
\end{equation}
where the repeated index $\mu$ indicates a sum from $0$ to $3$.  We
have intentionally written $S$ instead of $\s$ since we eventually
need to rescale the spacing to ensure $\ev\s=1$.  The desired spacing
distribution is proportional to the integral
\begin{align}
  I(S) & = \int dp\,da\,db\,\prod_{\nu=0}^3 dc_\nu\,
  P_0(p) P_a(a) P_b(b) P_{c_\nu}(c_\nu) \notag\\
  &\quad \times\delta\left(S-\sqrt{[a-(b+p/\lambda)]^2
      +4c_\mu c_\mu}\right),
  \label{generalspacingform}
\end{align}
where we have rescaled $S$ by $\lambda$ for simplicity and are not yet
concerned with the normalization.  The distributions
$P_\alpha(\alpha)$ of the random variables $\alpha=a, b, c_0, c_1,
c_2, c_3$ are Gaussian, with variances given by
\eq\eqref{standardnormform},
\begin{equation}
  \sigma^2_{a,b} = 2\sigma^2_{c_0, c_1, c_2, c_3} = 1 \,.
\end{equation}
Inserting this into \eq\eqref{generalspacingform} and shifting $b\to
b-p/\lambda$ gives
\begin{align}\label{generalspacingformagain}
  I(S)&\propto\int_0^\infty\!dp\int_{-\infty}^\infty\!da\,db\,\prod_{\nu=0}^3 dc_\nu\,
  e^{-p-\frac12a^2-\frac12(b-p/\lambda)^2-c_\mu c_\mu} \notag\\
  &\quad\times\delta\left(S-\sqrt{(a-b)^2 +4c_\mu c_\mu}\right).
\end{align}
The multi-dimensional integral in this expression is computed in
\app\ref{poisgseapp}.  Rescaling the spacing and normalizing the
distribution to satisfy \eq\eqref{eq:norm}, we obtain
\begin{align}
  \label{poissympresultform}
  &P_{0\to 4}(s;\lambda) = C s^4e^{-D^2s^2}\notag\\
  &\qquad\times\int_0^\infty dx \, e^{-\frac{x^2}{4\lambda^2}-x}
  \frac{z\cosh z-\sinh z}{z^3}
\end{align}
with $z=xDs/\lambda$ and
\begin{align}
  D(\lambda) &= \frac\lambda{2\sqrt{\pi}}
  \int_0^{\infty}d x\,e^{-2\lambda x}\label{eq:D}\\
  &\quad\times\frac{(4x^3 + 2x)e^{-x^2} +
    \sqrt{\pi}(4x^4+4x^2-1)\erf(x)}{x^3}\,,\notag\\
  C(\lambda) &= \frac{8D(\lambda)^5}{\sqrt\pi}\,\label{eq:C},
\end{align}
where $\erf$ is the error function \cite[\eq(7.1.1)]{Abram:1964}.  The
last term in the integrand of \eq\eqref{poissympresultform} is
proportional to $I_{3/2}(z)$ in agreement with \eqs\eqref{eq:intp} and
\eqref{eq:Xbeta}.

In the limiting cases of $\lambda\to0$ and $\lambda\to\infty$ we
find
\begin{align}\label{eq:cdlambdatozeropoissymp}
  D(\lambda) \sim 
  \begin{cases}
    1/(2\lambda) & \text{for }\lambda\to0\,,\\
    8/(3\sqrt\pi) & \text{for }\lambda\to\infty\,.
  \end{cases}
\end{align}
For $\lambda\to0$, manipulations analogous to those performed in
\eq\eqref{eq:poissonlimit} lead to the Poisson result $e^{-s}$.  For
$\lambda\to\infty$ the integral in \eq\eqref{poissympresultform}
becomes trivial and yields $1/3$ so that we obtain the Wigner surmise
$(64/9\pi)^3s^4e^{-64s^2/9\pi}$ for the GSE.

\begin{figure*}[t]
  \includegraphics[width=\threepicwidth,clip=true]
  {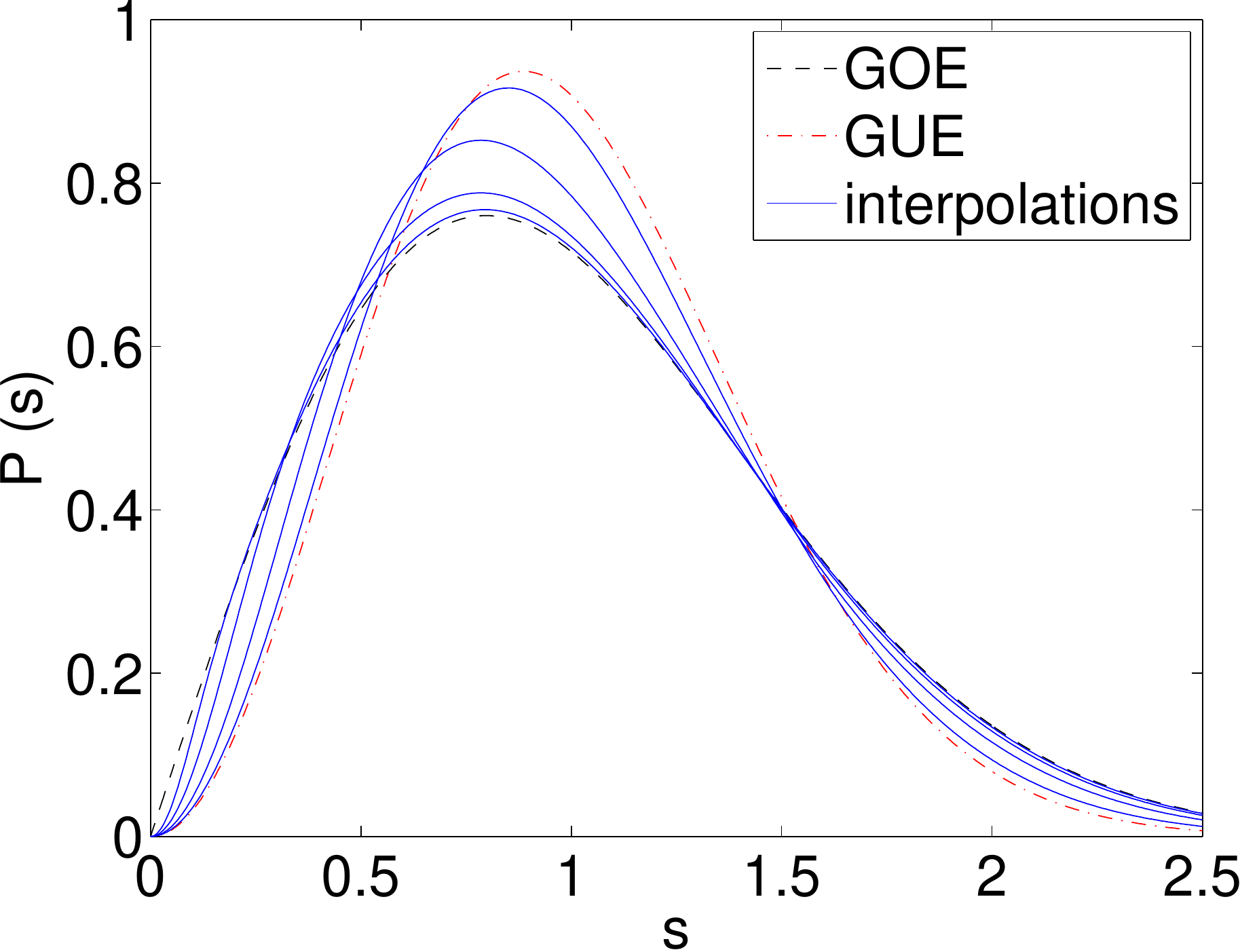}\hfill
  \includegraphics[width=\threepicwidth,clip=true]
  {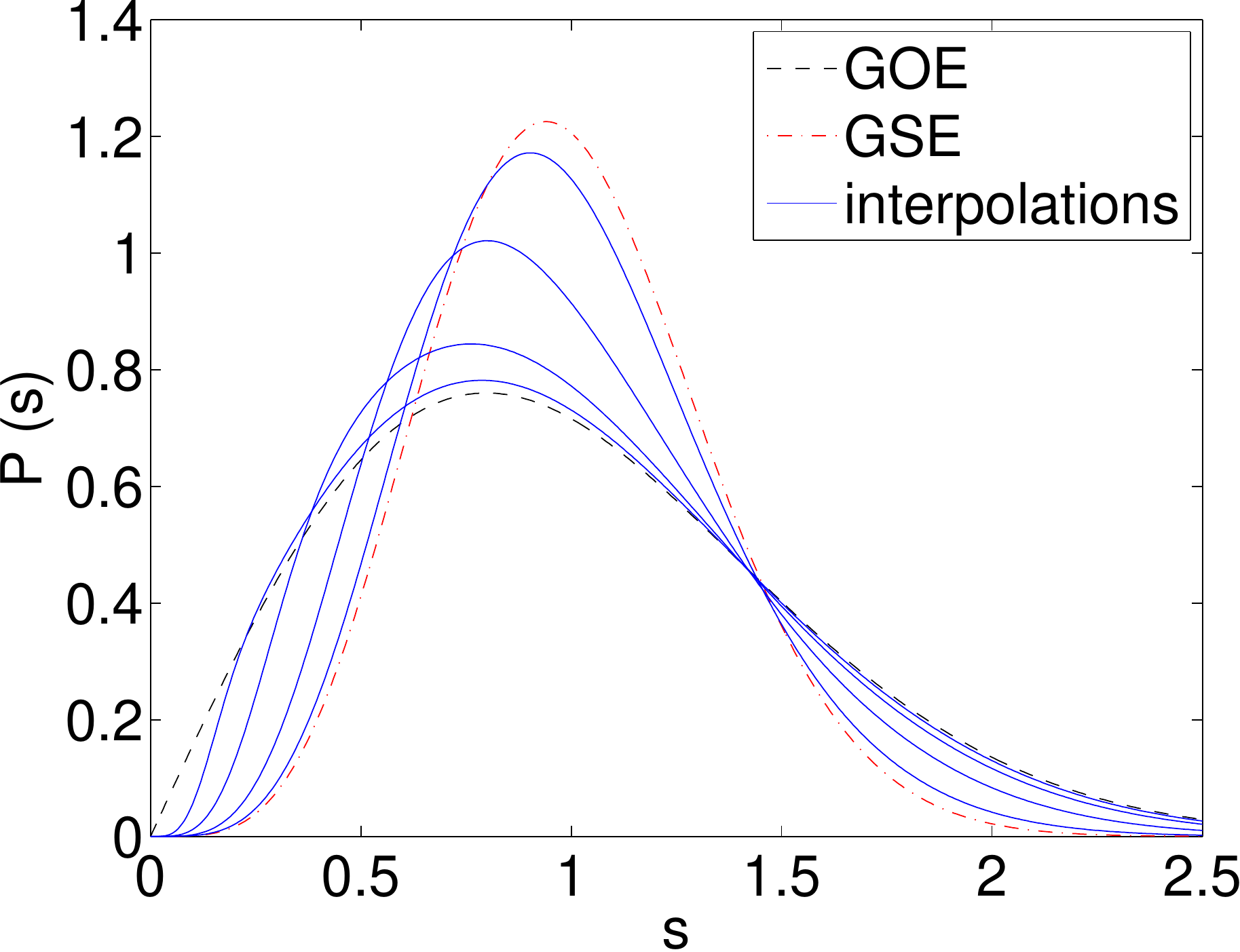}\hfill
  \includegraphics[width=\threepicwidth,clip=true]
  {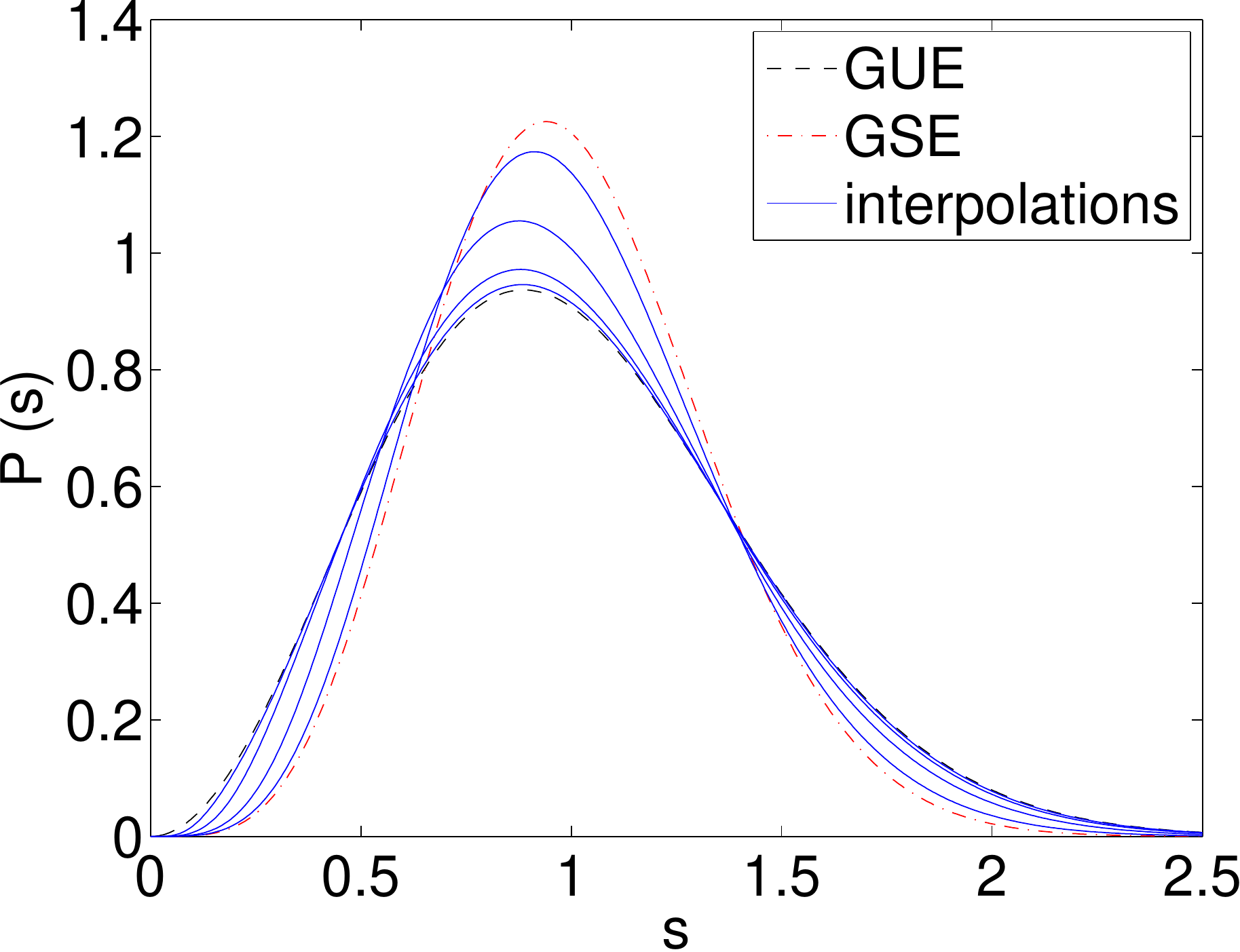}
  \caption{(Color online) Spacing distributions $P_{\beta\to\beta'}(\s; \lambda)$ of
    the transitions GOE $\to$ GUE (left, \eq\eqref{ortuniresultform}),
    GOE $\to$ GSE (middle, \eq\eqref{ortsympresultform}) and GUE $\to$
    GSE (right, \eq\eqref{unisympresultform}) for \two\ or \four\
    matrices with coupling parameters $\lambda = 0.1$, $0.2$, $0.4$,
    and $0.8$ (maxima increasing with $\lambda$).  In the cases
    involving the GSE, the GOE or GUE matrices were made \selfdual.
    All formulas were verified by comparison to numerically obtained
    spacing distributions of \two\ or \four\ random matrices.}
  \label{gausgaussimplefigure}
\end{figure*}

Equation~\eqref{poissympresultform} is plotted in
\fig\ref{poisgaussimplefigure} (right) and again displays a
discontinuity at $s=0$ as $\lambda\to 0$.  For $\lambda>0$ we now have
\begin{equation}\label{eq:poisgselambdatozero}
  P_{0\to 4}(\s;\lambda ) = c(\lambda)\s^4 + \mathcal O(\s^6)
\end{equation}
with
\begin{equation}\label{eq:poisgselambdatozero2}
  c(\lambda)\sim\frac1{12\lambda^4}
  \quad\text{for }\lambda\to0\,.
\end{equation}
For $\lambda\to0$, the maximum of the function is at
$\s_\text{max} = 3.76023\,\lambda$, with a value of
$P_{0\to 4}(\s_\text{max}; \lambda\to0) = 1.43453$
(see \app\ref{gibbsapprmt}).

The large-$s$ behavior of $P_{0\to 4}(\s;\lambda)$ is again
Poisson-like and given by \eq\eqref{eq:large_s}.

\subsection{GOE to GUE}\label{ortunisec}
With this subsection we start the investigation of transitions between
different chaotic ensembles using the smallest possible matrix size.

We consider the \two\ matrix
\begin{equation}
  H=H_1 + \lambda H_2\,.
\end{equation}
The spacing distribution for this transition was already computed in
\cite{Lenz:1991}. With the normalization of ensembles given in
\eq\eqref{standardnormform}, it reads 
\begin{equation}
  \label{ortuniresultform}
  P_{1\to 2}(s;\lambda) = C s\, e^{-D^2s^2}
  \erf\left(\frac{Ds}{\lambda}\right)
\end{equation}
with
\begin{align}
  D(\lambda) &= \frac{\sqrt{1+\lambda^2}}{\sqrt{\pi}}
  \left(\frac{\lambda}{1+\lambda^2} +
    \arccot\lambda\right),\\ 
  C(\lambda) &= 2\sqrt{1+\lambda^2}\,D(\lambda)^2\,.
\end{align}
This formula matches the result of \cite{Lenz:1991} up to a
rescaling of the coupling parameter $\lambda$ by a factor of $\sqrt2$,
which is due to a different normalization of the ensembles used there.

In the limiting cases of $\lambda\to0$ and $\lambda\to\infty$ we have
\begin{align} 
  D(\lambda) \sim
  \begin{cases}
    \sqrt\pi/2 & \text{for }\lambda\to0\,,\\
    2/\sqrt\pi & \text{for }\lambda\to\infty\,.
  \end{cases}
\end{align}
For $\lambda\to0$, the error function in \eq\eqref{ortuniresultform}
can be replaced by unity (for $s>0$), and we obtain the Wigner surmise
for the GOE.  For $\lambda\to\infty$, using the first-order Taylor
expansion of the error function yields the Wigner surmise for the GUE.

The result \eqref{ortuniresultform} is plotted in
\fig\ref{gausgaussimplefigure} (left).  In the small-$s$ region, we
now have for $\lambda>0$
\begin{align}
  P_{1\to 2}(s;\lambda) = c(\lambda)s^2+\mathcal O(s^4)
\end{align}
with
\begin{align}
  c(\lambda)\sim\frac{\pi}{2\lambda}\quad\text{for }\lambda\to0\,.
\end{align}
Similar to the previous subsections, a non-analytic transition between
weaker and stronger level repulsion develops as $\lambda\to0$, except
that now there is no jump in the function itself but rather in its
derivative at $s=0$.  Therefore, the stronger level repulsion takes
over immediately in the small $\s$-regime, if $\lambda>0$.  As we
shall see below, this also happens in the remaining transitions, GOE
to GSE and GUE to GSE, and seems to be a characteristic feature of the
mixed ensembles.

The large-$s$ behavor of $P_{1\to 2}(s;\lambda)$ is obtained
immediately from \eq\eqref{ortuniresultform} by noticing that
$\erf(x)\to1$ for $x\to\infty$.  In analogy to the transitions from
Poisson to RMT this implies that the large-$s$ behavior is dominated
by the ensemble with the smaller $\beta$.

\subsection{GOE to GSE}
\label{ortsympsec}

As the GSE is involved in this transition, we need matrices of size
\four.  Again there are two possibilities: The GOE matrix could be
made \selfdual, or it could be non-\selfdual\ (as it generically is).
Here we only consider the former case, while the latter case will be
discussed in \sect\ref{sec:gsetoother}.  As in \cite{Nieminen:2009}
we define a modified GOE matrix by
\begin{equation}
  H_1 \otimes \1_2 = 
  \begin{pmatrix} 
    a & 0 & c & 0 \\ 
    0 & a & 0 & c \\ 
    c & 0 & b & 0 \\ 
    0 & c & 0 & b
  \end{pmatrix}
\end{equation}
with real parameters $a,b,c$.  This matrix is \selfdual, so we can add
it to a matrix from the GSE without spoiling the symmetry properties
of the latter.  Thus we consider
\begin{equation}
  \label{eq:Hgoegse}
  H = H_1 \otimes \1_2 + \lambda H_4\,,
\end{equation}
where $H_1$ and $H_4$ are normalized according to
\eq\eqref{standardnormform}.  The eigenvalues of the sum are doubly
degenerate and can be calculated easily due to self-duality.

After some algebra (see \app\ref{goegseapp}) we obtain for the spacing
distribution of $H$
\begin{align}
  P_{1\to 4}(s;&\lambda) = C s^4 e^{-(1+2\lambda^2)D^2s^2}\notag\\
  &\times\int_{0}^{1}dx\,(1-x^2)\,e^{(xDs)^2}\left[I_0(z)-I_1(z)\right],
  \label{ortsympresultform}
\end{align}
where $z = (1-x^2)D^2s^2$, $I_0$ and $I_1$ are modified Bessel
functions, and
\begin{align}
  D(\lambda)&=\frac{\lambda-\lambda^3+(1+\lambda^2)^2\arccot\lambda} 
  {\sqrt{2\pi}\,\lambda\sqrt{1 + \lambda^2}}\,,\label{eq:Dgoegse}\\
  C(\lambda) &= \frac{2^{9/2}}{\sqrt{\pi}}\lambda^2(1+\lambda^2)^{3/2}
  D(\lambda)^5\,.\label{eq:Cgoegse}
\end{align}

In the limiting cases of $\lambda\to0$ and $\lambda\to\infty$ we have
\begin{align} 
  D(\lambda) \sim
  \begin{cases}
    \sqrt\pi/(2^{3/2}\lambda) & \text{for }\lambda\to0\,,\\
    8/(3\sqrt{2\pi}\lambda) & \text{for }\lambda\to\infty\,.
  \end{cases}
\end{align}
For $\lambda\to0$, we use the asymptotic expansion of the Bessel
functions to simplify the integral over $x$ in
\eq\eqref{ortsympresultform} and obtain the Wigner surmise for the
GOE.  For $\lambda\to\infty$, the exponential and the difference of
the Bessel functions in the integral over $x$ can be replaced by
unity, and the Wigner surmise for the GSE follows trivially.

The distribution $P_{1\to 4}(s;\lambda)$ is plotted for several values
of $\lambda$ in \fig\ref{gausgaussimplefigure} (middle) and displays a
continuous interpolation between the GOE and GSE curves. In the
small-$s$ region, the level repulsion is of fourth order for
non-vanishing $\lambda$. This is visible in the plots and can be shown
by expanding $P_{1\to 4}(s;\lambda)$ for $\lambda>0$ and small $s$,
\begin{equation}
  P_{1\to 4}(s;\lambda) = c(\lambda) s^4 + \mathcal O(s^6)
\end{equation}
with
\begin{equation}
  c(\lambda)\sim\frac{\pi^2}{12\lambda^3}\quad\text{for }\lambda\to0\,.
\end{equation}

The large-$s$ behavor of $P_{1\to 4}(s;\lambda)$ can be obtained
using the asymptotic expansion
\begin{align}
  I_0(z)-I_1(z)=e^z\left[\frac1{\sqrt{8\pi}z^{3/2}}
    +\mathcal O(z^{-5/2})\right]
\end{align}
in \eq\eqref{ortsympresultform}, resulting in 
\begin{align}
  P_{1\to 4}(s;\lambda)\sim \sqrt{\frac\pi{32}}\,\frac C{D^3}\,s\,
  e^{-2(\lambda Ds)^2}\quad\text{for }s\to\infty\,.
\end{align}
Again, the large-$s$ behavior is dominated by the ensemble with the
smaller $\beta$.

\subsection{GUE to GSE}
\label{unisympsec}

Again, due to the presence of the GSE, we have two possibilities for
the GUE: \selfdual\ or not.  The former case is simpler and analyzed
here, while the latter case will be considered in
\sect\ref{sympgaussec}.  We first have to clarify how to obtain a
\selfdual\ \four\ matrix whose eigenvalues have the same probability
distribution as those of a \two\ matrix from the GUE.  In analogy to
\sect\ref{ortsympsec}, one could try $H_2\otimes\1_2$, but the
resulting matrix is not \selfdual.  Instead, we consider the matrix
\begin{align}
  H_2^4=\begin{pmatrix} H_2 & 0 \\ 0 & H_2^T \end{pmatrix}
\end{align}
with $H_2$ given in \eq\eqref{poisuniensform}.  The eigenvalues of
$H_2^4$ are obviously equal to those of $H_2$, but twofold degenerate.
Interchanging the second and third row and column of $H_2^4$, we
obtain the matrix
\begin{equation}\label{seduGUEmatrform} 
  H_{2}^\text{sd} =
  \begin{pmatrix}
    a & 0 & c_0 + ic_1 & 0 \\
    0 & a & 0 & c_0 - ic_1 \\
    c_0 - ic_1 & 0 & b & 0 \\
    0 & c_0 + ic_1 & 0 & b
  \end{pmatrix},
\end{equation}
which is \selfdual\ and has the same eigenvalues as $H_2^4$. A matrix
of this form was already introduced in \cite{Nieminen:2009}.

The proper \selfdual\ matrix for the GUE to GSE transition is thus
\begin{equation}\label{unisympensform}
  H = H_{2}^{\text{sd}} + \lambda H_4
\end{equation}
with $H_4$ given in \eq\eqref{poissympensform}.  The calculation of
the corresponding spacing distribution proceeds in close analogy with
the one presented in \app\ref{goegseapp}, and we find the closed
expression
\begin{align}
  P_{2\to 4}(s;\lambda) &= Ce^{-(\lambda Ds)^2}\notag\\
  &\times\big[2(Ds)^2-\sqrt{\pi}Dse^{-(Ds)^2}\erfi(Ds)\big]
  \label{unisympresultform}
\end{align}
with the imaginary error function $\erfi(x)=-i\erf(ix)$ and
\begin{align}
  D(\lambda) &= \frac1{\lambda \sqrt{\pi}} \left(2+\lambda^2
    -\lambda^4\,\frac{\arccsch\lambda}{\sqrt{1+\lambda^2}}\right),\\ 
  C(\lambda) &= \frac{2\lambda^3}{\sqrt{\pi}}(1+\lambda^2)D(\lambda)\,,
\end{align}
where $\arccsch$ is defined in \cite[Eq.~(4.6.17)]{Abram:1964}.  

In the limiting cases of $\lambda\to0$ and $\lambda\to\infty$ we have
\begin{align}
  D(\lambda)\sim
  \begin{cases}
    2/(\lambda\sqrt\pi) & \text{for }\lambda\to0\,,\\
    8/(3\lambda\sqrt\pi) & \text{for }\lambda\to\infty\,.
  \end{cases}
\end{align}
For $\lambda\to0$, the asymptotic expansion of the second term in the
square brackets of \eq\eqref{unisympresultform} yields $-1$.  This can
be neglected compared to the first term in the square brackets, which
gives the Wigner surmise for the GUE.  For $\lambda\to\infty$, Taylor
expansion of the square brackets in \eq\eqref{unisympresultform}
yields the Wigner surmise for the GSE.

The result \eqref{unisympresultform} is plotted in \fig
\ref{gausgaussimplefigure} (right).  In the small-$s$ region, we have
for $\lambda\ne0$
\begin{align}
  P_{2\to 4}(s;\lambda) = c(\lambda)s^4+\mathcal O(s^6)
\end{align}
with
\begin{align}
  c(\lambda)\sim\frac{256}{3\pi^3\lambda^2}\quad\text{for }\lambda\to0\,.
\end{align}

The large-$s$ behavor of $P_{2\to 4}(s;\lambda)$ can be obtained by
noticing that for large $s$ the first term in the square brackets of
\eq\eqref{unisympresultform} dominates the second term so that
\begin{align}
  P_{2\to 4}(s;\lambda)\sim 2CD^2\,s^2\,e^{-(\lambda Ds)^2}
  \quad\text{for }s\to\infty\,.
\end{align}
Again, the large-$s$ behavior is dominated by the ensemble with the
smaller $\beta$.

\subsection{GSE to GUE without \selfdual\ symmetry}
\label{sympgaussec}

In this section, we consider a matrix taken from the GSE whose Kramers
degeneracy is lifted by a perturbation taken from the GUE without
\selfdual\ symmetry.  As we shall see, this case also gives a surmise
for other transitions involving the GSE and another ensemble without
\selfdual\ symmetry.  We will return to this point in
\sect\ref{sec:gsetoother}.

\begin{figure}[t]
  \includegraphics[width=0.9\linewidth,clip=true]{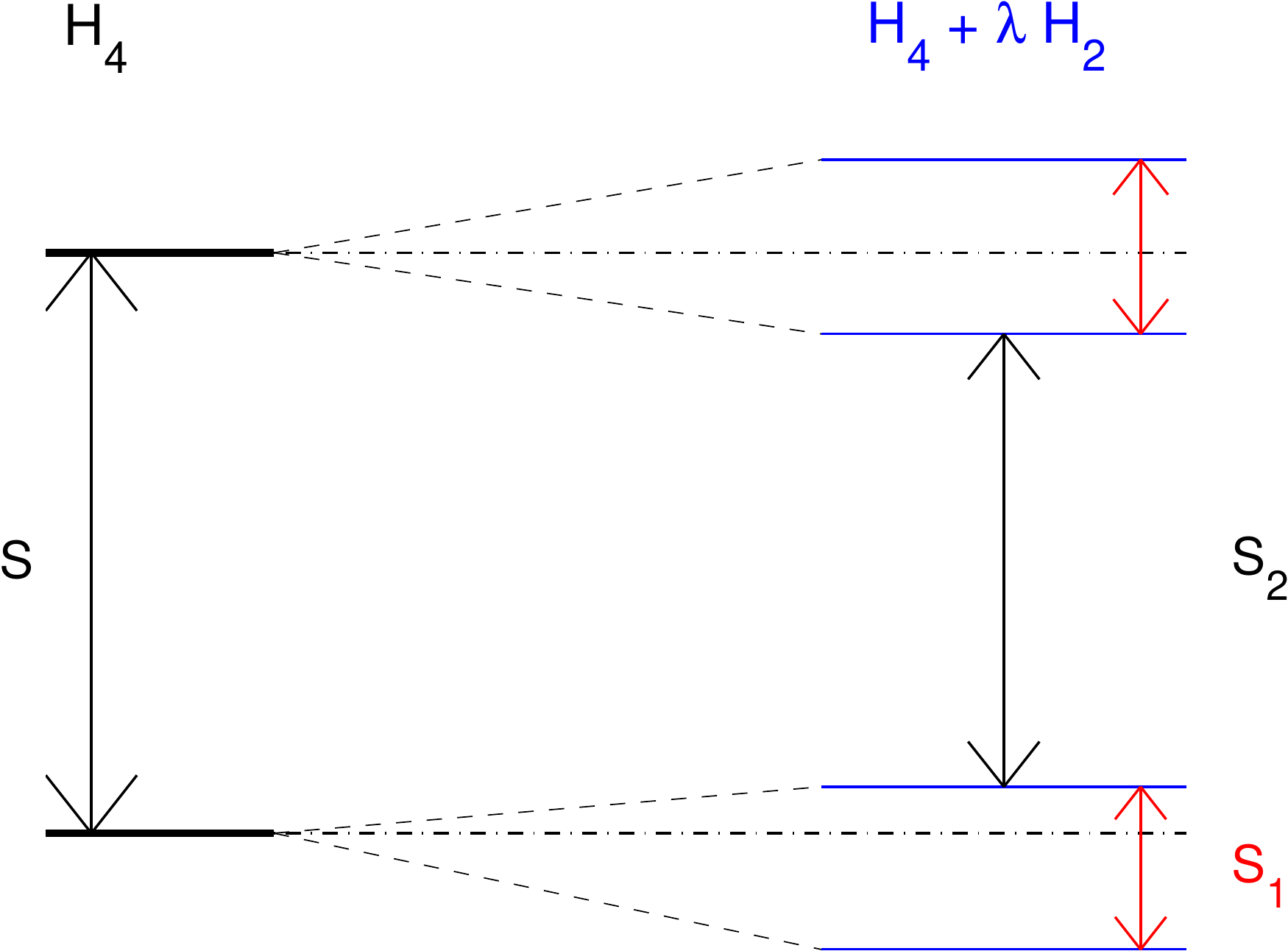}
  \caption{(Color online) Perturbation of GSE eigenvalues removing the degeneracy.}
  \label{gseperturbationfigure}
\end{figure}

\subsubsection{General considerations}
\label{sec:general}

The \four\ transition matrix is
\begin{equation}\label{sympunimatr4x4}
  H = H_4 + \lambda H_2
\end{equation}
with $H_4$ taken from the GSE and $H_2$ from the GUE, both in standard
normalization, \eq\eqref{standardnormform}. As $H_2$ has no
\selfdual\ symmetry, the two-fold degeneracy of the GSE spectrum is
removed and eigenvalue pairs split up.  If the perturbation is small,
there are two different spacing scales in this setup, as shown in
\fig\ref{gseperturbationfigure} where the perturbation of two
nearest-neighbor eigenvalues is sketched:
\begin{itemize}
\item $S_1$: The spacings between previously degenerate eigenvalues,
  which are of the same order of magnitude as the coupling parameter
  for small couplings. They are formed by the two smallest/largest
  eigenvalues of $H$.
\item $S_2$: The intermediate spacing, which is formed by the second
  and third largest eigenvalue of $H$.  In the limit $\lambda\to0$
  this is the original spacing of the GSE matrix $H_4$.
\end{itemize}

The joint probability density of the eigenvalues of $H$ is given, up
to a rescaling, by \cite[Eq.~(14.2.7)]{Mehta:2004}
\begin{align}\notag
  &P(\theta_1, \theta_2, \theta_3, \theta_4) = C_0\exp\left(-\sum_{i=1}^4 \theta_i^2\right)
  \Delta(\theta_1, \theta_2, \theta_3, \theta_4)\\\label{eq:jointsympherm}
  &\times\left[h(d_{21})h(d_{43}) + h(d_{32})h(d_{41}) - h(d_{31})h(d_{42})\right]
\end{align}
with
\begin{align}
 \Delta(\theta_1, \theta_2, \theta_3, \theta_4) &= \prod_{i<j} (\theta_j-\theta_i)\,,\\
 h(x) &= xe^{-x^2/\lambda^2}\,,\\
 d_{ij} &= \theta_i-\theta_j\,, \\
 C_0 &= \frac1{9\pi^{2}}\,\lambda^{-6}\,\left(2+\lambda^2\right)^5.
\end{align}
As we are only interested in spacings and thus in differences of
eigenvalues, we introduce new variables
\begin{align}
t_1 &= d_{21} = \theta_2 - \theta_1\,,\\
t_2 &= d_{32} = \theta_3 - \theta_2\,,\\
t_3 &= d_{43} = \theta_4 - \theta_3
\end{align}
and keep the original variable $\theta_1$.  The Jacobi determinant of
this transformation is $1$, and we can now perform the $\theta_1$
integration, which results (up to a constant factor) in
\begin{align}\label{eq:jointsympuni2}
  &P(t_1, t_2, t_3)= \Delta(-t_1, 0, t_2, t_2+t_3)\\\notag
  &\times\exp\left\{-\frac14\left[\left(t_1+2t_2+t_3\right)^2+2t_1^2+2t_3^2\right]\right\}\\\notag
  &\times\left[h(t_1)h(t_3) - h(t_1\!+\!t_2)h(t_2\!+\!t_3) + h(t_1\!+\!t_2\!+\!t_3)h(t_2)\right].
\end{align}

We now derive the distributions of the two different kinds of spacings
from this formula.  We assume
$\theta_1\le\theta_2\le\theta_3\le\theta_4$ and include the resulting
combinatorial factor of $4!$ explicitly.

\begin{figure*}[t]
  \includegraphics[width=\threepicwidth,clip=true]
  {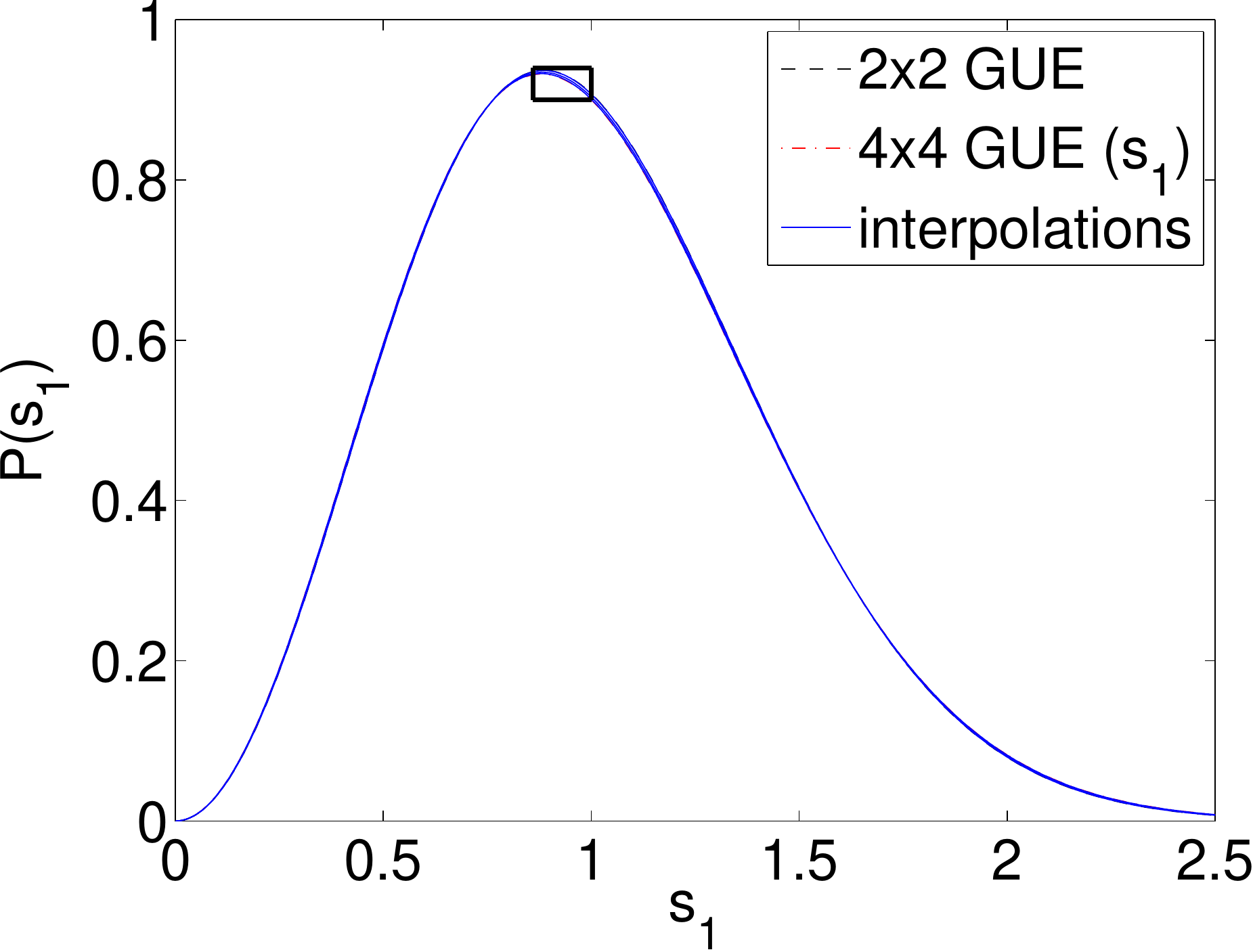}\hfill
  \includegraphics[width=0.97\threepicwidth,clip=true]
  {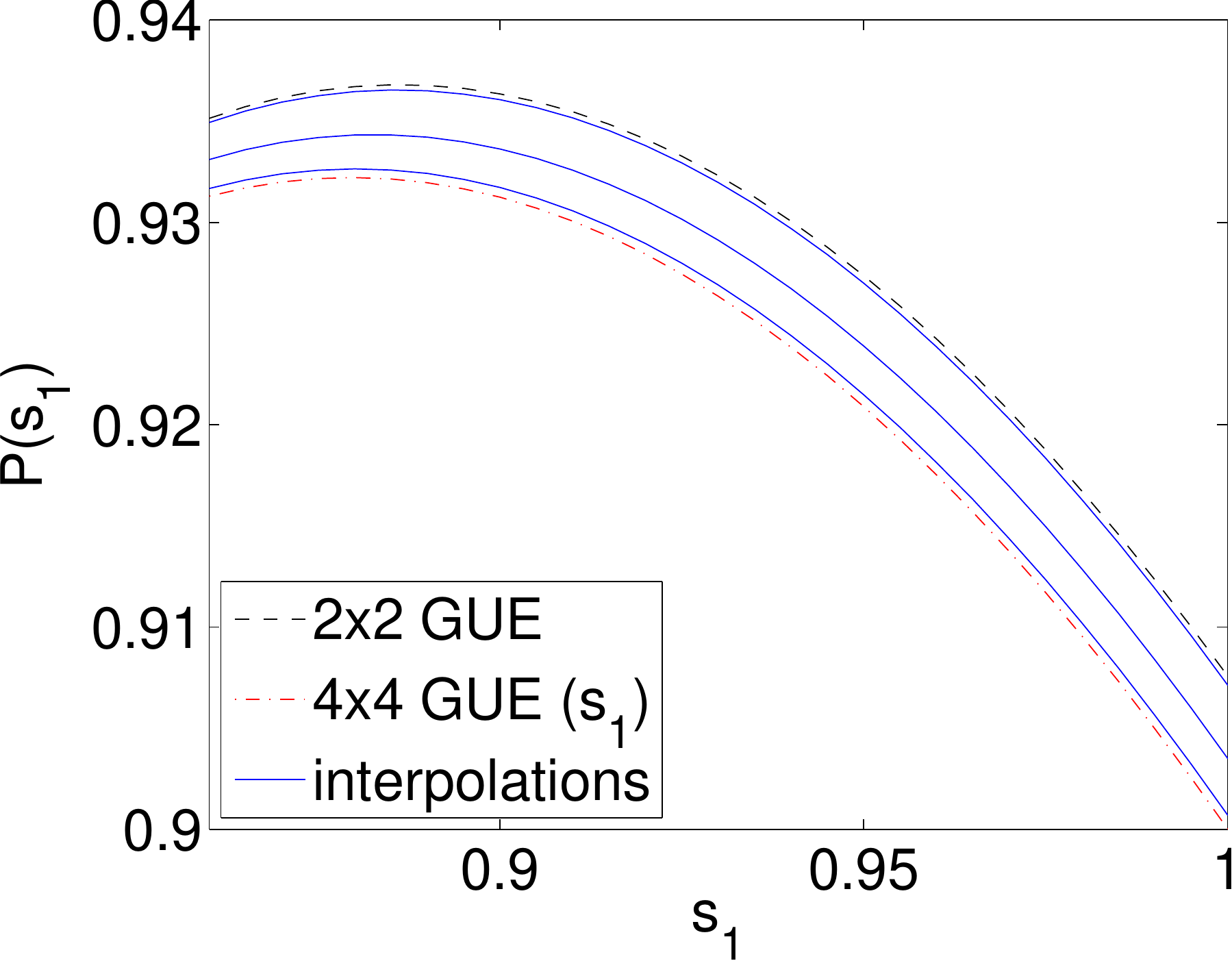}\hfill
  \includegraphics[width=\threepicwidth,clip=true]
  {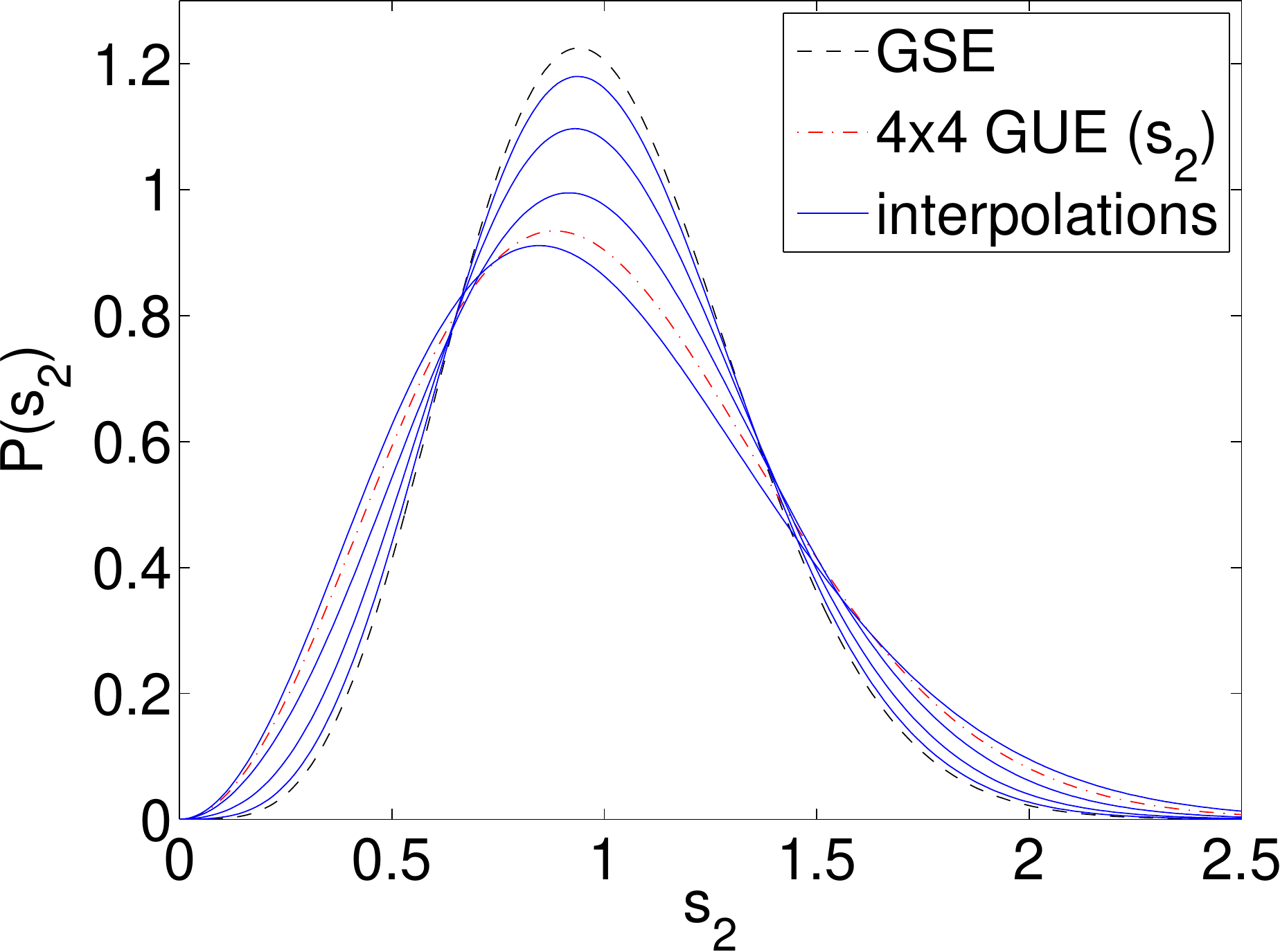}
  \caption{(Color online) Spacing distributions for the transition from GSE $\to$ GUE
    (non-\selfdual) for \four\ matrices and various values of the
    coupling parameter $\lambda$.  Left: spacings $s_1$ between
    previously degenerate eigenvalues.  Middle: also $s_1$, but zoomed
    in to the rectangular region indicated in the plot on the left;
    \two\ GUE stands for $\lambda\to0$; \four\ GUE stands for
    $\lambda\to\infty$; for the interpolation curves we chose
    $\lambda=0.4, 1, 2$ (maxima decreasing).  Right: intermediate
    spacing $s_2$; GSE stands for $\lambda=0$; \four\ GUE stands for
    $\lambda\to\infty$; for the interpolation curves we chose
    $\lambda=0.05, 0.15, 0.3, 1$ (maxima decreasing).}
  \label{fig:sympuni4x4}
\end{figure*}

\subsubsection{Spacings between originally degenerate eigenvalues}
\label{sympgauss1sec}

To obtain the distribution of the spacing between the two smallest
eigenvalues of $H$ (the two largest ones give the same result due to
symmetry), we set $t_1 = S_1$ and integrate over $t_2$ and $t_3$ from
$0$ to $\infty$.  This results in the spacing distribution
\begin{align}\label{eq:probsympunis1}
  P_{4\to2}^1(s_1; \lambda) =
  CD\int_{0}^{\infty}dt_2\,dt_3\,P(Ds_1, t_2, t_3) 
\end{align}
with
\begin{align}
 C(\lambda) &= \frac43\,\pi^{-3/2}\,\lambda^{-6}
 \left(2+\lambda^2\right)^{5},\\ 
 D(\lambda) &= C(\lambda)\,\int_0^{\infty}
 dS_1\,dt_2\,dt_3\,S_1\,P(S_1, t_2, t_3)\,. 
\end{align}
We replaced $S_1$ by $s_1$ to indicate that this is the spacing on the
unfolded scale, i.e., with a mean value of $1$.  One of the integrals
could in principle be done analytically, but this results in such a
lengthy expression that it seems more sensible to evaluate all
integrals numerically.  

The distribution in the limit $\lambda\to0$ can either be obtained by
perturbation theory, see \app\ref{pertmatrixgue}, or by directly
evaluating the spacing distribution in the limit $\lambda\to0$.
First note that
\begin{align}
 \lim_{\lambda\to0} \frac2{\sqrt{\pi} \lambda^3}\, x\, h(x) = \delta(x)\,,
\end{align}
where the $\lambda$-dependence of $h$, which is suppressed in our
notation, plays a crucial role.  As the mean value of the spacing
$S_1$ on the original scale has to become arbitrarily small in the
GSE-limit due to the Kramers degeneracy, we consider a rescaled
spacing $\tilde s_1 = S_1/\lambda$.  Therefore $h(S_1)$ becomes for
small $\lambda$
\begin{align}
 h(S_1) = h(\lambda\tilde s_1) \overset{\lambda\to0}\approx \lambda\tilde s_1
\, e^{-\tilde s_1^2}\,.
\end{align}
With these considerations we obtain from \eq\eqref{eq:jointsympuni2}
\begin{align}
  P(\lambda&\tilde s_1, t_2, t_3) \propto
  \lambda\,e^{-\frac14\left[\left(\lambda\tilde s_1+2t_2+t_3\right)^2
      +2\lambda^2\tilde s_1^2+2t_3^2\right]}\hfill\\\notag
  \times\bigg[&\frac{2\tilde s_1^2}{\sqrt{\pi}\lambda} e^{-\tilde s_1^2}
  \,\delta(t_3)(\lambda\tilde s_1+t_2)
  (\lambda\tilde s_1+t_2+t_3)t_2(t_2+t_3)\\\notag
  &-\delta(\lambda\tilde s_1+t_2)\,\delta(t_2+t_3)\,\lambda\tilde s_1 t_2 t_3
  (\lambda\tilde s_1+t_2+t_3)\\\notag
  &+\delta(\lambda\tilde s_1+t_2+t_3)\,\delta(t_2)\,\lambda\tilde s_1
  (\lambda\tilde s_1+t_2)(t_2+t_3)t_3\bigg]
\end{align}
as $\lambda\to0$. The last two terms in square brackets vanish upon
evaluation of the $t_2$ and $t_3$ integrals, because the zeros of the
arguments of their $\delta$-functions lie outside of the integration
region.  Performing the $t_3$ integration in the first term we obtain
for nonzero $\lambda$ and $\tilde s_1$
\begin{align}
 P_{4\to2}^1(\tilde s_1; \lambda) \overset{\lambda\to0}\propto
\tilde s_1^2\,e^{-\tilde s_1^2}\,.
\end{align}
Up to normalization and rescaling this is the spacing distribution of
a \two\ GUE matrix.

In the opposite limit $\lambda\to\infty$ the result
\eqref{eq:probsympunis1} reduces to the distribution of the first and
last spacings of a pure \four\ GUE matrix.  This distribution can be
obtained from similar considerations, starting from
\cite[Eq.~(3.3.7)]{Mehta:2004}.

The result \eqref{eq:probsympunis1} is shown in
\fig\ref{fig:sympuni4x4} (left and middle) for several values of
$\lambda$, along with the limiting distributions for $\lambda\to0$
and $\lambda\to\infty$.  All these curves are very similar and can
only be distinguished by the naked eye in the zoomed-in plot.

We have validated the result \eqref{eq:probsympunis1} by comparing it
to the spacing distribution of numerically obtained \four\ random
matrices.

\subsubsection{Perturbed GSE-spacing}
\label{sympgauss2sec}

We now consider the perturbed spacing of the original GSE matrix,
which was formed by the two degenerate eigenvalue pairs of $H_4$.  The
distribution of this spacing is obtained by setting $t_2 = S_2$ and
integrating $P$ defined in \eq\eqref{eq:jointsympuni2} over $t_1$ and
$t_3$ from $0$ to $\infty$.  With proper normalization as given in
\eq\eqref{standardnormform}, this yields
\begin{align}\label{eq:probsympunis2}
  P_{4\to2}^2(s_2; \lambda) =
  CD\int_{0}^{\infty}dt_1\,dt_3\,P(t_1, Ds_2, t_3)
\end{align}
with
\begin{align}
  C(\lambda) &= \frac43\,\pi^{-3/2}\,\lambda^{-6}
  \left(2+\lambda^2\right)^{5},\\ 
  D(\lambda) &= C(\lambda)\int_0^{\infty}
  dS_2\,dt_1\,dt_3\,S_2\,P(t_1, S_2, t_3)\,.
\end{align}
Again, the replacement of $S_2$ by $s_2$ means that this is the
intermediate spacing on the unfolded scale, i.e., with a mean value of
$1$.  

In the limit $\lambda\to0$ the result \eqref{eq:probsympunis2} reduces
to the Wigner surmise for the GSE, while in the opposite limit
$\lambda\to\infty$ it reduces to the spacing distribution of the
intermediate spacing of a pure \four\ GUE matrix, which can again be
obtained from similar considerations.  

The result \eqref{eq:probsympunis2} is shown in
\fig\ref{fig:sympuni4x4} (right) for several values of $\lambda$,
along with the limiting distributions for $\lambda\to0$ and
$\lambda\to\infty$.  The maximum of the interpolation first drops down
as $\lambda$ is increased from $0$, while at a value of $\lambda$
around $1$ it starts to rise again as the distribution approaches its
$\lambda\to\infty$ limit.  Note that the limiting distributions of
$s_1$ and $s_2$ for $\lambda\to\infty$, i.e., the red dashed curves in
\fig\ref{fig:sympuni4x4}, turn out to be almost identical to each
other and to the Wigner surmise for the GUE.

We have also validated the result \eqref{eq:probsympunis2} by
comparing it to the spacing distribution of numerically obtained
\four\ random matrices.

\section{Application to large spectra}
\label{largespectraapp}

In this section we will show numerically that the formulas derived in
\sect\ref{derspacsec} for small matrices describe the spacing
distributions of large random matrices very well.  This observation
should be viewed as our main result.  

When comparing the results obtained from large matrices to our
generalized Wigner surmises, a natural question is how the
corresponding coupling parameters, i.e., $\lambda$ in
\eq\eqref{eq:trans}, should be matched.  This question will be
addressed in the next subsection based on perturbation theory, while
the numerical results will be presented in the remaining subsections.

\subsection{Matching of the coupling parameters}
\label{largespectrapert}

The setup is most easily explained by means of the transition from
Poisson to the GUE.  The Poisson case is represented by a diagonal
$N\times N$ matrix $H_0$ with independent entries $\theta_i$
($i=1,\ldots,N$), each distributed according to the same distribution
$\mathcal P(\theta)$, which we choose independent of $N$.  The
eigenvalue density of $H_0$ is thus $\rho_0(\theta) = N\mathcal
P(\theta)$, and the local mean level spacing is $1/\rho_0(\theta)$.
We consider
\begin{equation}\label{defpoisgaus}
  H = H_0 + \alpha H_2\,,
\end{equation}
where $H_2$ is an $N\times N$ random matrix taken from the GUE,
subject to the usual normalization, \eq\eqref{standardnormform}.

As in the \two\ case, the eigenvalues $\theta_i$ will experience a
repulsion through $H_2$. We will show in first-order perturbation
theory that the relevant quantity for the repulsion is a combination
of the eigenvalue density of $H_0$ and the variance of the matrix
elements of $H_2$.

Ordinary perturbation theory in $\alpha$ yields a first-order
eigenvalue shift of the $\theta_i$ of
\begin{equation}
  \Delta \theta_i^{(1)} = \alpha (H_2)_{ii}\,. 
\end{equation}
This shift does not lead to a correlation of the eigenvalues, as it
just adds an independent Gaussian random number to each of
them. Therefore, the eigenvalues remain uncorrelated, and their
spacing distribution remains Poissonian.

However, if there is a small spacing of order $\alpha$ between
two\footnote{For small $\alpha$, we are unlikely to find three or more
  small (i.e., of order $\alpha$) consecutive spacings.}  adjacent
eigenvalues $\theta_k$ and $\theta_\ell$ of $H_0$, first-order
almost-degenerate perturbation theory \cite{Gottfried:1966} predicts
that the perturbed eigenvalues are the eigenvalues of the matrix
\begin{equation}\label{pertmatrixlargespectra}
  \begin{pmatrix} \theta_k & 0 \\ 0 & \theta_\ell \end{pmatrix} 
  + \alpha \begin{pmatrix} (H_2)_{kk} & (H_2)_{k\ell} \\ 
    (H_2)_{\ell k} & (H_2)_{\ell\ell} \end{pmatrix}.
\end{equation}
This matrix is almost identical to the \two\ matrix considered in
\sect\ref{poisunisec}, \eq\eqref{poisuniensform}, with two
differences: (i) The unperturbed eigenvalues $\theta_k$ and $\theta_l$
are shifted, but this does not affect the spacing distribution.  (ii)
The mean spacing of the unperturbed eigenvalues is not $1$, but
$1/\rho_0(\theta)$.  We dropped the subscript on the eigenvalue
$\theta$ here, because adjacent eigenvalues are very close for large
$N$, and therefore $\rho_0(\theta_k) \approx \rho_0(\theta_\ell) =
\rho_0(\theta)$.

To be able to match to the \two\ formulas, we have to correct for the
different mean spacing of the unperturbed matrix. We can do this by
multiplying the matrix in \eq\eqref{pertmatrixlargespectra} by
$\rho_0(\theta)$ without affecting the normalized spacing
distribution.  This results in the relation
\begin{equation}\label{lambeffform}
  \lambda(\theta) = \rho_0(\theta)\,\alpha
\end{equation}
between the coupling parameters of the \two\ and the $N\times N$ case.
Note that the \two\ parameter $\lambda$ has acquired a dependence on
the eigenvalue $\theta$ of $H$ through the local eigenvalue density of
$H_0$.  To be able to describe the spacing distribution of $H$ in the
spectral region around $\theta$ by the generalized Wigner surmise, we
assume that we have to insert this $\lambda(\theta)$ into the \two\
formulas. This choice of universal coupling parameter is in line with
an ``unfolded'' coupling parameter mentioned in
\cite{Guhr:1996wn,Pandey:1981} and a similar result from perturbation
theory \cite{French:1988}.  Appendix~\ref{pertcalc} contains a
calculation for large matrices in second-order perturbation theory,
also showing that the strength of the perturbation to be used in the
generalized Wigner surmise only depends on the combination
$\rho_0(\theta)\,\alpha$.

We now turn from the example ``Poisson to GUE'' to the general case,
which we write as
\begin{equation}
  \label{eq:alpha}
  H=H_{\beta}+\alpha H_{\beta'}\,.
\end{equation}
The same considerations hold with two modifications: (i) The
unperturbed matrix is not necessarily diagonal by
construction. However, it can be diagonalized by a transformation that
can be absorbed in the perturbation.\footnote{Note that we choose the
perturbations $H_{\beta'}$ such that their probability distribution
is always invariant under the transformations that diagonalize
$H_{\beta}$, just like in the Poisson to RMT cases.  However, this
does not work for some of the transitions between the GSE and ensembles
without \selfdual\ symmetry, which we discuss separately in
\sect\ref{sec:gsetoother}.}  We can therefore treat it as diagonal (with
eigenvalues correlated as dictated by the unperturbed ensemble).  (ii)
The mean spacing $\bar s_\beta$ of the unperturbed \two\ (\four)
matrix from \sect\ref{derspacsec} is
\begin{equation}
  \begin{split}
    \bar s_0 &= 1\text{ (Poisson)}\,,\quad
    \,\bar s_1 = \sqrt{\pi}\text{ (GOE)}\,,\\
    \bar s_2 &= \frac4{\sqrt\pi}\text{ (GUE)}\,,\quad
    \bar s_4 = \frac{16}{3\sqrt\pi}\text{ (GSE)}\,.
  \end{split}
\end{equation} 
Therefore, we now have to multiply \eq\eqref{eq:alpha} by $\bar
s_\beta \rho_\beta(\theta)$ to get the correct mean spacing $\bar
s_\beta$ for the unperturbed matrix. This results in a universal, but
$\theta$-dependent, coupling parameter
\begin{equation}\label{lambeffform2}
  \lambda(\theta) = \bar s_\beta \rho_\beta(\theta)\alpha
\end{equation}
with the eigenvalue density $\rho_\beta(\theta)$ of the unperturbed
matrix.  Equation~\eqref{lambeffform2} holds for all the transitions
we consider, and in each case $\beta$ is the Dyson index of the
unperturbed ensemble.

\begin{figure*}[t]
  \includegraphics[width=\onepicwidth,clip=true]
  {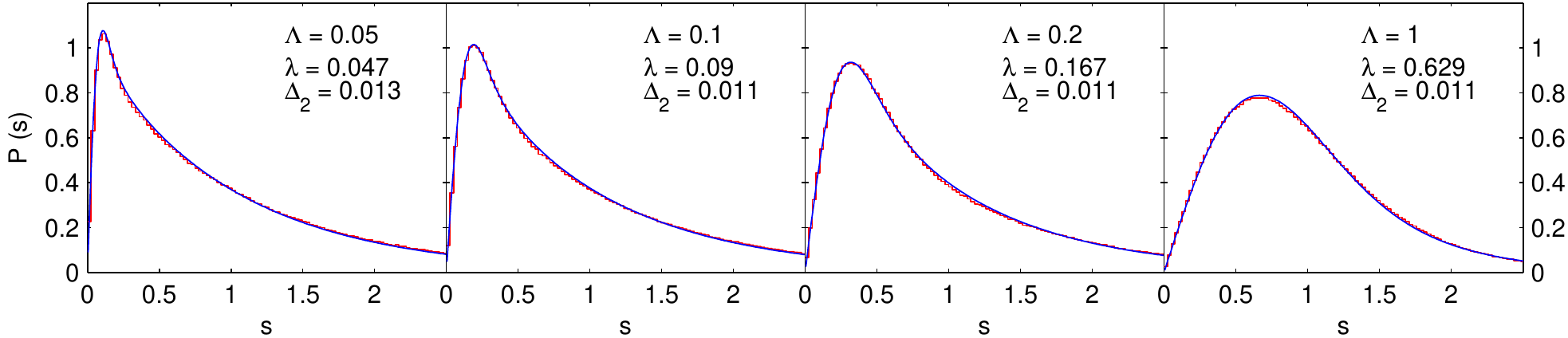}
  \includegraphics[width=\onepicwidth,clip=true]
  {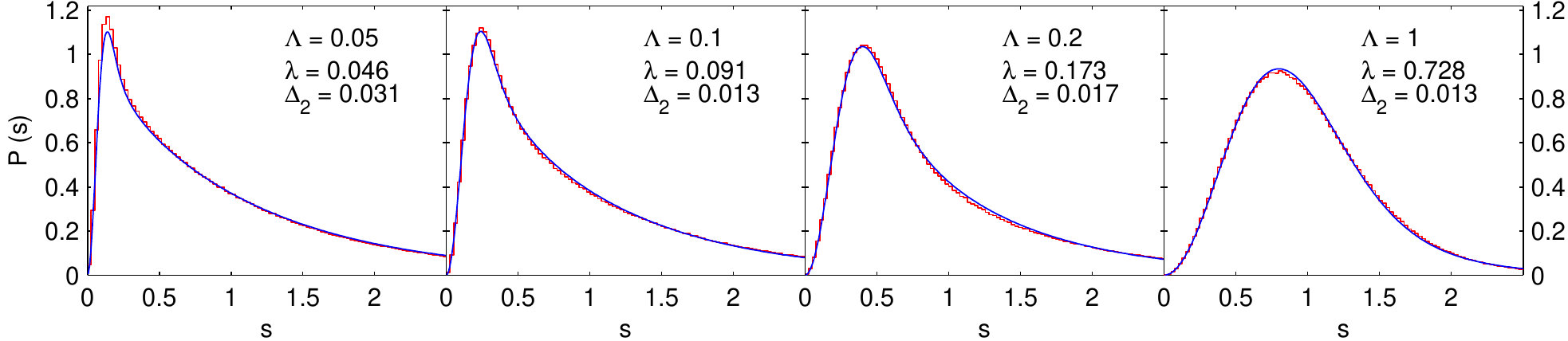}
  \includegraphics[width=\onepicwidth,clip=true]
  {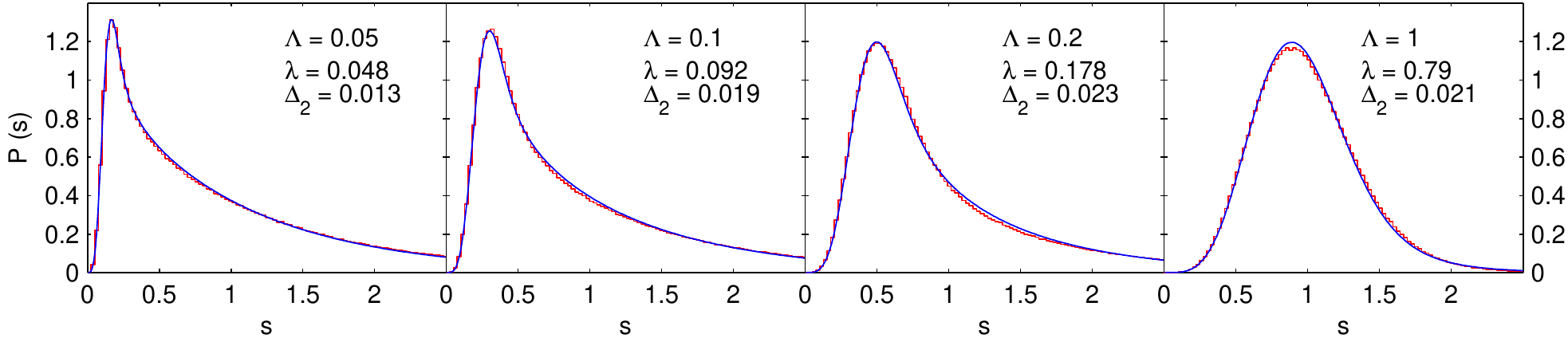}
  \caption{(Color online) Spacing distributions for the transition of large matrices
    from Poisson to GOE (top), GUE (middle), and GSE (bottom) with
    several values of the coupling $\Lambda$ in
    \eq\eqref{normalisePoissontransitionmatrices}.  The histograms
    show the numerical data, while the full curves are the analytical
    results for \two\ (\four) matrices with fitted coupling parameter
    $\lambda$, see \sects\ref{poisortsec} through \ref{poissympsec}.
    The quantity $\Delta_2$ defined in \app\ref{fittingmethod} is a
    measure of the fit quality, which is small for a good fit.  Each
    plot has been obtained by diagonalizing $50,000$ matrices with
    $400$ non-degenerate eigenvalues.}
  \label{poisgausfigure}
\end{figure*}

In turn, this perturbative argument provides us with a formula
of how to choose the coupling $\alpha$ in large matrices
in order to approximate the spacing distribution of $H$ by
\two\ (\four) formulas with parameter $\lambda$, i.e.,
\begin{equation}\label{suitedalphaform}
  \alpha = \frac{\lambda}{\rho_\beta(\theta) \bar s_\beta}\,,
\end{equation}
where $\rho_{\beta}(\theta)$ is the eigenvalue density in the spectral
region we wish to study.  In this way we can choose a value of
$\lambda$ resulting in a spacing distribution roughly in the middle of
the two limiting cases.  Choosing $\alpha$ in \eq\eqref{eq:alpha}
without this guidance is likely to result in a spacing distribution
that is dominated by one of the limiting cases.

\subsection{Transitions from integrable to chaotic}
\label{sec:poisgausslarge}

\begin{figure*}[t]
  \hspace{0.01cm}
  \includegraphics[width=0.95\threepicwidth,clip=true]
  {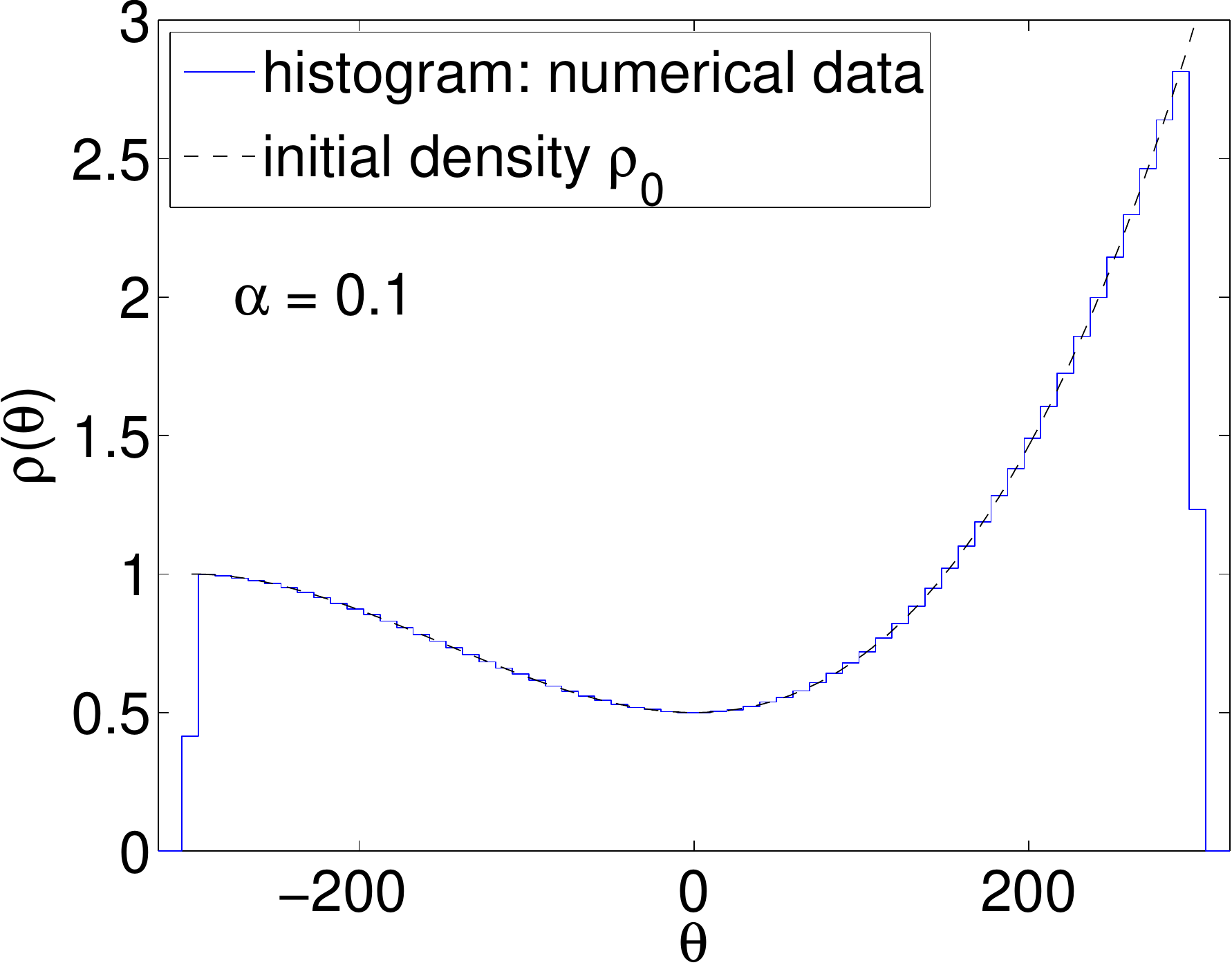}
  \hspace{0.17cm}
  \includegraphics[width=0.95\threepicwidth,clip=true]
  {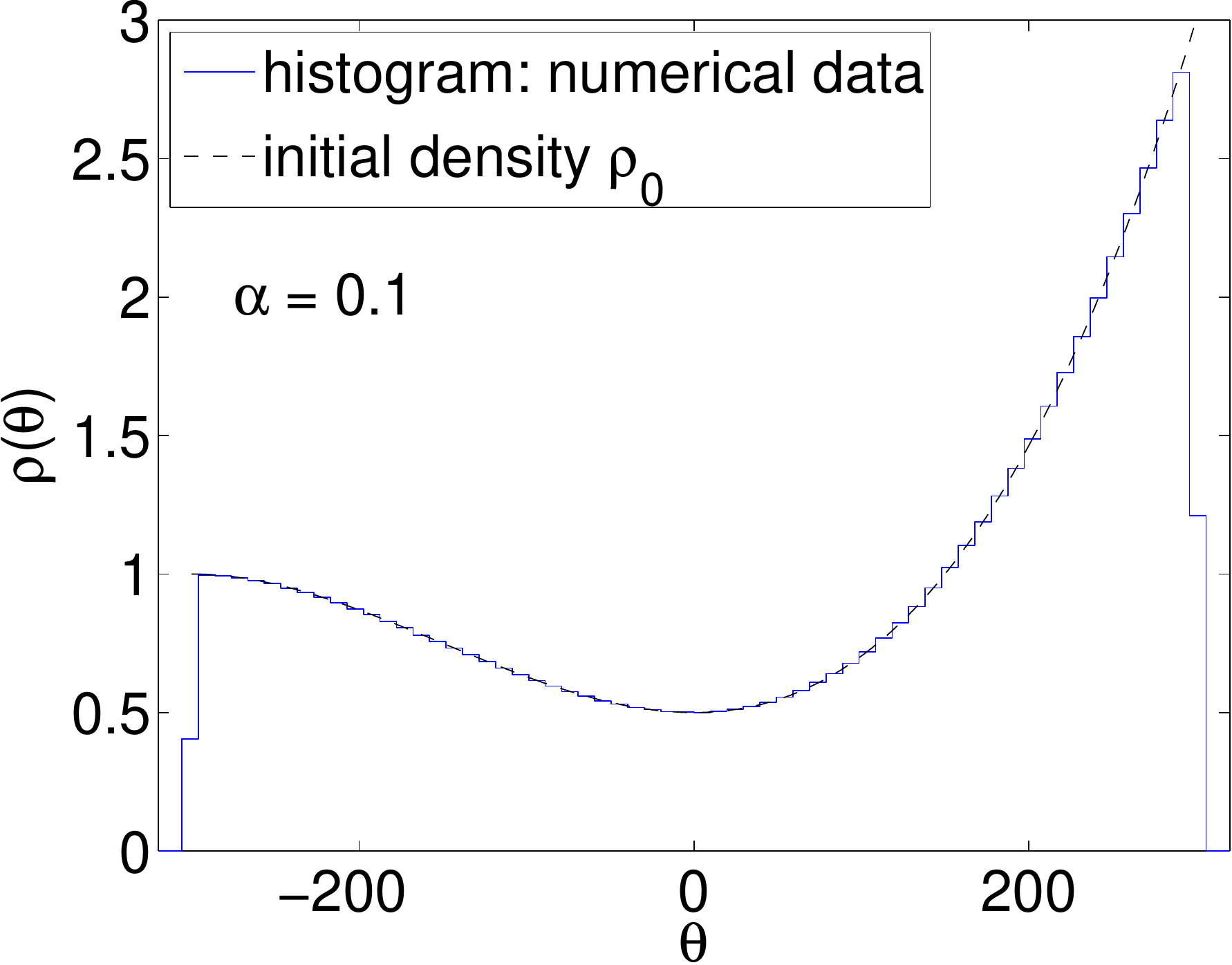}
  \hspace{0.17cm}
  \includegraphics[width=0.95\threepicwidth,clip=true]
  {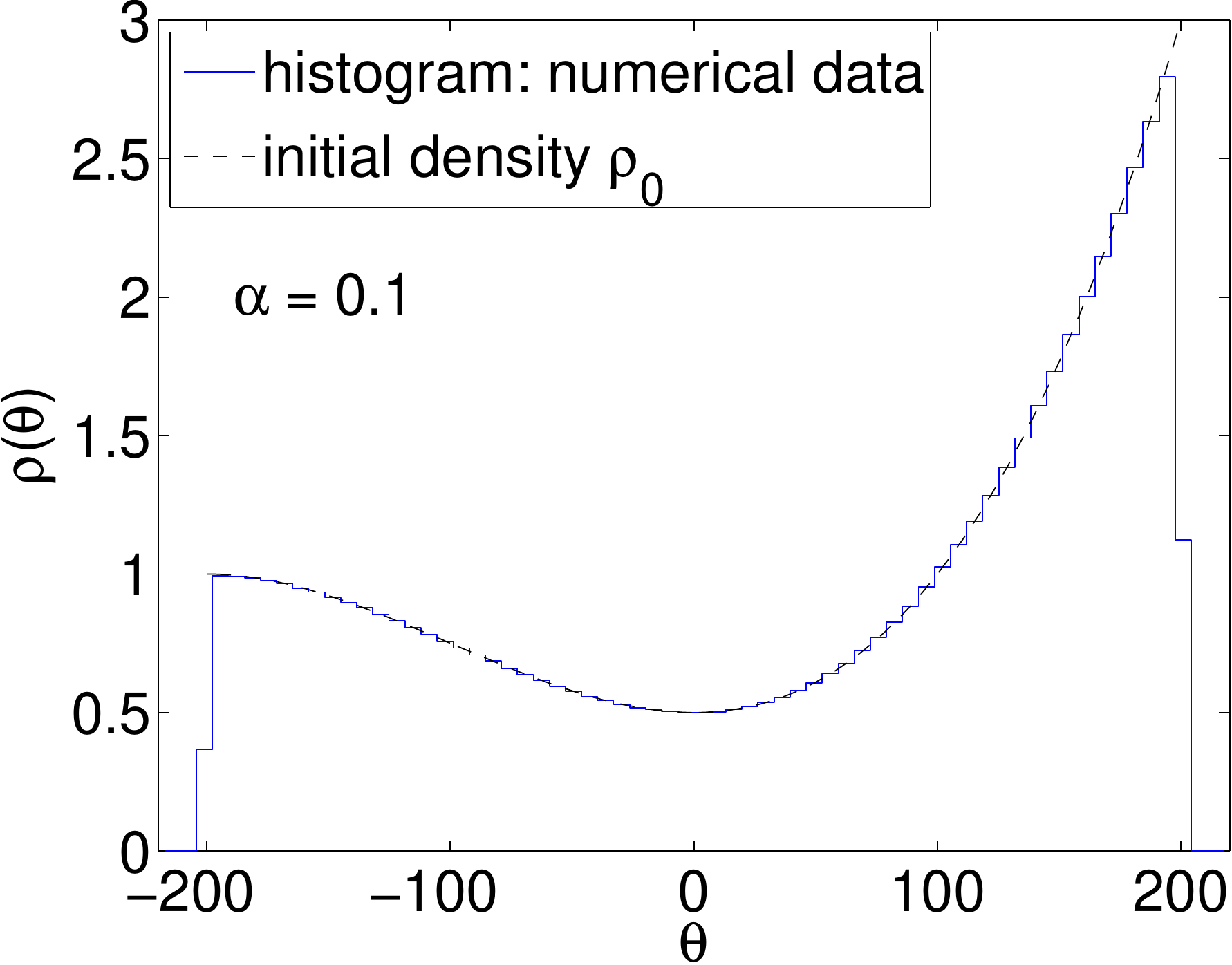}
  \includegraphics[width=\threepicwidth,clip=true]
  {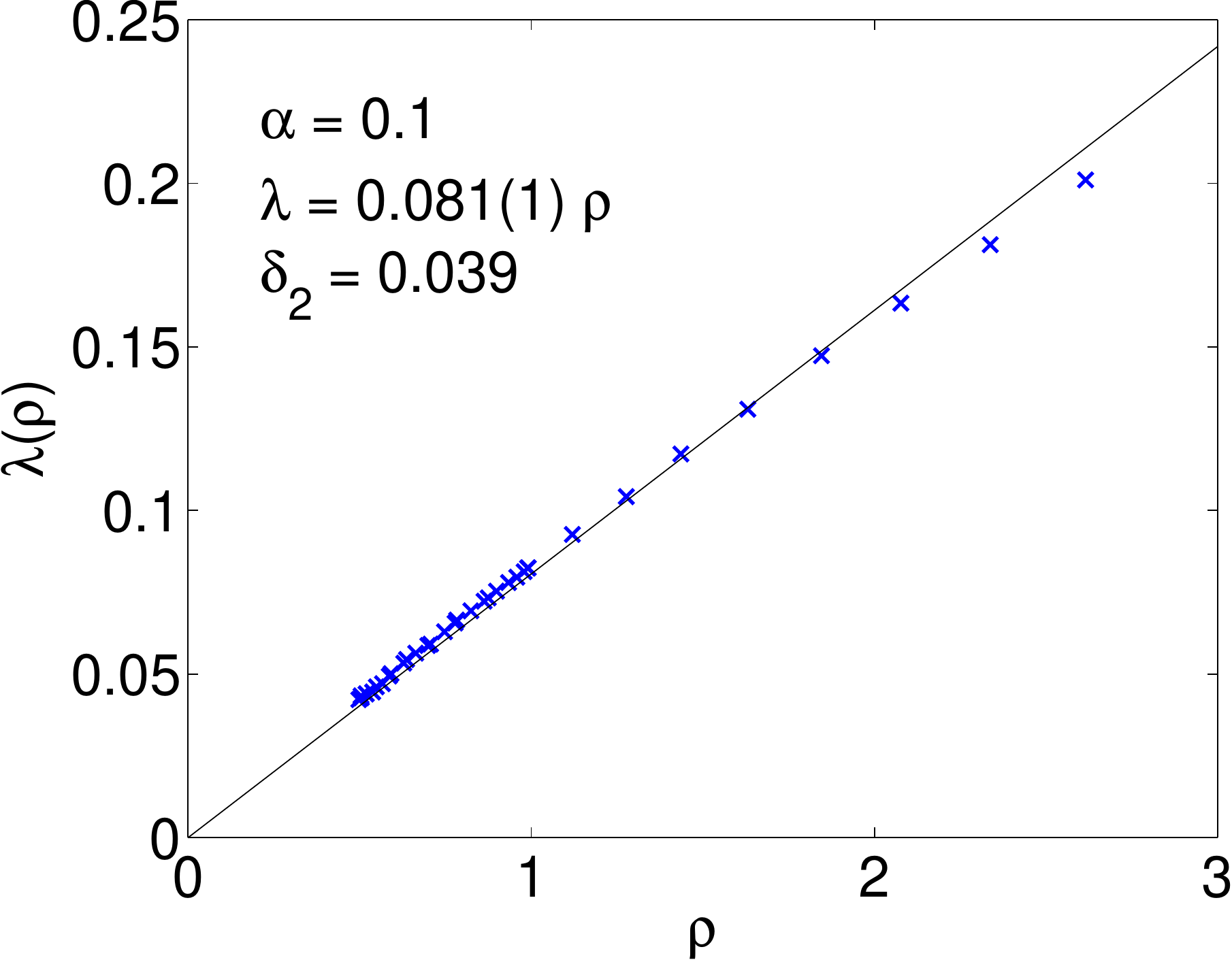}
  \includegraphics[width=\threepicwidth,clip=true]
  {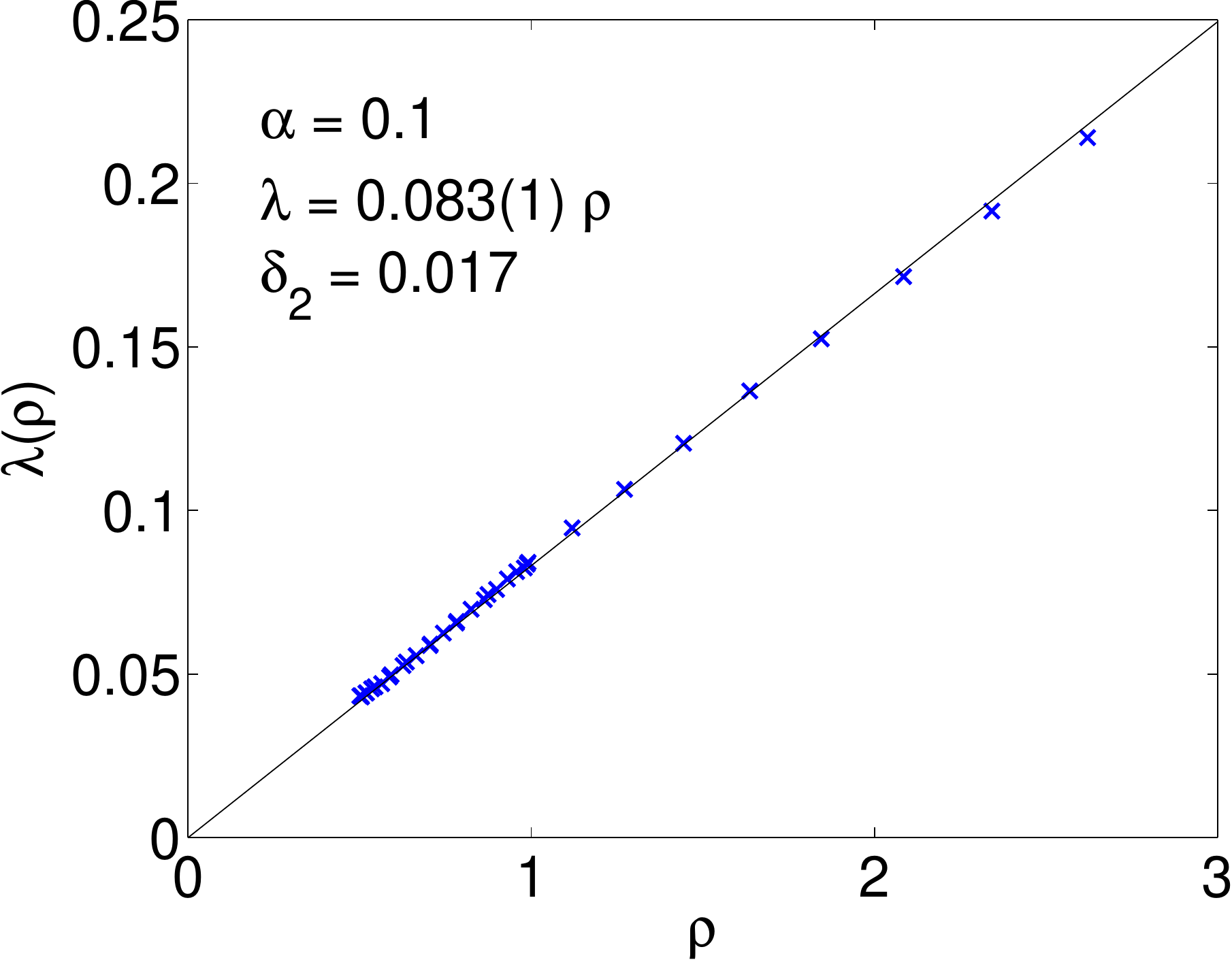}
  \includegraphics[width=\threepicwidth,clip=true]
  {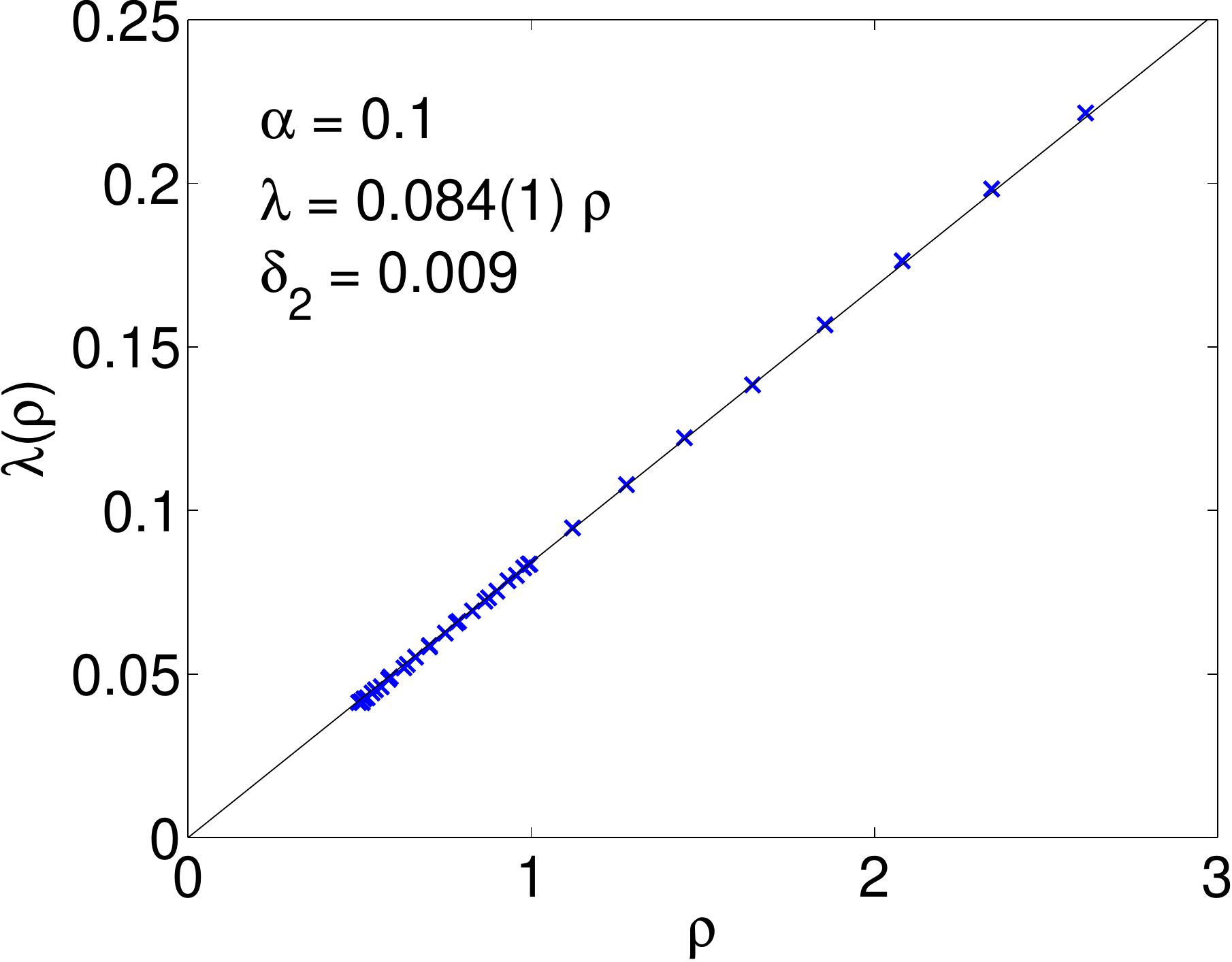}
  \caption{(Color online) Transitions Poisson $\to$ GOE (left), GUE (middle), GSE
    (right).  Top: unperturbed and perturbed eigenvalue density, the
    latter obtained numerically.  Bottom: effective coupling $\lambda$
    obtained from fits of the \two\ (\four) level spacings $P_{0 \to
      \beta'}(s)$ (see \eqs\eqref{poisortresultform},
    \eqref{poisuniresultform}, and \eqref{poissympresultform}) as a
    function of the local eigenvalue density in $35$ equally large
    windows of the spectrum.  Linear fit and proportionality factor
    with errors defined by the $95\%$-confidence interval are given in
    the plots.  The quantity $\delta_2$ defined in
    \eq\eqref{eq:definedelta2} is a measure of the fit quality, which
    is small for a good fit.  The numerical data were obtained from
    $10^5$ random matrices of dimension $600$ (GOE, GUE) or $800$
    (GSE).}
  \label{eigdenspoisgausfigure}
\end{figure*}

\subsubsection{Check of Wigner surmise}
\label{largespectra1}

We first consider transitions from Poisson to RMT for matrices with
$N$ non-degenerate eigenvalues.  The explicit numerical realization is
the Hamiltonian
\begin{equation}\label{normalisePoissontransitionmatrices}
 H = H_0 + \frac{\Lambda}{\rho_0(0)} H_{\beta'}\,,
\end{equation}
where $H_{\beta'}$ is a matrix taken from one of the Gaussian
ensembles, with normalization as given in \eq\eqref{standardnormform}.
$H_0$ is the same matrix as in \eq\eqref{defpoisgaus} for the
perturbation $H_{\beta'}$ in GOE or GUE, whereas a \selfdual\ $H_0$ is
constructed by a direct product with $\1_2$ as in
\sect\ref{poissympsec} if the perturbation is taken from the GSE.  We
choose a Gaussian for the distribution of the eigenvalues of $H_0$,
i.e., $\mathcal P(\theta) = (1/\sqrt{2\pi}) \exp(-\theta^2/2)$, so
$\rho_0(0) = N/\sqrt{2\pi}$.  From \eq\eqref{suitedalphaform} we would
then expect the spacing distribution in the center of the spectrum
around $\theta=0$ to be approximated by the corresponding \two\
formulas \eqref{poisortresultform}, \eqref{poisuniresultform}, and
\eqref{poissympresultform} with coupling $\lambda=\Lambda$.

As can be seen in \fig\ref{poisgausfigure}, the formulas for the \two\
matrices indeed describe the spectra of large matrices quite well in a
wide range of the coupling parameter $\Lambda$. The spacing
distribution was evaluated in the center of the spectrum, defined as
the interval $(-0.2,0.2)$, because the eigenvalue density is almost
constant and equal to $\rho_0(0)$ in this region so that no unfolding
is needed.  The analytical curve was obtained by a fit (see
\app\ref{fittingmethod} for details) of the \two\ (or \four) formula
to the numerical data with fit parameter $\lambda$.  As expected by
the perturbative considerations, $\lambda$ comes out on the same order
of magnitude as $\Lambda$, and almost matches for small $\Lambda$.
However, $\lambda$ is considerably smaller than $\Lambda$ for stronger
couplings.  Presumably, the repulsion of the many other eigenvalues in
the spectrum not present in the smallest matrices has a squeezing
effect on the spacing, which works against the repulsion caused by the
perturbation.  This would explain the smaller coupling parameter.

\subsubsection{Dependence of coupling parameter on eigenvalue density}
\label{eigdenstolambdapoisgaus}

\begin{figure*}[t]
  \includegraphics[width=\onepicwidth,clip=true]
  {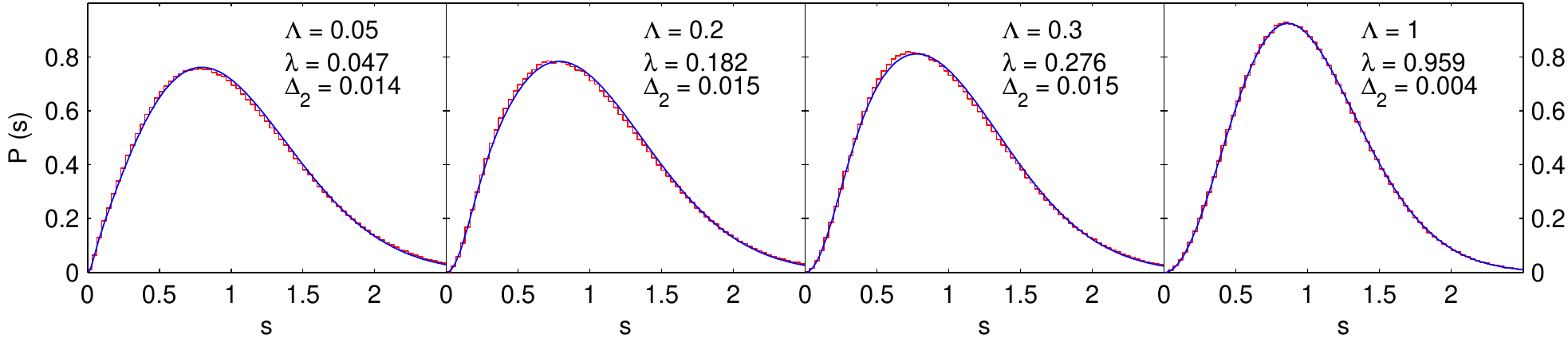}
  \includegraphics[width=\onepicwidth,clip=true]
  {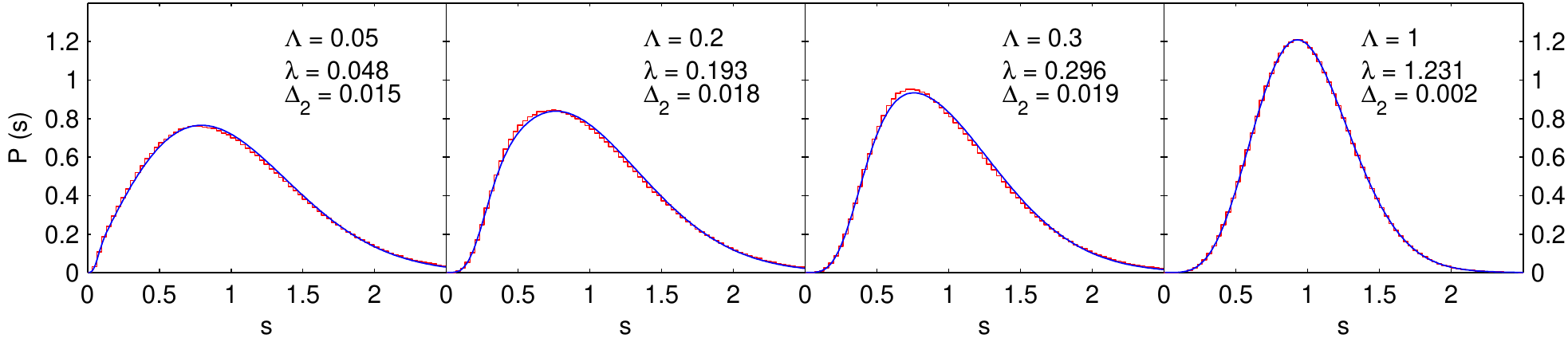}
  \includegraphics[width=\onepicwidth,clip=true]
  {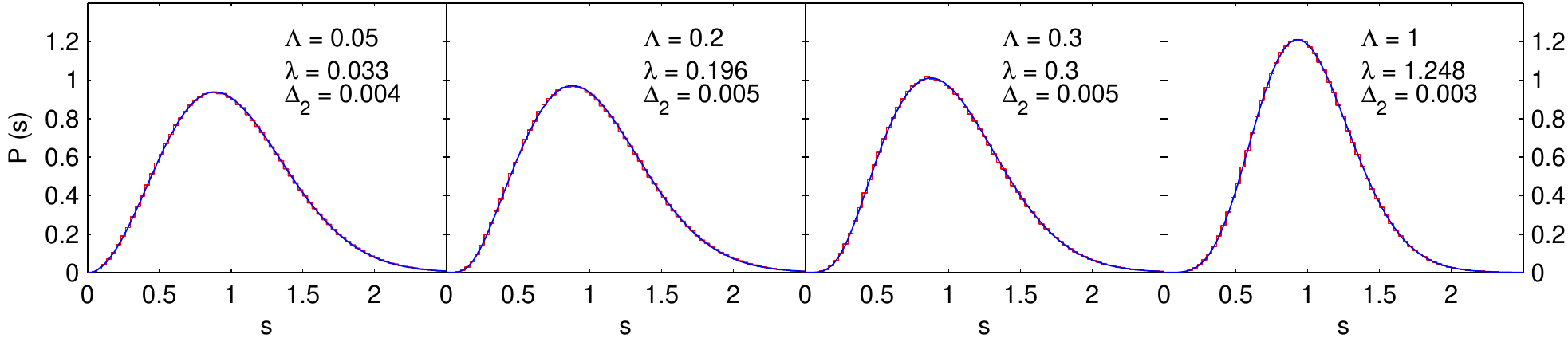}
  \caption{(Color online) Spacing distributions for the transition of large matrices:
    GOE $\to$ GUE (top), GOE $\to$ GSE (middle), and GUE $\to$ GSE
    (bottom), with several values of the coupling $\Lambda$ in
    \eq\eqref{normaliseGaussiantransitionmatrices}.  The histograms
    show the numerical data, while the full curves are the analytical
    results for \two\ (\four) matrices with fitted coupling parameter
    $\lambda$, see \sects\ref{ortunisec} through \ref{unisympsec}.
    The quantity $\Delta_2$ defined in \app\ref{fittingmethod} is a
    measure of the fit quality, which is small for a good fit.  Each
    plot has been obtained by diagonalizing $50,000$ matrices with
    $400$ non-degenerate eigenvalues.}
  \label{gausgausfigure}
\end{figure*}

The considerations in \sect\ref{largespectrapert} imply a linear
relation between local eigenvalue density and effective coupling
parameter, \eq\eqref{lambeffform2}, for matrices of the form given in
\eq\eqref{defpoisgaus}. This means that a perturbation should have a
different impact on the spacing distribution of a single matrix in
different regions of its spectrum (as qualitatively observed in
\cite{Lenz:1991}).  This subsection provides a detailed analysis of
this phenomenon.

Again, we consider a diagonal Poissonian matrix $H_0$ of large
dimension perturbed by a matrix taken from one of the Gaussian
ensembles $H_{\beta'}$,
\begin{equation}\label{mixingpoislambdaorig}
H = H_0 + \alpha H_{\beta'}\,.
\end{equation}
This time we will choose some fixed $\alpha$ and look separately at
different parts of the spectrum of $H$ with a varying eigenvalue
density.  According to \eq\eqref{lambeffform2} the effective \two\
(\four) coupling parameter $\lambda$ should be the product of
$\alpha$ and the local eigenvalue density of $H_0$.  In
\app\ref{pertcalc} we show in perturbation theory up to second order
that the local coupling parameter is in fact a function of this
product.

To treat such a system numerically one has to construct a Poissonian
ensemble with a varying eigenvalue density, perturb it, and measure
the coupling parameter in different parts of the spectrum. This is
done by cutting the spectrum into small windows with approximately
constant eigenvalue density and fitting (see \app\ref{fittingmethod}
for details) the spacing distributions inside the windows to the
formulas for the \two\ (\four) matrices.  We therefore obtain a fitted
coupling parameter $\lambda$ for each window.

For the numerical calculations, the eigenvalues $\theta_i$ of the
matrix $H_0$ were distributed in the interval $(-N/2,N/2)$ according
to the somewhat arbitrarily chosen distribution
\begin{equation}
  \mathcal P(\theta_i) =
  \frac1N \left[\frac12 + 6\left(\frac{\theta_i}N\right)^2 +
    8\left(\frac{\theta_i}N\right)^3\right], 
\end{equation}
$N$ being the number of independent eigenvalues of $H_0$.  The matrix
$H_{\beta'}$ is normalized in the usual way,
\eq\eqref{standardnormform}.

The eigenvalue density $\rho(\theta)$ of the total matrix $H$ is
plotted along with the analytical $\rho_0(\theta) = N\mathcal
P(\theta)$ of $H_0$ in the top row of
\fig\ref{eigdenspoisgausfigure}. One can see that the perturbation
only has a negligible effect on the spectral density.

The dependence of the coupling parameter on the eigenvalue density is
plotted in the bottom row of \fig\ref{eigdenspoisgausfigure} for
$\alpha = 0.1$.  No error bars are shown because the statistical
errors are negligibly small.  A linear fit through the origin with
minimized squared deviation was performed to obtain the
proportionality factor between the eigenvalue density and the coupling
parameter.  The quantity $\delta_2$ shown in the plots is a measure of
the fit quality and defined by
\begin{equation}
 \label{eq:definedelta2}
 \delta_2 = \sqrt{\sum_{i=1}^N \frac{(\lambda_i -
     \tilde\lambda_i)^2}N}\ \mbox{\LARGE{/}}\ 
 \sum_{j=1}^N \frac{\lambda_j}N\,,
\end{equation}
where the $\lambda_i$ are the numerically obtained coupling parameters
for each spectral window and the $\tilde\lambda_i$ are the
corresponding predictions from the linear fit at the given eigenvalue
density.  Because $\delta_2$ is a monotonically increasing function of
the squared deviation it is also minimized by our fitting procedure.

As can be seen, the linear dependence of the effective coupling
parameter on the eigenvalue density is confirmed very well by the
numerical data for all the transitions.  Note that the fit quality
gets better with increasing Dyson index $\beta'$, i.e., it is worst
for the GOE and best for the GSE.  This is most likely explained by
the fact that the spacing distributions change more rapidly with
respect to the coupling parameter for larger $\beta'$
(cf.~\fig\ref{poisgaussimplefigure}), which allows for a more precise
measurement of the coupling.

Although the linear dependence of the effective coupling on the
eigenvalue density has been demonstrated beyond reasonable doubt, the
proportionality factor is less clear. As can be read off from
\fig\ref{eigdenspoisgausfigure} the proportionality factor is smaller
than $\alpha$, i.e., the measured coupling parameter is smaller than
the expected one.  This agrees with the observation in
\sect\ref{largespectra1} where an explanation was given in terms of
the effect of other eigenvalues.

\subsection{Transitions from one symmetry class to another}

\subsubsection{Check of Wigner surmise}
\label{largespectra2}

We now consider chaotic systems composed of different symmetry
classes, the latter represented by pure Gaussian ensembles. If the GSE
is involved, we consider the case of a \selfdual\ perturbed ensemble in
this section (see \sect\ref{sympgauslargesec} for the case of a
non-\selfdual\ perturbed ensemble). A \selfdual\ GOE can be constructed
by taking the direct product with $\1_2$ as in \sect\ref{ortsympsec},
while the \selfdual\ GUE is more involved, see
\app\ref{selfdualGUE}. All ensembles are normalized as in
\eq\eqref{standardnormform}.  Again motivated by
\eq\eqref{suitedalphaform}, the Hamiltonian under consideration is
\begin{equation}
\label{normaliseGaussiantransitionmatrices} 
H = H_{\beta} + \frac{\Lambda}{\rho_{\beta}(0) \bar s_{\beta}}H_{\beta'}\,.
\end{equation}

For large matrix size, the eigenvalue density of $H_{\beta}$ is a
semicircle which extends to $r_{\beta} = \sqrt{2\beta N}$, and its
eigenvalue density in the center is
\begin{equation}
  \rho_{\beta}(0) = \frac{\sqrt{2N}}{\sqrt{\beta}\pi}\,.
\end{equation}

The results for the three transitions among the Gaussian ensembles are
shown in \fig\ref{gausgausfigure} for $N=400$.  Again, only the center
of the spectrum, defined as the interval $(-5,5)$, was evaluated (the
whole semicircle extends to about $\pm28$ for $H_\beta \in \text{GOE}$
and about $\pm40$ for $H_\beta\in \text{GUE}$).  The coupling
parameter $\lambda$ was obtained by a fit (see \app\ref{fittingmethod}
for details) to the corresponding \two\ (\four) formula, which yields
a good approximation to the numerical data throughout the transition
in each case. As in \sect\ref{largespectra1}, $\lambda$ is close to
$\Lambda$ as expected.

\begin{figure}[!b]
  \includegraphics[width=0.465\linewidth,clip=true]
  {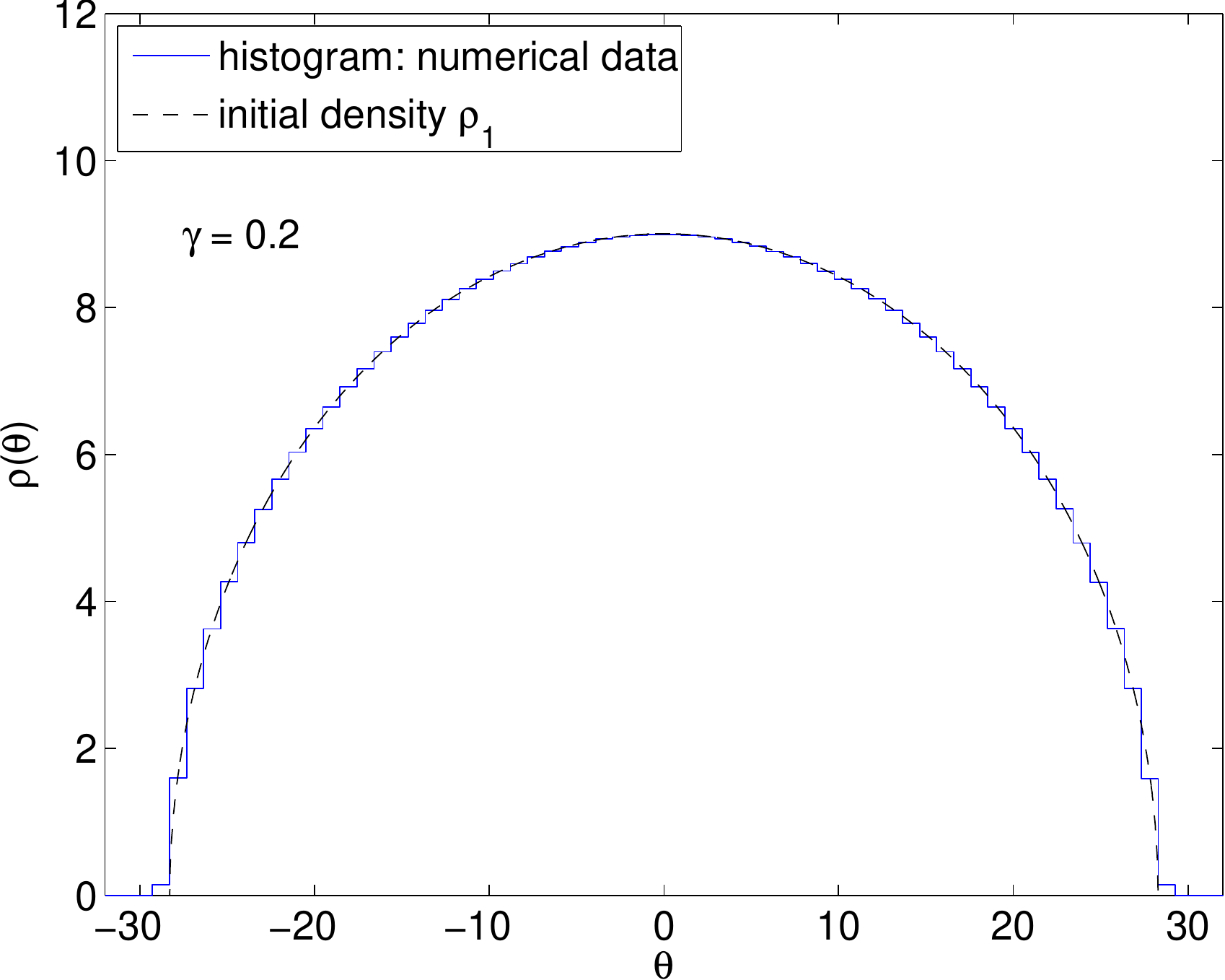}\hfill
  \includegraphics[width=0.53\linewidth,clip=true]
  {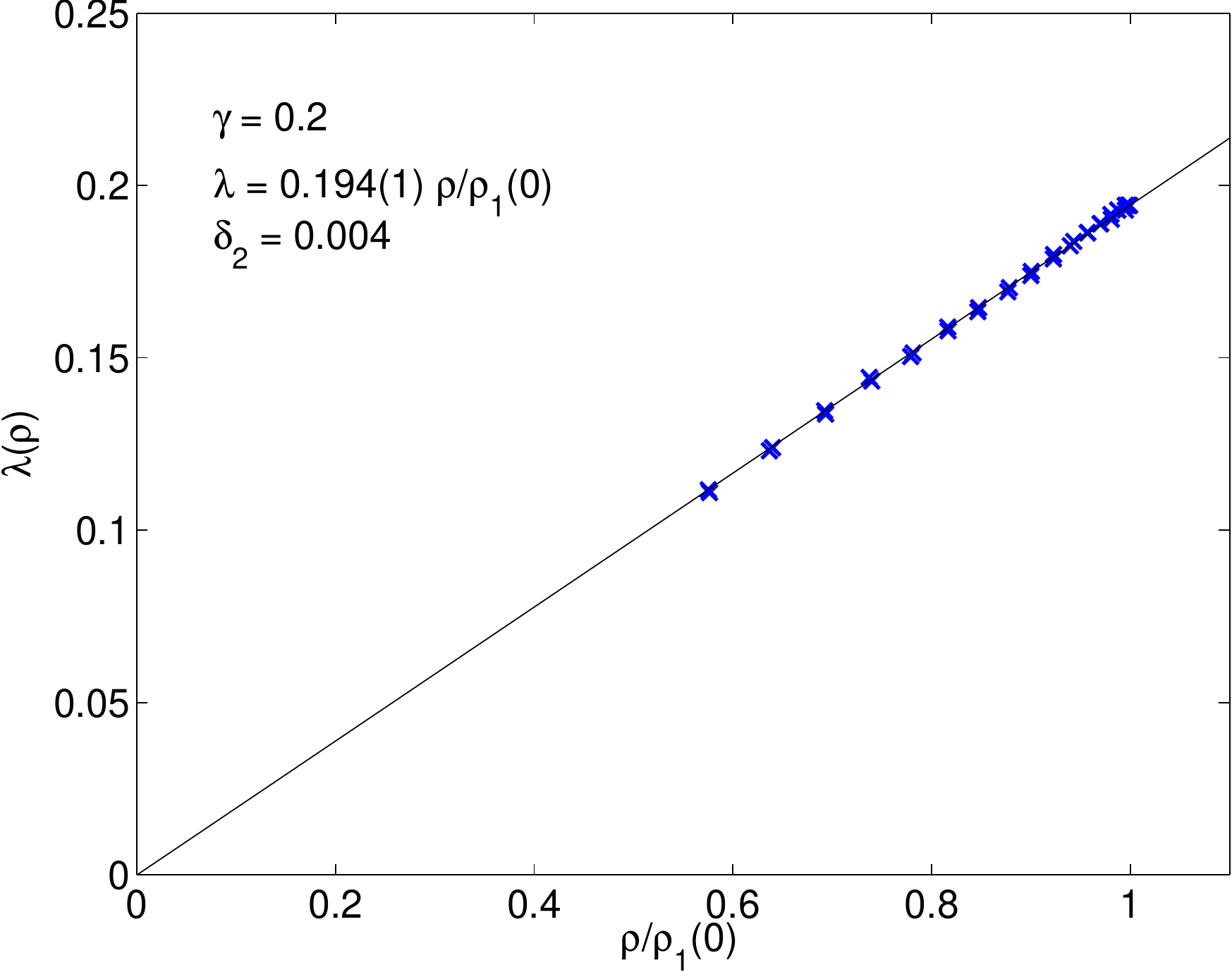}
  \caption{(Color online) Transition GOE $\to$ GSE. Left: unperturbed eigenvalue
    density (approximated by a semicircle) and perturbed eigenvalue
    density.  Right: effective coupling $\lambda$ obtained from fits
    of the \four\ level spacings $P_{1 \to 4}(s)$, see
    \eq\eqref{ortsympresultform}, as a function of the local
    eigenvalue density in $35$ equally large windows in the spectrum.
    Linear fit and proportionality factor with errors defined by the
    $95\%$-confidence interval are given in the plots.  The quantity
    $\delta_2$ defined in \eq\eqref{eq:definedelta2} is a measure of
    the fit quality, which is small for a good fit.  The numerical
    data were obtained from $10^5$ random matrices of dimension $800$.}
  \label{eigdensgoegsefigure}
\end{figure}

For $\Lambda=1$ and $N=400$ the mixed matrix is roughly given by $H =
H_\beta + \mathcal O(10^{-1})H_{\beta'}$.  From the $\lambda$-values
given in \fig\ref{gausgausfigure}, which should be compared to those
in \fig\ref{gausgaussimplefigure}, we see that the transition is
almost completed in this case and that the spacing distribution is
already very similar to the one of the perturbing ensemble.  What is
relevant for the transition is not the relative magnitude of
the matrix elements (which depends on $N$ through the local eigenvalue
density) but the rescaled coupling parameter $\Lambda$, i.e., the
transition occurs for $\Lambda=\mathcal O(1)$.  The same phenomenon
was found for the two-point function \cite{Pandey:1981}, which is
related to the spacing distribution for small $s$.

\begin{figure*}[t]
  \includegraphics[width=\onepicwidth,clip=true]
  {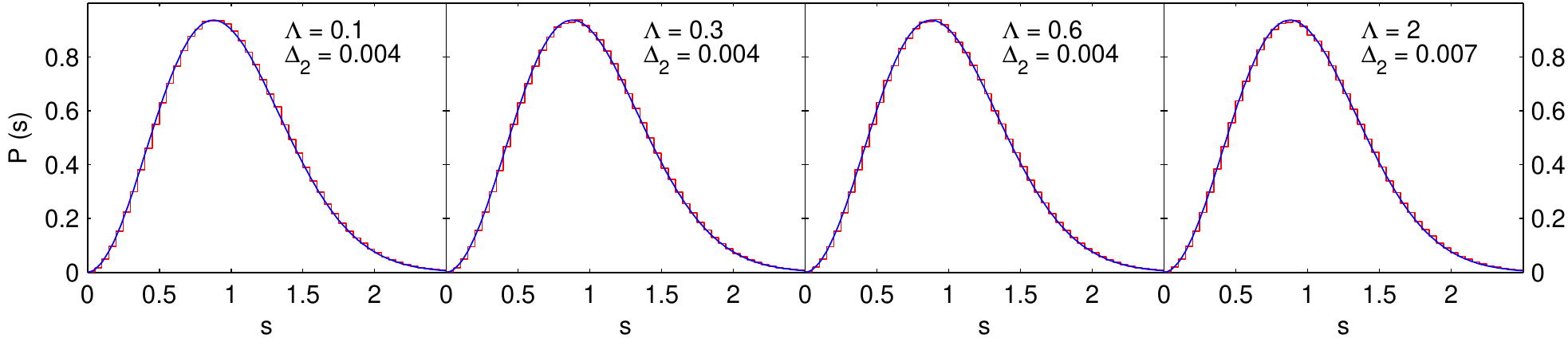}
  \includegraphics[width=\onepicwidth,clip=true]
  {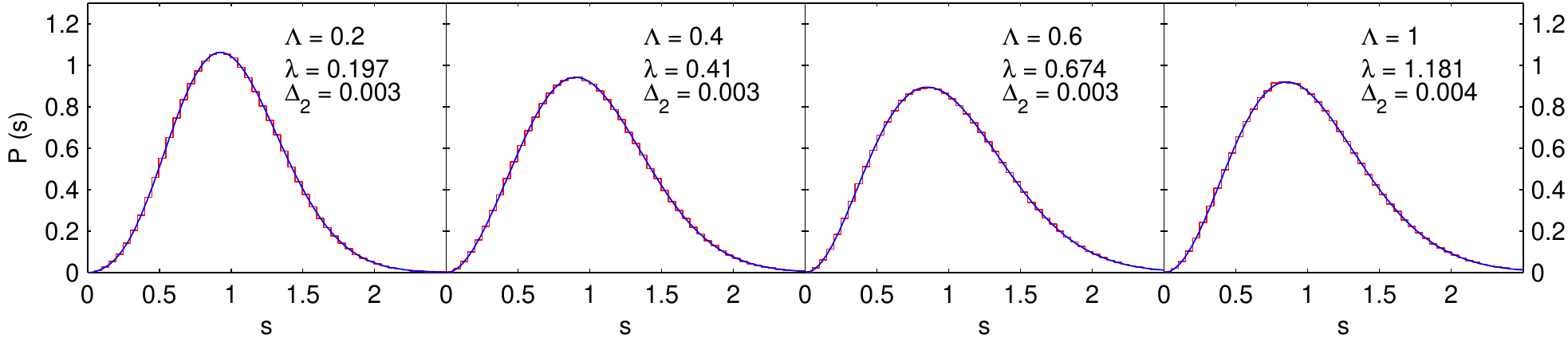}
  \caption{(Color online) Spacing distributions between previously degenerate
    eigenvalues $s_1$ (top) and previously non-degenerate eigenvalues
    $s_2$ (bottom) for the transition GSE $\to$ GUE without \selfdual\
    symmetry for various values of the coupling parameter $\Lambda$ in
    \eq\eqref{degenpertmatr}.  The histograms show the numerical data,
    while the full curves are the \two\ GUE surmise $P_2$ (top) and
    the surmise $P^{2}_{4\to2}(s_2; \lambda)$ given in
    \eq\eqref{eq:probsympunis2} (bottom), the latter with fitted
    coupling parameter $\lambda$.  The quantity $\Delta_2$ defined in
    \app\ref{fittingmethod} is a measure of the fit quality, which is
    small for a good fit.  The numerical data were obtained by
    diagonalizing $50,000$ random matrices of dimension $400$ for each
    plot.}
  \label{sympgauslargesfigure}
\end{figure*}

\subsubsection{Dependence of coupling parameter on eigenvalue density}
\label{eigdenstolambdagausgaus}

We now consider the dependence of the coupling parameter on the local
eigenvalue density as in \sect\ref{eigdenstolambdapoisgaus}, but now
for transitions between Gaussian ensembles.  In these cases, the
fitting procedure of the effective coupling becomes less precise,
because the functions of the spacing distributions change only very
slowly with $\lambda$, as can be seen in
\fig\ref{gausgaussimplefigure}. Therefore, we restrict ourselves to
the case of a \selfdual\ GOE matrix $H_1$ that is perturbed by a GSE
matrix $H_4$ as the level repulsion differs the most in these two
ensembles.

In \fig\ref{eigdensgoegsefigure} we show results from the mixed matrix
\begin{equation}
  H = H_1 + \alpha H_4\,, \qquad 
  \alpha = \frac{\gamma}{\rho_1(0)\bar s_1}
\end{equation}  
with $\gamma = 0.2$ (for details about the \selfdual\ GOE and the
normalization, see \sect\ref{largespectra2}). According to
\eq\eqref{suitedalphaform} the effective \four\ coupling parameter
$\lambda$ should be $\alpha\rho(\theta)\bar
s_1=\gamma\rho(\theta)/\rho_1(0)$, i.e., proportional to the local
eigenvalue density normalized by the density $\rho_1(0)$ in the
center, with proportionality factor given by the input parameter
$\gamma$.  As one can see, there is again a linear dependence of the
fitted coupling parameter $\Lambda$ on the local density (and again,
the perturbation has no measurable effect on the eigenvalue density).
The proportionality factor is almost compatible with the expected
value $\gamma$.

\subsection{Perturbation of a GSE matrix by a non-\selfdual\ GUE matrix}
\label{sympgauslargesec}

\begin{figure*}[t]
  \includegraphics[width=\onepicwidth,clip=true]
  {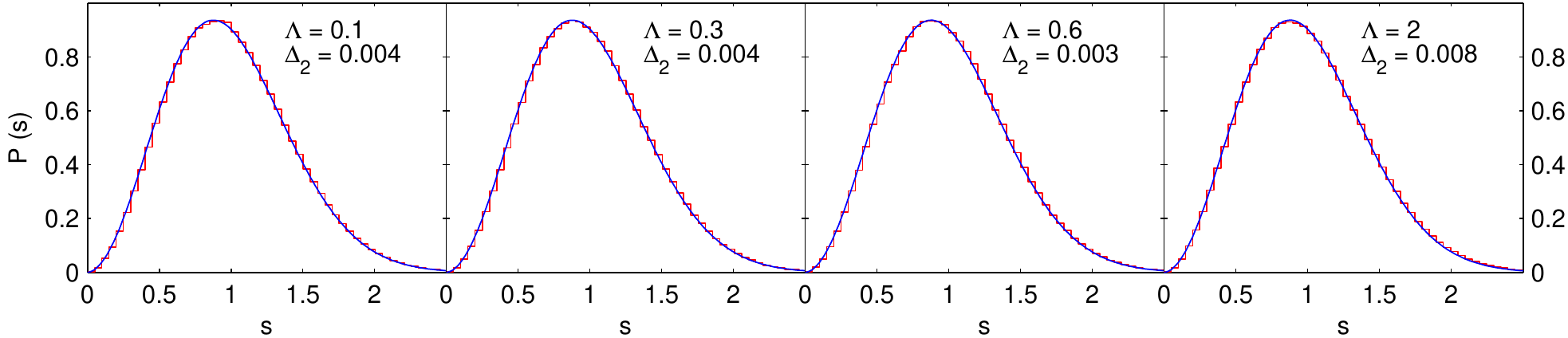}
  \includegraphics[width=\onepicwidth,clip=true]
  {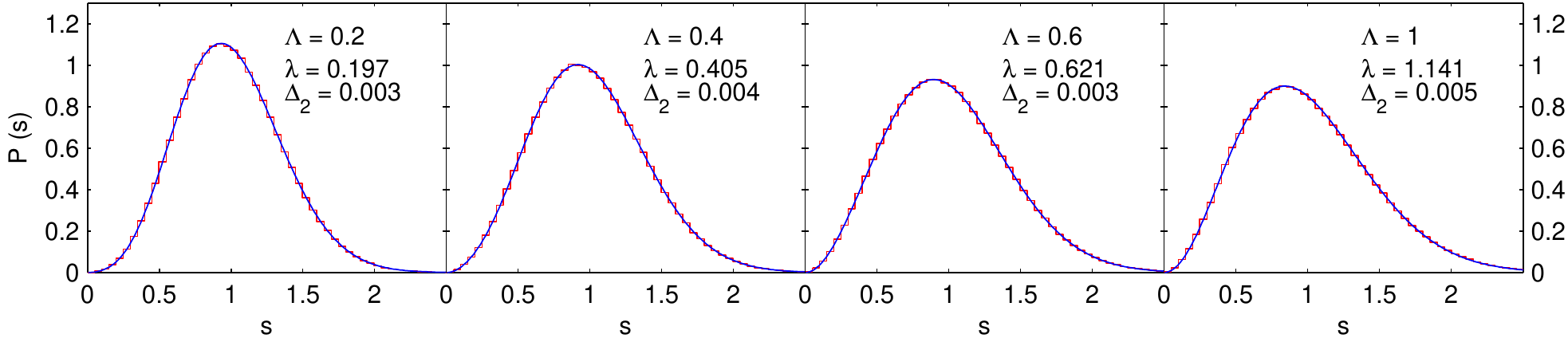}
  \includegraphics[width=\onepicwidth,clip=true]
  {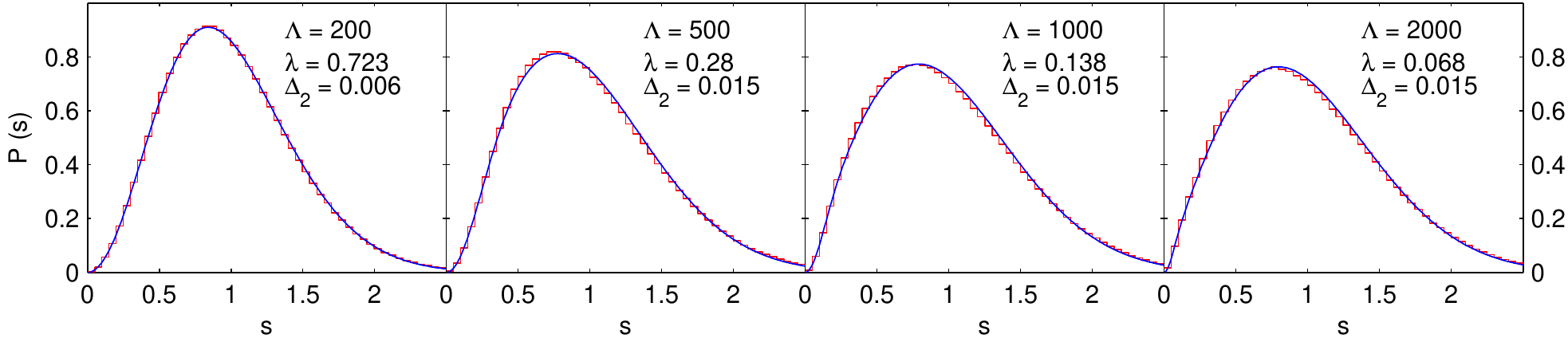}
  \caption{(Color online) Spacing distributions for the transition GSE $\to$ GOE
    without \selfdual\ symmetry for various values of the coupling
    parameter $\Lambda$ in \eq\eqref{eq:matrsedusymmlarge}.  Top:
    spacings $s_1$ between previously degenerate eigenvalues (for
    small $\Lambda$).  Middle: spacings $s_2$ between previously
    non-degenerate eigenvalues (also for small $\Lambda$).  Bottom:
    all spacings (for large $\Lambda$).  The histograms show the
    numerical data, while the full curves are the \two\ GUE surmise
    $P_2$ (top), the surmise $P^{2}_{4\to1}(s_2; \lambda)$ given in
    \eq\eqref{sympgausgeneralform} (middle), and the surmise
    $P_{1\to2}(s; \lambda)$ given in \eq\eqref{ortuniresultform}, the
    latter two with fitted coupling parameter $\lambda$.  The quantity
    $\Delta_2$ defined in \app\ref{fittingmethod} is a measure of the
    fit quality, which is small for a good fit.  The numerical data
    were obtained by diagonalizing $50,000$ random matrices of
    dimension $400$ for each plot.}
  \label{sympgoelargesfigure}
\end{figure*}

In this section, we apply the formulas derived in
\sect\ref{sympgaussec} for the spacing distributions of a \four\
matrix from the GSE perturbed by a matrix from the GUE, this time
without \selfdual\ symmetry, to large matrices.  We consider a
$2N\times 2N$ matrix
\begin{equation}\label{degenpertmatr}
  H = H_4 + \frac{\Lambda}{\rho_4(0) \bar s_4} H_2\,,
\end{equation}
where $H_4$ is taken from the GSE and $H_2$ is the perturbation from
the GUE.  Both $H_4$ and $H_2$ are normalized in the usual way, see
\eq\eqref{standardnormform}, and for the prefactor of $H_2$ see
\sect\ref{largespectrapert}.  To ensure a constant eigenvalue density,
we again restrict the measurements to the center of the spectrum,
defined by the interval ($-5,5$). The numerically obtained spacing
distributions were rescaled to a mean value of $1$.

As in \sect\ref{sympgaussec}, we will separately consider the spacings
between originally degenerate eigenvalues and the remaining ones.  The
distributions of the former were obtained by measuring every second
spacing, starting with the first one of each random matrix.  They are
plotted in \fig\ref{sympgauslargesfigure} (top) and show perfect
agreement with the \two\ GUE surmise, which is practically
indistinguishable from the exact result derived for \four\ matrices,
$P^1_{4\to2}(s_1; \lambda)$ given in \eq\eqref{eq:probsympunis1}.  As
in \sect\ref{sympgauss1sec}, this distribution is almost independent
of the coupling parameter.  

The distribution of the spacings between previously non-degenerate
eigenvalues is shown in \fig\ref{sympgauslargesfigure} (bottom).
Again, every second spacing was measured, but starting with the second
one this time.  We get an almost perfect agreement of the numerical
data with the surmise $P^2_{4\to\beta'}(s_2; \lambda)$ defined in
\eq\eqref{eq:probsympunis2} throughout the transition.  The parameter
$\lambda$ was again determined by a fit (see \app\ref{fittingmethod})
and approximately matches the perturbative prediction from
\sect\ref{largespectrapert}.

\subsection{Other transitions between the GSE and ensembles without
  \selfdual\ symmetry}
\label{sec:gsetoother}

Let us now consider the transition from the GSE to either the GOE or
Poisson, both without \selfdual\ symmetry.  These two cases are more
complicated than the cases discussed so far because, as we shall
discuss now, the transitions proceed via an intermediate transition
to the GUE.  

Let us first focus on the case
\begin{align}\label{eq:matrsedusymmlarge}
  H=H_4+\frac\Lambda{\rho_4(0)\bar s_4} H_1\,,
\end{align}
where $H_4$ is from the GSE, $H_1$ is from the GOE without \selfdual\
symmetry, and we again concentrate on the central part of the spectrum
(near zero).  For small $\Lambda$, we show in \app\ref{pertmatrix} in
first-order perturbation theory that the perturbation by the GOE has
exactly the same effect on the eigenvalues as the perturbation by the
GUE considered in \sect\ref{sympgaussec}, modulo a rescaling of the
coupling parameter, i.e.,
\begin{align}
  \label{sympgausgeneralform1}
  P^1_{4\to1}(s_1;\lambda) &= P^1_{4\to2}(s_1;\lambda/\sqrt2)
  \simeq P_2(s_1)\,,\\
  P^2_{4\to1}(s_2; \lambda) &= P^2_{4\to2}(s_2; \lambda/\sqrt{2})\,.
  \label{sympgausgeneralform}
\end{align}
Therefore, we first expect a transition from the GSE to the GUE,
corresponding to the breaking of the \selfdual\ symmetry.  This
expectation is confirmed in \fig\ref{sympgoelargesfigure} (top and
middle).

As $\Lambda$ is increased to very large values, a transition to GOE
behavior must eventually occur.  The question is whether this
transition is described by the surmise of \sect\ref{ortunisec}.  We
show in \fig\ref{sympgoelargesfigure} (bottom) that this is indeed the
case.  Note that a rising $\Lambda$ amounts to a shrinking fitted
coupling parameter $\lambda$ because the direction of the transition
is turned around compared to \sect\ref{ortunisec}. Here,
$\Lambda\to\infty$ means that $H$ is a pure GOE matrix, which is
described by the surmise with $\lambda=0$.

The case of GSE to Poisson without \selfdual\ symmetry is analogous.
For small values of the coupling parameter, the \selfdual\ symmetry of
the GSE is broken by the perturbation so that we expect a GSE to GUE
transition for the spacings $s_1$ and $s_2$ as in the GSE to GOE case
considered above.  For very large values of the coupling parameter we
should eventually find a transition to Poisson behavior, described by
the surmise of \sect\ref{poisunisec}.  We have confirmed these
expectations numerically but do not show the corresponding plots here.

\begin{figure}
  \includegraphics[width=\columnwidth]{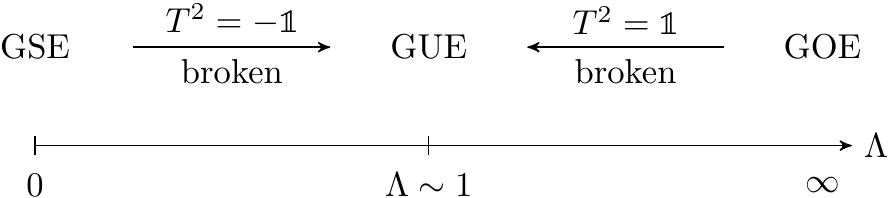}
  \caption{Schematic picture of the transition from GSE to
    non-\selfdual\ GOE, which proceeds via an intermediate transition
    to the GUE.  An analogous picture applies to the transition from
    GSE to non-\selfdual\ Poisson.}
  \label{fig:transitions}
\end{figure}

Note that in the transitions considered in
\sects\ref{sec:poisgausslarge} through \ref{sympgauslargesec} a single
anti-unitary symmetry (or integrability in the case of Poisson) was
broken or restored.  In contrast, we now have two transitions.  As
$\Lambda$ increases from zero, an anti-unitary symmetry $T$ with
$T^2=-\1$ gets broken.  As $\Lambda$ decreases from infinity, either
an anti-unitary symmetry with $T^2=\1$ gets broken (in the case of
GOE) or integrability gets broken (in the case of Poisson).  For
intermediate values of $\Lambda$ the system follows GUE statistics
because all anti-unitary symmetries and/or integrability are broken.
This is illustrated in \fig\ref{fig:transitions}.

\section{Summary}
\label{summarysec}

We have derived generalized Wigner surmises for the nearest-neighbor
spacing distributions of various mixed RMT ensembles from \two\ and
\four\ matrices.  If the GSE was involved in the transition, we have
distinguished two cases: (i) perturbations of the GSE by a \selfdual\
ensemble, and (ii) perturbations of the GSE by a non-\selfdual\
ensemble, for which we separately considered two different kinds of
spacings.

We have shown that all of these distributions yield a good description
of the spectra of large mixed matrices when restricted to a range of
constant spectral density.  The coupling parameters in the
generalized Wigner surmise and in the large mixed matrices are related
via the local eigenvalue density of the latter.  This relation is well
approximated by \eq\eqref{lambeffform2}.

We expect that the results for $P(s)$ derived in this paper will be
useful in numerical and/or experimental studies of systems with mixed
symmetries, such as those mentioned in the introduction.  $P(s)$ is a
convenient quantity that is easily analyzed numerically or
experimentally and typically does not suffer from serious unfolding
issues.  In particular, the properties of the level spacings should
help us to clarify whether the mixing of the symmetry classes in a
given physical system is of the additive type (\ref{eq:trans}) we have
investigated here. If so, fits to the generalized Wigner surmises
provide estimates of the coupling parameter in terms of the local
eigenvalue density.  In turn, the coupling parameter could quantify
other properties of the mixed systems.

\acknowledgments

We thank Thomas Guhr for communication at an early stage of this work.
We also acknowledge DFG (BR 2872/4-2 and SFB/TR-55) and EU (StrongNet)
for financial support.

\appendix

\section{\boldmath Analysis of the large-$s$ behavior}
\label{app:large_s}

We consider the large-$s$ behavior of the spacing distributions for
the three transitions from Poisson to RMT.  For simplicity, we first
consider the non-normalized spacing $S$ and convert to the normalized
spacing $s$ at the end.  We start with the initial ansatz for the
distributions,
\begin{align}
  I(S) & =  \int dp\,da\,db\,\prod_{\nu=0}^{\beta-1} dc_\nu\,
  P_0(p) P_a(a) P_b(b) P_{c_\nu}(c_\nu) \notag\\
   &\!\!\!\times\delta\bigg(S-\sqrt{[a-(b+p/\lambda)]^2
      +4\,\textstyle{\sum_{\mu=0}^{\beta-1}}\:c_\mu c_\mu}\,\bigg)\,,
  \label{eq:generalansatz}
\end{align}
which is the generalization of \eq\eqref{generalspacingform} to
$\beta=1$, $2$, and $4$.  As in \app\ref{poisgseapp}, we introduce new
variables $u=a+b$ and $t=a-b$, transform the $c_\nu$ to spherical
coordinates, and eliminate the $\delta$-function by integrating out the
radius.  This yields
\begin{align}
 I(S) &\sim \int_0^\infty dp\int_{-S}^S dt\,
(S^2-t^2)^{\frac\beta2 -1} S
e^{-\frac{p^2}{4\lambda^2}-p-\frac{pt}{2\lambda}
-\frac14 S^2}\notag\\
&\sim S^\beta e^{-\frac14 S^2}\!\!\int_0^\infty\! dp\,
e^{-p^2-2\lambda p}
\!\int_{-1}^1\! dx\,(1-x^2)^{\frac\beta2 -1} e^{-pxS}\notag\\
&\sim S^\beta e^{-\frac14 S^2}\int_0^\infty dp\, e^{-p^2-2\lambda
  p}\, X_\beta(pS)\,,\label{eq:intp}
\end{align}
where we substituted $t=xS$, rescaled $p\to 2\lambda p$, and expressed
the $x$-integral (up to normalization) as
\begin{align}\label{eq:Xbeta}
  X_\beta(pS) = (pS)^{-\frac{\beta-1}2} I_\frac{\beta-1}2(pS)\,,
\end{align}
where $I$ is a modified Bessel function
\cite[\eq(9.6.18)]{Abram:1964}.  We now compute the integral in
\eq\eqref{eq:intp} in saddle-point approximation, assuming $S$ to be
large.  The asymptotic expansion of $X_\beta$ for $\beta=1,2,4$ reads
\begin{align}
  X_\beta(pS) = \frac1{\sqrt{2\pi}}\, (pS)^{-\frac\beta2}\, e^{pS}\,.
\end{align}
For $p=\mathcal O(S^{-1})$ we cannot use this expansion, but the
contribution of this region to the integral can be shown to be
negligible compared to the leading order we consider here.  The
exponential in the integrand is now
\begin{align}\label{eq:exponent}
  e^{-p^2-2\lambda p+pS}\,,
\end{align}
with a maximum at $p_\text{max} = S/2-\lambda$.  Standard
manipulations then yield the saddle-point result
\begin{align}
 I(S) \sim e^{-\lambda S} \left[1+\mathcal O(S^{-1})\right].
\end{align}
The normalized distributions are obtained from $I(S)$ by rescaling the
spacing and restoring the normalization factors that were omitted in
the calculation above, resulting in
\begin{align}\label{eq:large_s}
 P_{0\to\beta}(s;\lambda) = e^{-2\lambda Ds}
 \big[2\lambda De^{\lambda^2}+\mathcal O(s^{-1})\big]\,,
\end{align}
where we replaced $S$ by $2Ds$ with $D$ given in
\eqs\eqref{eq:poisgoeD}, \eqref{eq:poisgueD} and \eqref{eq:D} for
$\beta=1,2,4$, respectively.  The meaning of this result is that for
arbitrarily large (but finite) $\lambda$, i.e., arbitrarily close to
the pure Gaussian ensemble, the large-$s$ behavior is Poisson-like.
This is in contrast to the small-$s$ behavior, which is dominated by
the Gaussian ensemble for arbitrarily small (but non-zero) $\lambda$.
The findings for the large-$s$ behavior were also confirmed
numerically.

\section{Analysis of the Gibbs Phenomenon}
\label{gibbsapp}

The spacing distribution of ensembles interpolating between Poisson
and RMT reveal a Gibbs-like phenomenon close to the Poisson limit,
i.e., for small $\lambda$: $P(s;\lambda)$ does not converge uniformly
to the Poisson curve $e^{-s}$ at $s=0$.  Rather, there is an overshoot
whose amount does not vanish in the $\lambda\to 0$ limit and whose
position $s$ approaches $0$ in this limit. In this appendix we work
out the value and position of this maximum. We start with a brief
review of the Gibbs phenomenon in the Fourier transform, as known from
textbooks such as \cite{walker:1988} (which, however, mostly discuss
the Gibbs phenomenon only in the Fourier series).

\subsection{Gibbs Phenomenon in the Fourier transform}

The Gibbs phenomenon is related to the convergence of the inverse
Fourier transform with a cut-off in the integral (or to the
convergence of the Fourier series with a cut-off in the sum) towards
the original function $f$. Let us denote its Fourier transform by
$\f$,
\begin{align}
  \f(\omega)= \frac{1}{2\pi}\int_{-\infty}^{\infty}
  ds\, e^{-i\omega s} f(s)\,,
\end{align}
and the result of the cut-off inverse transform by $f$ with two
arguments
\begin{align}
  f(s;\lambda)=\int_{-1/\lambda}^{1/\lambda}
  d\omega\, e^{i\omega s} \f(\omega)\,.
\end{align}
These formulas can be combined into a convolution
\begin{align}\label{eq:convolution_Dirichlet}
  f(s;\lambda)=\int_{-\infty}^{\infty}
  ds'f(s')\delta_\lambda(s-s')
\end{align}
of the original function with the Dirichlet kernel
\begin{align}
  \delta_\lambda(s-s')=
  \frac{1}{2\pi}\int_{-1/\lambda}^{1/\lambda}
  d\omega\, e^{i\omega (s-s')}
  =\frac{\sin[(s-s')/\lambda]}{\pi (s-s')}\,.
\end{align}
The question is how in the $\lambda\to 0$ limit\footnote{The
  $\lambda\to0$ limit of $f(s;\lambda)$ is also denoted as the
  principal value.} $f(s;\lambda)$ is related to the original function
$f(s)$. If $f(s)$ is smooth and absolutely integrable, $f(s;\lambda)$
approaches it everywhere. Accordingly, the Dirichlet kernel approaches
the delta distribution in the sense of acting on smooth test
functions.

At discontinuities of the original function $f(s)$, however,
$f(s;\lambda)$ approaches the average of the left and right limit of
$f(s)$. Intuitively, this comes from the nonzero width of the
Dirichlet kernel,\footnote{This nonzero width is relevant in many
  areas of physics such as band-limited signals, ringing, and
  diffraction of waves at slits.} which in the convolution
\eqref{eq:convolution_Dirichlet} probes both sides of the
discontinuity. Furthermore, the functions $f(s;\lambda)$ for fixed
$\lambda$ possess maxima and minima whose positions move, in the limit
$\lambda\to 0$, towards the discontinuity and whose values over-
and undershoot the function. This is the Gibbs phenomenon.

\begin{figure}[t]
 \includegraphics[width=0.95\linewidth,clip=true]{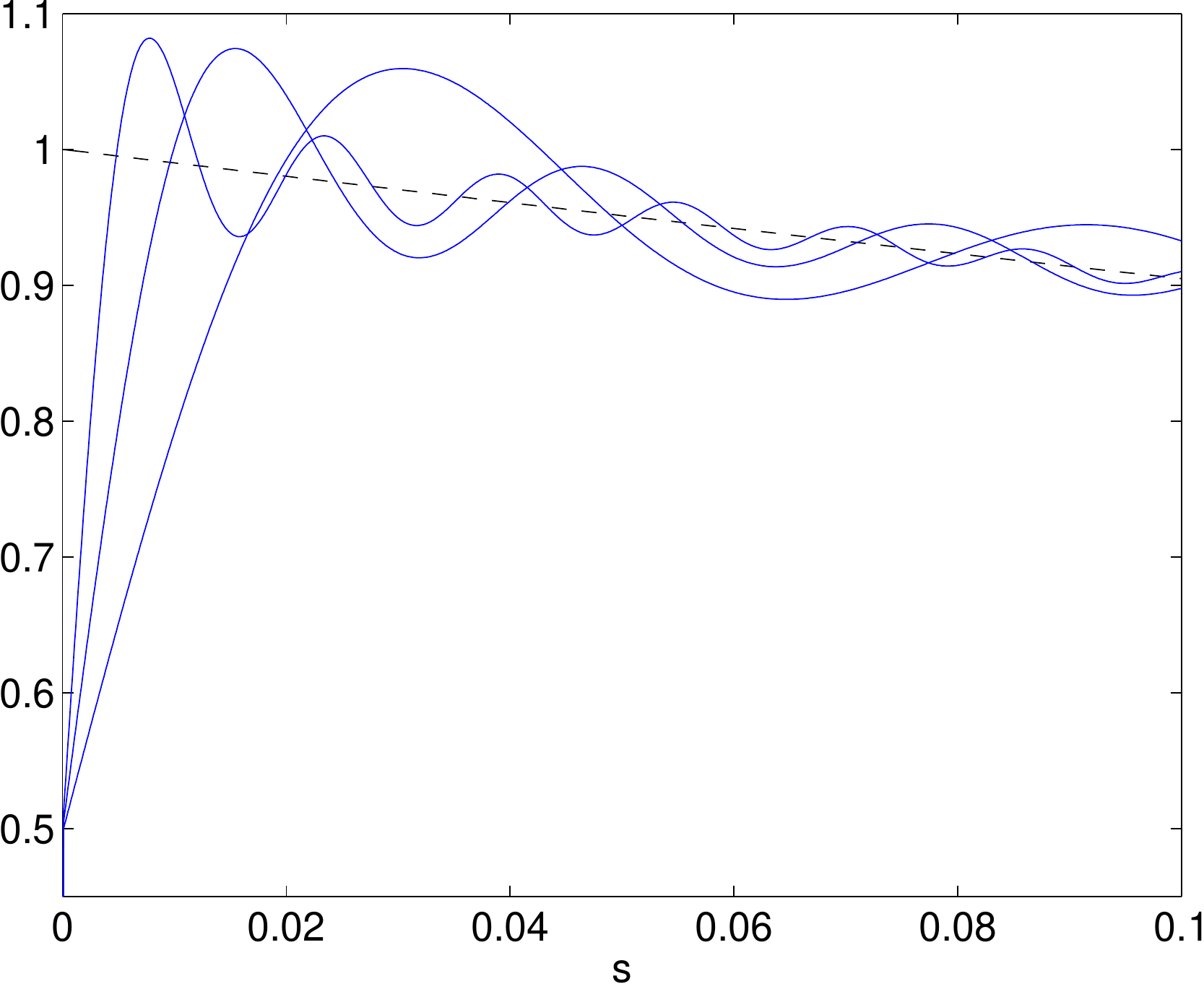}

 \includegraphics[width=0.95\linewidth,clip=true]{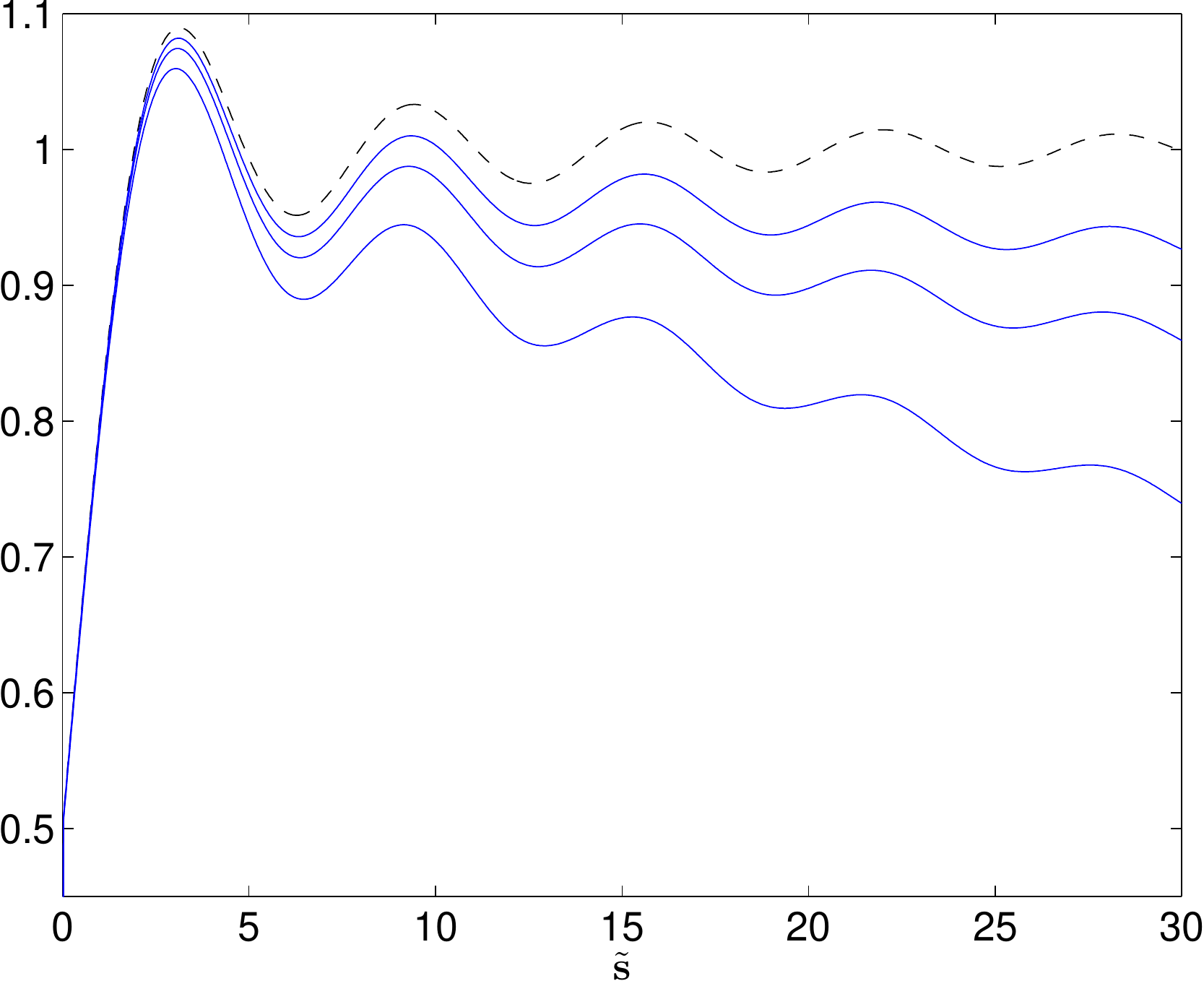}
 \caption{(Color online) The Gibbs phenomenon in the Fourier transform of the Poisson
   curve, with $\lambda=0.01,0.005,0.0025$, respectively. Top: the
   functions $f(s;\lambda)$ approaching the original function $e^{-s}$
   (dashed), first maximum moving to the left as $\lambda$
   decreases. Bottom: the rescaled functions $\A(\tilde s;\lambda)$
   approaching the limiting function $\B(\tilde s)$ (dashed) with
   decreasing $\lambda$, see text.}
 \label{fig:gibbsfourier}
\end{figure}

For definiteness let us consider a set of exponentially decaying
functions with a jump discontinuity of unit size\footnote{As all
  formulas are linear in $f(s)$, the case of arbitrary jumps is
  completely analogous.} at $s=0$,
\begin{equation}
 f(s)=\begin{cases}
   0 & \text{for }s<0\,, \\ 
   e^{-\g s} & \text{for }s>0\,,
 \end{cases}
\end{equation}
that include the Poisson curve ($\g=1$) and the Heaviside function
($\g=0$).  The Fourier transforms are
\begin{equation}
 \f(\omega)=\frac{1}{2\pi}\frac{1}{\g+i\omega}\,,
\end{equation}
and the cut-off inverse transforms read
\begin{equation}
  f(s;\lambda)=\frac{i}{2\pi}e^{-\g s}\left[
    \Ei\left(\g s-i\frac{s}{\lambda}\right)
    -\Ei\left(\g s+i\frac{s}{\lambda}\right)\right].
\end{equation}
For the Poisson curve these functions are plotted for three small
values of $\lambda$ in \fig\ref{fig:gibbsfourier} (top), where several
maxima above and several minima below $e^{-s}$ are clearly visible.

To analyze the limit $\lambda\to 0$ we can zoom into the region of
small $s$, of size proportional to $\lambda$. This amounts to
considering functions of a rescaled argument
\begin{equation}
  \tilde s=\frac{s}{\lambda}
\end{equation}
in a constant $\tilde s$-range. We define
\begin{align}
 \A(\tilde s;\lambda)&= f(\tilde s\lambda;\lambda)\notag\\
 &=\frac{i}{2\pi}e^{-\g \tilde s\lambda}\left[
   \Ei(\g \tilde s\lambda-i\tilde s)
   -\Ei(\g \tilde s\lambda+i\tilde s)\right].
 \label{eq:func_A}
\end{align}
Keeping $\tilde s$ fixed, these functions have a well-defined limit
$\lambda\to 0$,
\begin{align}
 \B(\tilde s)&= \lim_{\lambda\to 0}\A(\tilde s;\lambda)\notag\\
 &= \frac{i}{2\pi}\left[\Ei(-i\tilde s)-\Ei(i\tilde s)\right]
 =\frac{1}{2}+\frac{\Si(\tilde s)}{\pi}
 \label{eq:func_B}
\end{align}
with the sine integral $\Si(\tilde s)=\int_0^{\tilde s}dx\, \sin x/x$.
As \fig\ref{fig:gibbsfourier} (bottom) shows, this limiting function
captures infinitely many maxima at $\tilde s=\pi,3\pi,\ldots$ and
infinitely many minima at $\tilde s=2\pi,4\pi,\ldots$ The overshoot at
the first maximum is the well-known number
\begin{equation}
 \frac{1}{2}+\frac{\Si(\pi)}{\pi}-1=0.0894899\,.
\end{equation}
Concerning the convergence of the Fourier transform, we conclude that
in the limit $\lambda\to 0$ the functions $f(s;\lambda)$ have a
maximum at $s=\pi\lambda$, with an overshoot approaching $8.9\%$.

Note that the limiting function $\B$ is the same for all these
functions independently of the decay constants $d$, i.e., it is solely
determined by the discontinuity. In other words, the smooth part of
the function $f(s)$ drops out when going from $\A(\tilde s;\lambda)$
to $\B(\tilde s)$ in the $\lambda\to 0$ limit, see
\eq\eqref{eq:func_A} vs \eqref{eq:func_B}.

This can be shown to be universal. Rescaling the integration variable
in \eq\eqref{eq:convolution_Dirichlet} and using
$\lambda\delta_\lambda(\lambda x)=\delta_1(x)$ one has
\begin{align}
  f(s;\lambda)&=\int_0^{\infty}
  ds''f(\lambda s'')\delta_1(\frac{s}{\lambda}-s'')\,,\\
  \A(\tilde s;\lambda)&=\int_0^{\infty}
  ds''f(\lambda s'')\delta_1(\tilde s-s'')\,,
\end{align}
where we still assume $f(s<0)=0$ for simplicity. The limiting function
is
\begin{align}
 \B(\tilde s)=f(0^+)\int_{-\infty}^{\tilde s} dt\, \delta_1(t)
 =f(0^+)\left[\frac{1}{2}+\frac{\Si(\tilde s)}{\pi}\right],
\end{align}
which agrees with \eqref{eq:func_B} for all functions with $f(0^+)=1$.

The last equation in particular relates the Dirichlet kernel
$\delta_\lambda$ and the limiting function $\B$. Therefore,
$f(s;\lambda)$ can also be reconstructed by a convolution with (the
derivative of) $\B$,
\begin{align}
  \label{eq:conv}
  f(s;\lambda)=\int_0^\infty ds'\,
  f(s')\,\frac{1}{\lambda}\,g'\left(\frac{s-s'}{\lambda}\right) ,
\end{align}
where $g'(x)=dg/dx$.

\subsection{Gibbs Phenomenon in the Poisson to RMT transitions}
\label{gibbsapprmt}

\begin{figure}[t]
 \includegraphics[width=0.95\linewidth,clip=true]{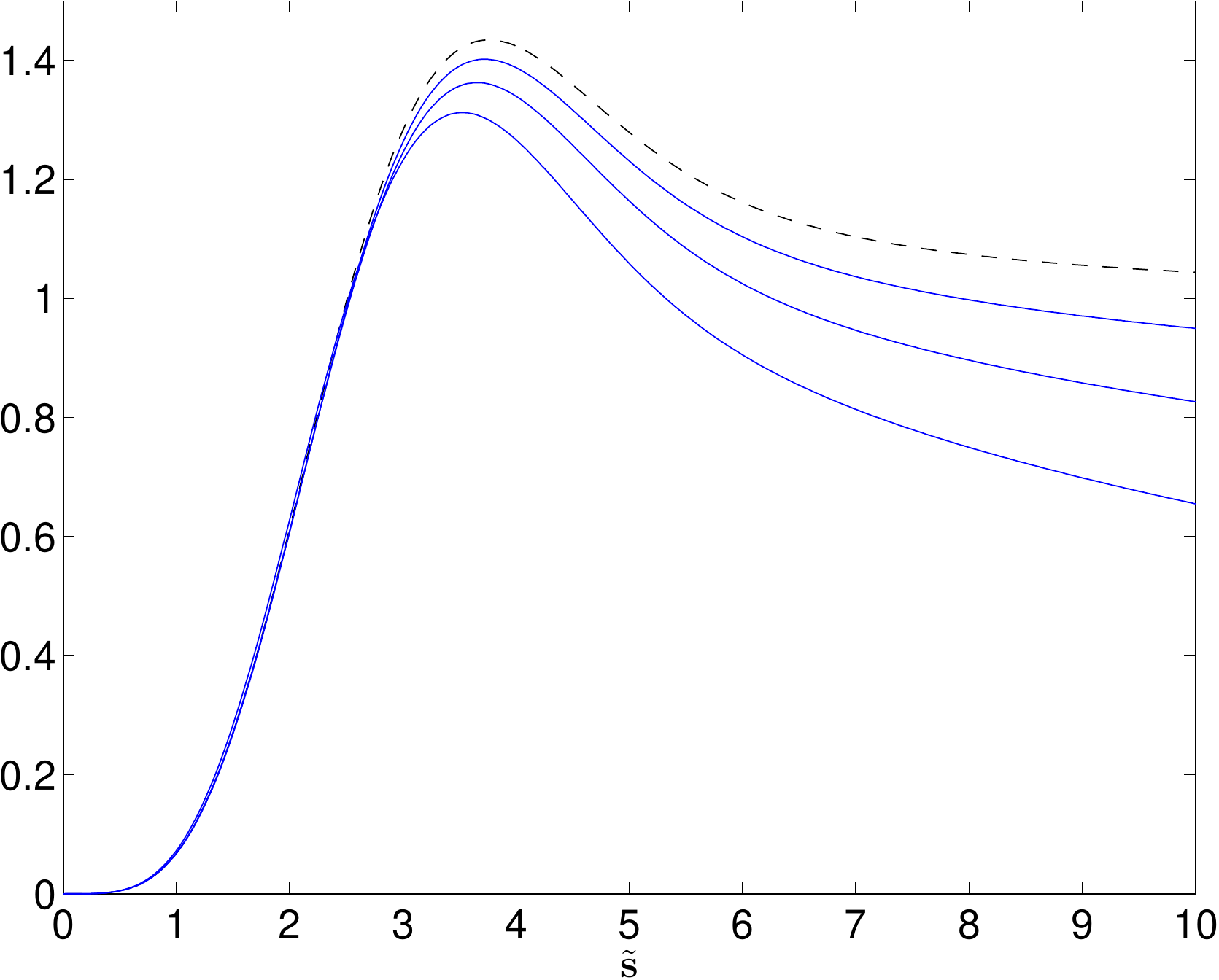}
 \caption{(Color online) The rescaled spacing distributions $\hat{P}_{0\to 4}(\tilde
   s;\lambda)$ in the Poisson to GSE transition,
   \eq\eqref{eq:rescaled_P04}, for $\lambda=0.05, 0.025, 0.01$ (maxima
   increasing) approaching the limiting function $g_{0\to 4}(\tilde
   s)$, \eq\eqref{eq:limiting_P04_final} (dashed).}
 \label{fig:gibbspgse}
\end{figure}

For the Gibbs-like phenomenon in the mixed spacing distributions we
start with the Poisson to GSE case. For this transition we found the
spacing distribution \eq\eqref{poissympresultform}.
In analogy to the previous subsection we rescale the argument and define
\begin{align}
  \label{eq:rescaled_P04}
  &\hat{P}_{0\to 4}(\tilde s;\lambda)
  = P_{0\to 4}(\tilde s\lambda;\lambda)
  =C\lambda^4 \tilde s^4 e^{-D^2\lambda^2 \tilde s^2}\\
  &\times\int_{-1}^{1}dx \, (1-x^2)\,
  e^{(xD\lambda \tilde s)^2 + 2\lambda^2 xD\tilde s}
  \erfc(xD\lambda \tilde s + \lambda)\,.\notag
\end{align}
In the limit $\lambda\to 0$ we make use of the behavior of
$C(\lambda)$ and $D(\lambda)$,
\begin{align}
  D(\lambda)\sim \frac1{2\lambda}\quad\text{and}\quad 
  C(\lambda) \sim \frac1{(2\lambda)^4}\,,
\end{align}
to arrive at the limiting function
\begin{align}
  g_{0\to 4}(\tilde s)&= \lim_{\lambda\to 0}\hat{P}_{0\to
    4}(\tilde s;\lambda)\notag\\ 
  &= \frac{\tilde s^4}{16}\, e^{-\frac{\tilde
      s^2}4}\int_{-1}^{1}dx \, (1-x^2)\,e^{(x\tilde
    s/2)^2} \erfc\frac{x\tilde s}2\notag\\
  &= \frac{\tilde s}8 \left[\left(2+\tilde s^2\right)\sqrt{\pi}
    e^{-\frac{\tilde s^2}4}\erfi(\tilde s/2) - 2\tilde s\right]. 
\label{eq:limiting_P04_final}
\end{align}
In \fig\ref{fig:gibbspgse} we plot this function, together with
$\hat{P}_{0\to 4}(\tilde s;\lambda)$ approaching it for small
$\lambda$.

Again, this limiting function captures a maximum, which can
numerically be determined to be at $\tilde s=3.76023$ with a value of
$1.43453$. As before, the phenomenon is solely determined by the
discontinuity of the Poisson curve at $s=0$ as the original Poisson
distribution $e^{-p}$ can be shown to drop out from
\eqs\eqref{eq:rescaled_P04} to \eqref{eq:limiting_P04_final}.

So in the transition from Poisson to GSE, $P_{0\to 4}(\s;\lambda)$ in
the limit of small $\lambda$ has a maximum at $s=3.76 \lambda$
overshooting the Poissonian $e^{-\s}$ by $43.5\%$.  Likewise, in the
other transitions $P_{0\to 1}(\s;\lambda)$ and $P_{0\to
  2}(\s;\lambda)$ we have equivalently defined limiting functions
\begin{align}
  g_{0\to1}(\tilde s) &= \frac{\sqrt\pi}2 \tilde s\,e^{-\tilde
    s^2/8}\,I_0(\tilde s^2/8)\,,\\ 
  g_{0\to2}(\tilde s) &= \frac{\sqrt\pi}2 \tilde s\,e^{-\tilde
    s^2/4}\,\erfi(\tilde s/2)\,.
\end{align}
These have maxima at $\tilde s=2.51$ and $\tilde s=3.00$,
overshooting the Poisson curve by $17.5\%$ and $28.5\%$, respectively,
as quoted in the body of the paper.  We observe that these numbers
grow with the Dyson index $\beta'$ of the perturbing ensemble.

From the small-$\tilde s$ behavior $g_{0\to 4}(\tilde s)=\tilde
s^4/12+\mathcal O(\tilde s^6)$ we conclude that $P_{0\to
  4}(s;\lambda)=s^4/(12\lambda^4)+\mathcal O(s^6)$ for small
$\lambda$, which reproduces our observation in
\eqs\eqref{eq:poisgselambdatozero} and
\eqref{eq:poisgselambdatozero2}.  Analogous agreement is obtained with
\eqs\eqref{eq:poisgoelambdatozero}, \eqref{eq:poisgoelambdatozero2}
and \eqs\eqref{eq:poisguelambdatozero},
\eqref{eq:poisguelambdatozero2} for the other two cases.  This
concludes our empirical results on the Gibbs-like phenomenon.

Concerning the analogies at a more fundamental level, the spacing
distribution $P_{0\to4}(s;\lambda)$ is related to the integral
\eqref{generalspacingform}
\begin{equation}
 \frac1{\lambda}\,I(S/\lambda)=\int_0^\infty dp\,
 e^{-p}\,\delta_\lambda(S,p)\label{eq:convolution_PGSE} 
\end{equation}
of the unperturbed Poisson distribution with the kernel 
\begin{align}
  \delta_\lambda(S,p)&=\int da\ldots dc_3\, P_a(a)\ldots
  P_{c_3}(c_3)\label{eq:kernel_PGSE}\\ 
  &\quad\times\delta\left(S-\lambda\sqrt{(a-b-p/\lambda)^2+4(c_\mu
      c_\mu)}\right).\notag
\end{align}
The nonzero width of this kernel causes the Gibbs phenomenon in the
spacing distribution near the discontinuity of the Poisson
distribution $e^{-p}$ at $p=0$. Note that in the limit $\lambda\to 0$
the second line of \eq\eqref{eq:kernel_PGSE} approaches $\delta(S-p)$,
thus decoupling from the integrals over $a,\ldots,c_3$.  The latter are
normalized by construction so that the kernel $\delta_\lambda(S,p)$
approaches $\delta(S-p)$.

There are (at least) two features that are different from the Fourier
case.  First, the kernel is not a function of $S-p$, and thus
\eq\eqref{eq:convolution_PGSE} is not a convolution, in contrast to
the Fourier case, \eq\eqref{eq:conv}.  Second, at the discontinuity
$P_{0\to 4}(0;\lambda)=0$ is not the average (equal to $1/2$) of the
left and right limit of the original Poisson curve $e^{-p}$ (put to
zero for negative $p$).

\section{Explicit calculation of spacing distributions}
\label{app:calc}

\subsection{Poisson to GSE}
\label{poisgseapp}

We start from \eq\eqref{generalspacingformagain}, transform $c_0,
\ldots, c_3$ to spherical coordinates with $c^2=c_\mu c_\mu$, and
introduce $u = a + b$ and $t = a - b$.  This yields
\begin{align}
  I(S)&\propto\int_0^\infty\!\!
  dp\,dc\,c^3\int_{-\infty}^\infty\!\! du\,dt\, 
  e^{-p-\frac14(u^2+t^2)+\frac{p(u-t)}{2\lambda}
    -\frac{p^2}{2\lambda^2}-c^2}\notag\\
  &\quad\times\delta\left(S-\sqrt{t^2+4c^2}\right)\notag\\
  &\propto\int_0^\infty dp\,dc\,c^3\int_{-\infty}^\infty dt\,
  e^{-p-\frac{pt}{2\lambda}-\frac{p^2}{4\lambda^2}-\frac14(t^2+4c^2)}\notag\\
  &\quad\times\delta\left(S-\sqrt{t^2+4c^2}\right),
\end{align}
where in the last step we have integrated over $u$.  We now use the
$\delta$-function to integrate over $c$, resulting in
\begin{align}
  I(S)&\propto S\,e^{-\frac{S^2}4}
  \int_{0}^{\infty}dp\,e^{-\frac{p^2}{4\lambda^2}-p}
  \int_{-S}^{S}dt\,\left(S^2-t^2\right)e^{-\frac{pt}{2\lambda}}\notag\\
  &\propto S^4\,e^{-\frac{S^2}4}
  \int_{0}^{\infty}dp\,e^{-\frac{p^2}{4\lambda^2}-p}
  \,\frac{z\cosh{z}-\sinh{z}}{z^3}\notag\\
  &\equiv J(S)\,,\label{eq:IS1}
\end{align}
where $z=pS/2\lambda$.
  
The normalized level spacing distribution $P_{0\to 4}(\s)$,
\eq\eqref{poissympresultform}, is obtained from $J(S)$ by rescaling
and normalization, i.e., $P_{0\to4}(\s)= C\cdot J(2D\s)/(2D)^4$.
Defining the moments of the distribution,
\begin{align} 
  \mI_n=\int_0^\infty dS\,S^n J(S)\,,
\end{align}
we obtain from \eq\eqref{eq:norm}
\begin{align}
 D = \frac{\mI_1}{2\mI_0}\quad\text{and}\quad C = \frac{(2D)^5}{\mI_0}\,.
\end{align}
Explicit evaluation of $\mI_0$ and $\mI_1$ gives
\begin{align}
  \mI_0&=4\sqrt\pi\,, \\
  \mI_1&= 4\lambda \int_0^{\infty}d x\,e^{-2\lambda x}\notag\\
  &\quad\times\frac{(4x^3 \!+\! 2x)e^{-x^2} \!+\!
    \sqrt{\pi}(4x^4\!+\!4x^2\! -\! 1)\erf(x)}{x^3}\,,
\end{align}
from which we obtain \eqs\eqref{eq:D} and \eqref{eq:C}. 

\subsection{GOE to GSE}
\label{goegseapp}

We consider the matrix $H$ in \eq\eqref{eq:Hgoegse}.  With a small
change in notation for $H_1$, we have
\begin{align}
  &H=H_1\otimes\1_2+\lambda H_4=
  \begin{pmatrix}
    A&0&C&0\\
    0&A&0&C\\
    C&0&B&0\\
    0&C&0&B
  \end{pmatrix}\notag\\
  &+\lambda
  \begin{pmatrix}
    a&0&c_0+ic_3&c_1+ic_2\\
    0&a&-c_1+ic_2&c_0-ic_3\\
    c_0-ic_3&-c_1-ic_2&b&0\\
    c_1-ic_2&c_0+ic_3&0&b
  \end{pmatrix},
  \label{eq:HO+HS}
\end{align}
where the variances of the random variables are given by
\eq\eqref{standardnormform}.  If two variables are Gaussian
distributed with variances $\sigma_1^2$ and $\sigma_2^2$, their sum is
again Gaussian distributed with variance $\sigma_1^2+\sigma_2^2$.
Since $H$ depends on $A,B,C$ and $a,b,c_0$ only through the
combinations $A+\lambda a$, $B+\lambda b$, $C+\lambda c_0$, we can
immediately integrate out $A,B,C$, with the corresponding change in
the variances of $a,b,c_0$.  To simplify the notation, we absorb
$\lambda$ in $H_4$ and divide all matrix elements of $H$ by
$\sqrt{1+\lambda^2}$.  This yields a problem equivalent to
\eq\eqref{eq:HO+HS},
\begin{align}
  \label{ortsympensform}
  H\to
  \begin{pmatrix}
    a&0&c_0+ic_3&c_1+ic_2\\
    0&a&-c_1+ic_2&c_0-ic_3\\
    c_0-ic_3&-c_1-ic_2&b&0\\
    c_1-ic_2&c_0+ic_3&0&b
  \end{pmatrix}
\end{align}
with 
\begin{align}
  \label{eq:defsigma}
  \sigma_{a,b}^2=2\sigma_{c_0}^2=1\,,\quad
  2\sigma_{c_i}^2=\frac{\lambda^2}{1+\lambda^2}\equiv\sigma^2\,,
\end{align}
where $i=1,2,3$.  The matrix in \eq\eqref{ortsympensform} has two
non-degenerate eigenvalues whose spacing is given by
\begin{equation}
  S = \bigg[(a-b)^2 + 4\sum_{\nu=0}^3 c_\nu c_\nu\bigg]^{1/2},
\end{equation}
where we have again written $S$ instead of $s$ since we still need to
enforce the normalizations \eqref{eq:norm}.
The spacing distribution is proportional to the integral
\begin{align}
  I(S) &= \int_{-\infty}^{\infty} da\,db\,dc_0dc_1dc_2dc_3\,
  e^{-\frac12(a^2+b^2+2c_0^2)-\frac{c_ic_i}{\sigma^2}}\notag\\       
  &\quad\times\delta\left(S-\sqrt{(a-b)^2 +4c_0^2+4c_j c_j}\right), 
\end{align}
where repeated indices indicate a sum over $i$ and $j$ from $1$ to $3$.
We now transform $c_1$, $c_2$, $c_3$ to spherical coordinates with
$c^2 = c_i c_i$ and introduce $u = a + b$ and $t = a - b$.  This yields
\begin{align}
  I(S)&\propto\int_{-\infty}^{\infty}du\,dt\,dc_0dc \,
  c^2\,e^{-\frac14(u^2+t^2+4c_0^2)-\frac{c^2}{\sigma^2}}\notag\\
  &\quad\times\delta\left(S-\sqrt{t^2 + 4c_0^2 + 4c^2}\right).
\end{align}
The integral over $u$ can be performed trivially and only results in a
prefactor.  Using the $\delta$-function to integrate over $c$, we
obtain
\begin{align}
  I(S)&\propto\int_{0}^{\infty}dt\,dc_0\,
  e^{-\frac14(t^2+4c_0^2)-\frac{1}{4\sigma^2}(S^2-t^2-4c_0^2)}\notag\\
  & \quad\times S\sqrt{S^2-t^2-4c_0^2}\:
  \theta\left(S^2-t^2-4c_0^2\right),
  \label{eq:IStc0}
\end{align}
where we have used the symmetries of the integrand to raise the lower
limit of the integrations to zero.  We now perform the transformation
\begin{align}
  t=Sx\quad\text{and}\quad  c_0=\frac12Sy\sqrt{1-x^2}
\end{align}
with Jacobian $\frac12 S^2\sqrt{1-x^2}$.  Since $t$ and $c_0$ are
non-negative, so are $x$ and $y$.  The $\theta$-function in
\eq\eqref{eq:IStc0} then implies $0\le x,y\le1$.  Reinserting the
definition of $\sigma^2$ from \eq\eqref{eq:defsigma}, we obtain
\begin{align}
  I(S) \propto S^4 &\int_{0}^{1}dx\,dy\,(1-x^2)\sqrt{1-y^2}\notag\\
  &\times e^{-\frac{S^2}{4\lambda^2}[\lambda^2+(1-x^2)(1-y^2)]}\,.
  \label{ortsympbeforeform} 
\end{align}
We now substitute $y = \cos\phi$, note that $\cos^2\phi =
\frac12(1-\cos2\phi)$, and use the integral representation
\cite[\eq(9.6.19)]{Abram:1964} of the modified Bessel functions $I_0$
and $I_1$ to obtain after some algebra
\begin{align}  
  I(S) &\propto S^4 e^{-\frac{1+2\lambda^2}{8\lambda^2}S^2}
  \int_{0}^{1}dx\,(1-x^2)\, e^{\frac{S^2x^2}{8\lambda^2}}
  \left[I_0(z)-I_1(z)\right]\notag\\
  &\equiv J(S)\label{ortsympresultlambdaform} 
\end{align}
with $z=(1-x^2)S^2/(8\lambda^2)$. This corresponds to
\eq\eqref{ortsympresultform} with $S=\sqrt8\lambda Ds$. The properly
normalized spacing distribution is therefore given by
$P_{1\to4}(s)=CJ(\sqrt8\lambda Ds)/(\sqrt8\lambda D)^4$.  Defining
\begin{align}
  \mI_n&=\int_0^\infty dS\,S^n J(S)
\end{align}
we obtain from \eq\eqref{eq:norm}
\begin{align}
  D=\frac{\mI_1}{\sqrt8\lambda \mI_0}\quad\text{and}\quad
  C=\frac{(\sqrt8\lambda D)^5}{\mI_0}\,.
\end{align}
Explicit evaluation of $\mI_0$ and $\mI_1$ gives
\begin{align}
  \mI_0&=8\sqrt\pi\left(\frac{\lambda^2}{1+\lambda^2}\right)^{3/2},\\
  \mI_1&=16\lambda^3\left[\frac{\lambda(1-\lambda^2)}{(1+\lambda^2)^2}
    +\arccot\lambda\right],
\end{align}
from which we obtain \eqs\eqref{eq:Dgoegse} and \eqref{eq:Cgoegse}.

\section{Perturbation of a large GSE matrix by a non-\selfdual\ matrix}
\label{pertmatrix}

We consider a mixed $2N\times 2N$ matrix that interpolates between the
GSE and one of the other Gaussian ensembles,
\begin{equation}\label{degenpertmatrapp}
H = H_4 + \frac{\Lambda}{\rho_4(0) \bar s_4} H_{\beta'}\,,
\end{equation}
where $H_4$ is taken from the GSE and $H_{\beta'}$ from the GOE or
GUE.  We study this matrix for large $N$ in first-order degenerate
perturbation theory to show similarities between the two different
perturbations and to make a connection to the case of GSE to
non-\selfdual\ GUE for $N=2$, which was treated in
\sect\ref{sympgaussec}.

Degenerate perturbation theory predicts that each of the $N$ previously
degenerate eigenvalue pairs splits up and that the shifts of the two
members of the pair are the eigenvalues of the matrix
\begin{equation}
  \frac{\Lambda}{\rho_4(0) \bar s_4}\,M_{ij}\text{ with } M_{ij} = \langle
  \psi_i \left| H_{\beta'} \right| \psi_j \rangle\,;\; i, j = 1, 2\,.
\end{equation}
The $|\psi_{1,2}\rangle$ are the orthonormal eigenvectors of the
unperturbed matrix $H_4$ that span the degenerate subspace of the
eigenvalue pair under consideration.

We show in the following that $M$ is a \two\ GUE matrix for $\beta' =
2$ as well as for $\beta' = 1$, in the latter case with a
normalization different from \eq\eqref{standardnormform}.

\subsection[GUE]{GUE}
\label{pertmatrixgue}

This case is very simple, because the GUE is invariant under unitary
transformations, which contain the symplectic transformations. This
means that the transformation diagonalizing the GSE matrix $H_4$ can
be absorbed in $H_2$ without loss of generality, and therefore one can
choose $|\psi_i\rangle_k = \delta_{ik}$ with $i = 1,2$ and
$k=1,\ldots,2N$.  Thus, we obtain
\begin{equation}
  M_{ij} = \sum_{k,l = 1}^{2N} \delta_{ik}\ (H_2)_{kl}\ \delta_{lj} =
  (H_2)_{ij}\,, 
\end{equation}
which is obviously a \two\ matrix from the GUE with the usual
normalization, \eq\eqref{standardnormform}.  As this holds also for
$N=2$, it is a perturbative explanation for the fact that in the limit
$\lambda\to0$ the spacings between previously degenerate eigenvalues
are distributed exactly like the ones of \two\ GUE matrices.

\subsection[GOE]{GOE}
\label{pertmatrixgoe}
We will show that in this case $M$ is again a matrix from the GUE with
the only difference that the variances of its elements are only half
as large as in the previous subsection. This case is a bit more
involved because one cannot generally diagonalize a \selfdual\ matrix
by an orthogonal transformation (which would preserve the probability
distribution of $H_1$), and thus it is impossible to choose the
eigenvectors of $H_4$ as in the previous subsection.  Explicitly, the
matrix elements read
\begin{align}
  \!\!\! M_{ij} = \langle \psi_i| H_1 |\psi_j\rangle
  &= \langle \psi_i\re| H_1 |\psi_j\re\rangle + \langle \psi_i\im| H_1
  |\psi_j\im\rangle\\ 
  &\quad+ i\left(\langle \psi_i\re| H_1 |\psi_j\im\rangle - \langle
    \psi_i\im| H_1 |\psi_j \re\rangle\right),\notag
\end{align}
where we split the eigenvectors $|\psi_i\rangle$ in real and imaginary
parts: $|\psi_i\rangle = |\psi_i\re\rangle + i|\psi_i\im\rangle$, and
$H_1$ is real.

We will now show that the four vectors $|\psi_1\re\rangle$,
$|\psi_1\im\rangle$, $|\psi_2\re\rangle$, and $|\psi_2\im\rangle$ are
orthogonal in the limit of infinite matrix size. For some combinations
of them one can show this also for finite $N$ using the quaternionic
structure of the eigenvectors,
\begin{equation}
\left(\begin{matrix} \langle\psi_1| \\ \langle\psi_2|\end{matrix}\right) =
\left(\begin{matrix} q_1 & q_2 & \cdots & q_N\end{matrix}\right),
\end{equation}
with quaternions in matrix representation
\begin{equation}
q_k = \left(\begin{matrix} q^{(0)}_k + iq^{(3)}_k & q^{(1)}_k + iq^{(2)}_k \\
-q^{(1)}_k + iq^{(2)}_k & q^{(0)}_k - iq^{(3)}_k\end{matrix}\right)\,.
\end{equation}
One can read off immediately that
\begin{align}\label{scalarquat1}
  \langle\psi_1\re|\psi_2\re\rangle &= \langle\psi_1\im|\psi_2\im\rangle = 0\,, \\
  \langle\psi_1\re|\psi_1\im\rangle &= -\
  \langle\psi_2\re\left.\right|\psi_2\im\rangle\,, \\
  \langle\psi_1\re|\psi_2\im\rangle &=
  \langle\psi_1\im|\psi_2\re\rangle\,, \label{scalarquat4} 
\end{align}
i.e., there are only two independent scalar products.

Let us assume that for large $N$ the $q_k^{(\rho)}$ can be treated as
independent random variables with mean value zero.  Then the mean
values of those scalar products are zero as well, e.g.,
\begin{align}
 \big\langle\langle\psi_1\re|\psi_1\im\rangle\big\rangle = 
 \sum_{k=1}^N\big\langle q^{(0)}_k q^{(3)}_k + q^{(1)}_k q^{(2)}_k\big\rangle = 0\,,
\end{align}
where the outer angular brackets indicate an average over the random
matrix ensemble.  From the normalization of the eigenvectors
$|\psi_i\rangle$ the variances of the $q_k^{(\rho)}$ are proportional
to $1/N$. This yields for the variances of the scalar products
\begin{align}
  \big\langle\langle\psi_1\re|\psi_1\im\rangle^2\big\rangle
  &=\sum_{k=1}^N \Big(\!\big\langle [q^{(0)}_k]^2\big\rangle
  \big\langle[q^{(3)}_k]^2\big\rangle +
  \big\langle[q^{(1)}_k]^2\big\rangle
  \big\langle[q^{(2)}_k]^2\big\rangle \!\Big)\notag\\
  &\propto \sum_{k=1}^N \frac1{N^2} = \frac1N\label{eq:scalarprod}
\end{align}
and likewise for $\langle\psi_1\re|\psi_2\im\rangle$.  Since in the
$N\to\infty$ limit both the mean values and the variances of the
scalar products vanish, the four vectors become orthogonal in this
limit for every single realization of the random matrix.  We have
checked this numerically, which implies that the assumption of the
independence of the $q_k^{(\rho)}$ was valid.

As for the normalization of the four vectors, the squared norms of the
real and imaginary parts agree on average and sum up to 1 due to the
normalization of the eigenvectors $|\psi_i\rangle$.  Invoking the
central limit theorem, we observe that in the limit $N\to\infty$ the
norms of the real and imaginary parts equal $1/\sqrt2$ even for a
single realization of the random matrix.  Hence, multiplying the four
real vectors $|\psi_1\re\rangle$, $|\psi_1\im\rangle$,
$|\psi_2\re\rangle$, and $|\psi_2\im\rangle$ by $\sqrt{2}$, one
obtains, in the limit $N\to\infty$, an orthonormal real basis in the
subspace under consideration.

Finally, we use the fact that the matrix elements of a GOE matrix
$H_1$ are independent random numbers \emph{in every orthonormal (real)
  basis}, with variances 1 and $1/2$ on and off the diagonal,
respectively. Thus we conclude that the $M_{ij}$ are also independent
random numbers with variances
\begin{align}
\big\langle \left[M_{11}\re\right]^2\big\rangle &=
\big\langle \left[M_{22}\re\right]^2\big\rangle = \frac12\,,\\
\big\langle \left[M_{12}\re\right]^2\big\rangle &=
\big\langle \left[M_{12}\im\right]^2\big\rangle = \frac14\,.
\end{align}
These are half the variances of a GUE matrix, which is equivalent to a
multiplication of each element of $M$ by $1/\sqrt{2}$. This explains
the rescaling of the coupling parameter in the definitions of
$P^1_{4\to1}(s_1; \lambda)$ and $P^2_{4\to1}(s_2; \lambda)$,
\eqs\eqref{sympgausgeneralform1} and \eqref{sympgausgeneralform}.

For small $N$ the argumentation in this section does not work.
Presumably, this is the reason why the spacing distributions for the
transition from GSE to GOE differ from those for the transition from
GSE to GUE in the case of \four\ matrices (not shown in this paper,
but checked numerically), whereas they match very well for large
matrices.

\section{Perturbative calculation of the relation between eigenvalue
  density and coupling parameter}
\label{pertcalc}

We consider a diagonal Poissonian matrix $H_0$ perturbed by a matrix
taken from one of the Gaussian ensembles $H_{\beta'}$,
\begin{equation}\label{eigdenstolambdamatrix}
H = H_0 + \alpha H_{\beta'}\,,
\end{equation}
where $H_{\beta'}$ is chosen in the usual normalization, see
\eq\eqref{standardnormform}.  The calculations are done for arbitrary
matrix dimension, which will be sent to infinity at the end. We denote
the number of generically non-degenerate eigenvalues by $N$, i.e., we
consider $N\times N$ matrices.  If $H_{\beta'}$ is taken from the GSE,
these are quaternion valued and correspond to complex $2N\times 2N$
matrices. 

To obtain an $N$-independent eigenvalue density of the Poissonian
ensemble, we define the probability distribution of the individual
eigenvalues $\theta_i$ of $H_0$ by
\begin{equation}
  \mathcal P_0(\theta_i) = \frac{1}N \hat{\mathcal P}_0 (\theta_i/N)\,, 
\end{equation}
where $\hat{\mathcal P}_0$ is some $N$-independent probability
distribution.  Both $\mathcal P_0$ and $\hat{\mathcal P_0}$ are
normalized to one.  The eigenvalue density of the Poissonian ensemble
is thus
\begin{align}
  \rho_0(\theta)=N\mathcal P_0(\theta)=\hat{\mathcal P}_0(\theta/N)
  =\hat{\mathcal P}_0(\hat\theta)\,,
\end{align}
where we have defined $\hat\theta=\theta/N$.  Generically we have
$\theta_i=\mathcal O(N)$ and $\hat\theta_i=\mathcal O(1)$.

We now consider a fixed spacing $S$ between two adjacent eigenvalues
of $H_0$, $\theta_1$ and $\theta_2 = \theta_1 + S$.  The remaining
eigenvalues have to reside outside the interval $(\theta_1,
\theta_2)$.  This results in the conditional probability distribution
\begin{align}
  \mathcal P_0^{\text{out}}(\theta_i) &= \frac{1}N \hat{\mathcal
  P}^\text{out}_0 (\theta_i/N)\notag\\ 
  &=
  \begin{cases} 0 & \text{for } \theta_i\in (\theta_1, \theta_2)\,,\\
    \frac{\mathcal P_0(\theta)}{1-\int_{\theta_1}^{\theta_2} d
      \theta'\, \mathcal P_0(\theta')} & \text{otherwise}\,.
\end{cases}
\end{align}
The eigenvalue density is assumed to be almost unaffected by the
perturbation, which is confirmed in \fig\ref{eigdenspoisgausfigure}
(top).  Of course, this assumption is expected to hold only for
small values of the coupling parameter.

We want to calculate the effect of the perturbation on the spacing
$S$.  If the remaining eigenvalues of $H_0$ are close to $\theta_1$ or
$\theta_2$ we have to apply almost-degenerate perturbation theory.  Up
to second order in $\alpha$ we obtain for the perturbation of the
spacing
\begin{align}
  \Delta S &= \underbrace{\left(\EVD\left[\left(H_0+\alpha
          H_{\beta'}\right)_{kl|\theta_k,\theta_l \in W}\right]
      - S\right)}_{\text{first-order almost-degenerate pert. theory}}\notag\\
  &\quad+ \underbrace{\alpha^2 \sum_{i=3|\theta_i\notin W}^N
    \left(\frac{\left|(H_{\beta'})_{2i}\right|^2}{\theta_2-\theta_i}
      -\frac{\left|(H_{\beta'})_{1i}\right|^2 }
      {\theta_1-\theta_i}\right)}_{\text{second-order perturbation
      theory}}\,,
  \label{eq:deltaS}
\end{align}
where the absolute values are taken with respect to the
real/complex/quaternionic standard norm, $\EVD$ denotes the difference
of the two eigenvalues of the matrix $\left(H_0+\alpha
  H_{\beta'}\right)_{kl}$ that correspond to the unperturbed
eigenvalues, and $W$ is the interval in which eigenvalues have to be
considered almost degenerate with $\theta_1$ or $\theta_2$.  This is
defined by the eigenvalue range $(\theta_1 - C_W,\theta_2 + C_W)$,
where we choose $C_W = C_W^{(0)}N^\varepsilon \alpha$ with
$0<\varepsilon<1$ and $C_W^{(0)} > 1$. This choice ensures that the
closest possible eigenvalue outside $W$ cannot give a second-order
contribution of lower order in $\alpha$ than the almost-degenerate
part.\footnote{An eigenvalue $\theta_i = \theta_2+C_W$ at the border
  of $W$ yields a second-order shift of $\Delta \theta_2 =
  -\alpha^2|(H_{\beta'})_{i2}|^2/(C_W^{(0)}N^\varepsilon\alpha) =
  -\alpha|(H_{\beta'})_{i2}|^2/(C_W^{(0)}N^\varepsilon)$.} Note that
the ``degenerate window'' $W$ grows with $N$.  Therefore arbitrarily
distant eigenvalues are considered almost degenerate in the limit
$N\to\infty$, which is justified because almost-degenerate
perturbation theory is valid for any difference of eigenvalues.

Considering the first-order contribution, we have to deal with the
matrix
\begin{equation}
  M_{kl} = (H_0)_{kl}+\alpha (H_{\beta'})_{kl} = \theta_k
  \delta_{kl}+\alpha (H_{\beta'})_{kl}\,, 
\end{equation}
where the indices $k$ and $l$ run over all values for which the
eigenvalues $\theta_k$ and $\theta_l$ are localized in $W$, which
includes at least $\theta_1$ and $\theta_2$. This is a matrix taken
from the Poissonian ensemble perturbed by a matrix taken from one of
the Gaussian ensembles, but unlike $H$ defined in
\eq\eqref{eigdenstolambdamatrix} it has a constant eigenvalue density
in the limit $N\to \infty$. To show this, we first consider the
density at the lower end of the interval $W$,
\begin{align}
  &\lim_{N\to\infty} N \mathcal P_0(\theta_1-C_W) 
  = \lim_{N\to\infty} \hat{\mathcal P}_0\left(\frac{\theta_1}N
    -\frac{C_W}N\right)\\
  &= \lim_{N\to\infty} \hat{\mathcal P}_0
  \left(\hat\theta_1-C_W^{(0)}N^{\varepsilon-1}\alpha\right) 
  = \hat{\mathcal P}_0(\hat\theta_1) = \rho_0(\theta_1)\,.\notag
\end{align}
This is the same as the eigenvalue density at the other end of $W$,
\begin{align}
  &\!\!\!\lim_{N\to\infty} N \mathcal P_0(\theta_2+C_W) 
  = \lim_{N\to\infty} \hat{\mathcal P}_0
  \left(\frac{\theta_1}N+\frac{S+C_W}N\right)\\ 
  &\!\!\!= \lim_{N\to\infty} \hat{\mathcal P}_0\left(\hat\theta_1+\frac{S}N +
    C_W^{(0)}N^{\varepsilon-1}\alpha\right)  
  = \hat{\mathcal P}_0(\hat\theta_1) = \rho_0(\theta_1)\,.\notag
\end{align}
Thus the spectrum of $M$ can be unfolded by multiplying with the
local eigenvalue density,
\begin{equation}
  \rho_0(\theta_1)\,M_{kl} =
  \underbrace{\rho_0(\theta_1)\theta_k\delta_{kl}}_{\text{unfolded}}
  \quad+\quad 
  \underset{\makebox[0mm]{\scriptsize$\text{effective coupling}$}}
  {\underbrace{\alpha \rho_0(\theta_1)}} (H_{\beta'})_{kl}\,.
\end{equation}
Therefore we can define a new effective coupling parameter
that solely determines the magnitude of the perturbation as in
\sect\ref{largespectrapert}.

The second-order contribution to $\Delta S$ in \eq\eqref{eq:deltaS} is
a sum of at most $N-2$ independent random numbers.  As all of these
random numbers have the same distribution we pick out one of them,
\begin{equation}
  x = \alpha^2
  \left(\frac{b}{\theta_2-\theta_i} -
    \frac{a}{\theta_1-\theta_i}\right)\text{ with }\theta_i \notin W\,,
\end{equation}
where we defined $a=\left|(H_{\beta'})_{1i}\right|^2$ and
$b=\left|(H_{\beta'})_{2i}\right|^2$.  Its probability distribution is
given by
\begin{align}
  &\mathcal P_x(x) = \left[\int_{-\infty}^{\theta_1-C_W} +
    \int_{\theta_2+C_W}^{\infty}\right] 
  d\theta\,\frac1N\, \hat{\mathcal P}_0^\text{out}(\theta/N)\\
  &\!\times\int_{0}^{\infty} \! da\, db\,
  \mathcal P_{\beta'}(a)\mathcal P_{\beta'}(b)\,\delta\!
  \left[x- \alpha^2 \!\left(\frac{b}{\theta_2\!-\!\theta} -
      \frac{a}{\theta_1\!-\!\theta}\right)\right],\notag 
\end{align}
where we renamed $\theta_i=\theta$ for convenience. The distribution
$\mathcal P_{\beta'}$ depends on the symmetry class of the perturbing
ensemble ($a$ and $b$ are squared sums of $\beta'$ Gaussian random
variables).  The moments of this distribution are
\begin{equation}
  p_m = \int_{-\infty}^{\infty} dx\, \mathcal P_x(x)\, x^m\,. 
\end{equation} 
After a short calculation, we obtain
\begin{align}
  p_m &= \int_{-\infty}^{0} d\theta\int_{0}^{\infty} da\, db\,
  \frac1N \Bigg[\hat{\mathcal P}_0^\text{out}
  \left(\frac{\theta-C_W^{(0)}\alpha}N +
    \hat\theta_1\right)\notag\\ 
  &\quad+\hat{\mathcal P}_0^\text{out}\left(\frac{S-\theta-C_W^{(0)}\alpha}N +
    \hat\theta_1\right)\Bigg] \mathcal P_{\beta'}(a)
  \mathcal P_{\beta'}(b)\notag\\ 
  &\quad\times \left[\alpha^2
    \left(\frac{b}{S+C_W^{(0)}N^\varepsilon\alpha-\theta} -
      \frac{a}{C_W^{(0)}N^\varepsilon\alpha-\theta}\right)\right]^m.
\end{align}
In the limit $N\to\infty$, all terms that are divided by $N$ in the
arguments of $\hat{\mathcal P}^{\text{out}}_0$ can be neglected. This
can be done in spite of $\theta$ being integrated to $\infty$, because
the last part of the integrand (in square brackets) suppresses the
large-$\theta$ region and because $\hat{\mathcal P}^{\text{out}}_0$ is
a probability density that has to converge to $0$ for large argument.
Also, $\lim_{N\to\infty}\hat{\mathcal
  P}^{\text{out}}_0(\hat\theta_1)=\hat{\mathcal
  P}_0(\hat\theta_1)=\rho_0(\theta_1)$.  We thus obtain
\begin{align}
  p_m &= \frac{2\rho_0(\theta_1)}N \int_{-\infty}^{0}
  d\theta\int_{0}^{\infty} da\, db \,\,
  \mathcal P_{\beta'}(a)\,\mathcal P_{\beta'}(b)\\
  &\quad\times\left[\alpha^2
    \left(\frac{b}{S+C_W^{(0)}N^\varepsilon\alpha - \theta} -
      \frac{a}{C_W^{(0)}N^\varepsilon\alpha -
        \theta}\right)\right]^m. \notag
\end{align}
Let us denote the second line of \eq\eqref{eq:deltaS} by $\Delta
S^{(2)}$.  It is $\mathcal O(N p_1)$, and therefore its mean value
becomes zero for $N\to\infty$, as it is suppressed by
$N^{-\varepsilon}$. The same holds for the second moment of $\Delta
S^{(2)}$, which goes like $N^{-2\varepsilon}$. Thus the distribution
of $\Delta S^{(2)}$ is a delta function at zero, and we can neglect
its contribution to the perturbation of the spacing. The linear
relation between eigenvalue density and coupling parameter could hence
be shown up to second-order perturbation theory.

\section{Method for fits to the surmises}
\label{fittingmethod}

Since most of the analytical formulas for the small matrices contain
integrals, it takes some time to compute them numerically. In order to
get good fits to data in a reasonable time, a list of $1000$
$\lambda$-values in the interval $(0.01,10)$ was created, with
\begin{equation}
\lambda_i = 0.01 \cdot 1000^{\frac{i-1}{999}}\,;\quad i = 1,\ldots,1000\,.
\end{equation}
For each $\lambda_i$ and each surmise, the corresponding spacing
distribution was stored.  The pure cases $\lambda=0$ and
$\lambda=\infty$ were included as well.

As a measure of the fit quality, we use the $L_2$-distance
\begin{align}
  \Delta_2=\bigg\{\int dx\,[f(x)-g(x)]^2\bigg\}^{1/2}
\end{align}
between the fit and the numerical data.  The fitting was done by
calculating the $\Delta_2$ value of each spacing distribution in the
list.  From the one resulting in the smallest $\Delta_2$ we read off
the coupling $\lambda$.  Note that the largest $\Delta_2$ we encounter
in all the fits is $0.019$.  For comparison, the $L_2$-norms of the
pure Wigner surmises $P_{\beta}(s)$ range from $0.71$ for $\beta=0$ to
$0.94$ for $\beta=4$.  We give no error bars, because the statistical
errors of $\lambda$ obtained by methods such as Jackknife were
negligibly small.  This is also the reason why we use $\Delta_2$
instead of a statistical quantity like chi-squared as a measure of
the fit quality.

\section{Construction of a \selfdual\ GUE}
\label{selfdualGUE}

In the following we construct a Hermitian, \selfdual\ $2N \times 2N$
matrix whose eigenvalues are twofold degenerate and whose
non-degenerate eigenvalues correspond to those of a matrix from the
GUE.  We start with a matrix $M$ that contains an $N \times N$ GUE
matrix $H$ and its complex conjugate (equal to the transpose),
\begin{equation}
M = \begin{pmatrix} H & \0_N \\ \0_N & H^* \end{pmatrix}.
\end{equation}
The eigenvalues of $M$ are obviously those of $H$, but now twofold
degenerate as desired.  However, $M$ is not self-dual.  To transform
$M$ into a self-dual matrix without changing its eigenvalues, we apply
an orthogonal transformation
\begin{equation}
O = \begin{pmatrix}\begin{smallmatrix} 1&0&0&0&\cdots&0&0 \\ 0&0&0&0&\cdots&0&0 \\ 0&0&1&0&\cdots&0&0 \\ 0&0&0&0&\cdots&0&0
\\\vdots&\vdots&\vdots&\vdots&\ddots&&\vdots\\0&0&0&0&&1&0\\0&0&0&0&\cdots&0&0\end{smallmatrix}
& \begin{smallmatrix} 0&0&0&0&\cdots&0&0 \\ 1&0&0&0&\cdots&0&0 \\ 0&0&0&0&\cdots&0&0 \\ 0&0&1&0&\cdots&0&0
\\\vdots&\vdots&\vdots&\vdots&\ddots&&\vdots\\0&0&0&0&&0&0\\0&0&0&0&\cdots&1&0\end{smallmatrix} \\ &
\\\begin{smallmatrix} 0&1&0&0&\cdots&0&0 \\ 0&0&0&0&\cdots&0&0 \\ 0&0&0&1&\cdots&0&0 \\ 0&0&0&0&\cdots&0&0
\\\vdots&\vdots&\vdots&\vdots&\ddots&&\vdots\\0&0&0&0&&0&1\\0&0&0&0&\cdots&0&0\end{smallmatrix} &
\begin{smallmatrix} 0&0&0&0&\cdots&0&0 \\ 0&1&0&0&\cdots&0&0 \\ 0&0&0&0&\cdots&0&0 \\ 0&0&0&1&\cdots&0&0
\\\vdots&\vdots&\vdots&\vdots&\ddots&&\vdots\\0&0&0&0&&0&0\\0&0&0&0&\cdots&0&1\end{smallmatrix}\end{pmatrix} = O^T = O^{-1}\,,
\end{equation}
which transforms a matrix by exchanging every $2n$-th row and column
with the $(N + 2n - 1)$-th one.  Each of the four blocks is a square
matrix of dimension $N$.  This is in complete analogy to the
construction of a \selfdual\ \four\ GUE matrix in \sect\ref{unisympsec}.

We now show that the transformed matrix $O^TMO$ is \selfdual, the
condition for which is
\begin{align}
  O^T M O&\overset!=\ J \left(O^T M O\right)^T J^T = J O^T M^T O J^T\notag\\
  \to\quad M &\overset!= O J O M^T O J^T O \label{eq:selfdual}
\end{align}
with
\begin{equation}
  \label{eq:J}
  J = \1_N \otimes \left(\begin{matrix}0 & -1 \\ 1 & 0\end{matrix}\right).
\end{equation}
Multiplying $J$ by $O$ from the left and the right interchanges the
second, forth, \ldots\ with the $(N+1)$-th, $(N+3)$-th, \ldots\ column
and row.  We thus obtain
\begin{align}
  OJO &=\begin{pmatrix} \0_N & -\1_N \\ \1_N & \0_N \end{pmatrix},\notag\\
  OJ^TO &= -OJO = \begin{pmatrix} \0_N & \1_N \\ -\1_N & \0_N \end{pmatrix} 
\end{align}
and hence
\begin{align}
  OJOM^TOJ^TO &= \begin{pmatrix} \0_N & -\1_N \\ \1_N & \0_N \end{pmatrix} 
  \begin{pmatrix} H^T & \0_N \\ \0_N & H \end{pmatrix}
  \begin{pmatrix} \0_N & \1_N \\ -\1_N & \0_N \end{pmatrix}\notag\\
  &=\begin{pmatrix} H & \0_N \\ \0_N & H^T \end{pmatrix} = M\,,
\end{align}
which proves \eq\eqref{eq:selfdual}. $O^T M O$ can therefore be
written as a quaternion matrix with real quaternions and their
conjugates at the transposed position. Each of these quaternions
stands for a matrix of the form
\begin{equation}
  \begin{pmatrix} c_0 + ic_3 & c_1 + ic_2 \\ -c_1 + ic_2 & c_0 -
    ic_3 \end{pmatrix}  
  = \begin{pmatrix} q & p \\ -p^* & q^* \end{pmatrix}
\end{equation}
with complex numbers $p$ and $q$. Evidently, $p$ has to be zero for
each quaternion in $O^T M O$, because our original $M$ generically
contains no element which is the negative complex conjugate of any
other, and we only exchanged elements by applying $O$. This means that
at least half of the matrix elements are zero. In the original $M$,
exactly half of the matrix elements were zero, while the other half
were random variables which depended on a total of $N^2$ real
parameters, so the same has to hold for $O^T M O$. From this and
Hermiticity if follows that every off-diagonal $q$ has to be an
independent complex random number, while the $q$ on the diagonal are
real, so that there are again $N^2$ real degrees of freedom.

With this equivalence proven, one can construct a \selfdual\ GUE
matrix by taking a matrix from the GSE and set its $c_1$ and $c_2$
components to zero.  This matrix has the same joint probability
density of the eigenvalues as an $N\times N$ matrix taken from the
GUE, as it is related to a matrix of the form of $M$ by a fixed basis
transformation.

\smallskip

\bibliography{mixedrmt}

\end{document}